\documentclass[11pt,fleqn]{article}
\usepackage{amsmath,amssymb}

\newcommand{\C}{{\mathbb C}}
\newcommand{\R}{{\mathbb R}}
\newcommand{\N}{{\mathbb N}}
\newcommand{\p}{{\partial}}
\newcommand{\rank}{\mathop{\rm rank}\nolimits}
\newcommand{\vspaceafter}{\raisebox{0ex}[0ex][1.2ex]{\null}}
\newcommand{\vspacebefore}{\raisebox{0ex}[2.5ex][0ex]{\null}}
\newcommand{\dotline}{\vspaceafter\\[-1.7ex]\multicolumn{4}{|c|}{\hspace*{-1.8ex}\dotfill\hspace*{-2ex}}\\[-1ex]\vspacebefore}
\newcommand{\myhline}{\\[0.5ex]\hline\vspacebefore}

\begin{document}

\begin{flushleft}
\Large \bf Realizations of Real Low-Dimensional Lie Algebras
\end{flushleft}

\begin{flushleft} \bf
Roman O. POPOVYCH~$^\dag$,
Vyacheslav M. BOYKO~$^\dag$,\\
Maryna O. NESTERENKO~$^\dag$, and
Maxim W. LUTFULLIN~$^\ddag$
\end{flushleft}

\noindent $\dag$~Institute of Mathematics of NAS of Ukraine, 3
Tereshchenkivska Str., Kyiv 4, 01601 Ukraine\\
$\phantom{\dag}$~E-mail: rop@imath.kiev.ua, boyko@imath.kiev.ua,
maryna@imath.kiev.ua\\
$\phantom{\dag}$~URL: http://www.imath.kiev.ua/\~{}rop/ \\
$\phantom{\dag\mbox{~URL:}}$~http://www.imath.kiev.ua/\~{}boyko/\\
$\phantom{\dag\mbox{~URL:}}$~http://www.imath.kiev.ua/\~{}maryna/

\noindent
$\ddag$~Poltava State Pedagogical University, 2
Ostrogradskoho Str., Poltava, Ukraine\\
$\phantom{\dag}$~E-mail: MWL@pdpu.septor.net.ua

\bigskip

\begin{abstract}
Using a new powerful technique based on the notion of
megaideal, we construct a complete set of inequivalent
realizations of real Lie algebras of dimension no greater than
four in vector fields on a space of an arbitrary (finite) number
of variables. Our classification amends and essentially
generalizes earlier works on the subject.

Known results on classification of low-dimensional real Lie algebras,
their automorphisms, differentiations, ideals, subalgebras and realizations are reviewed.
\end{abstract}

\section{Introduction}

The description of Lie algebra representations by vector
fields was first considered by S.~Lie.
However, this problem
is still of great interest and widely applicable e.g., in
particular, to integrating of ordinary differential
equations~\cite{lie1883,lie1891} (see also some new results and
trends in this area, e.g.,
in~\cite{Anderson&Davison1974,Cerquetelli&Ciccoli&Nucci2002,Edelstein,mahomed&leach1989,
Mahomed&Leach1990,Schmucker&Czichowski1998,
Schwarz2000,Soh&Mahomed2000a,soh&mahomed2001jphysa,soh&mahomed2001jnlm}),
group classification of partial differential
equations~\cite{basarab-horwath&lahno&zhdanov2001,heredero&olver1996,
lahno&spichak&stognii2004,zhdanov&lahno1999},
classification of gravity fields of a general form under the
motion groups or groups of conformal
transformations~\cite{kruchkovich1954,kruchkovich1957,kruchkovich1977,Makaruk1993,petrov1966}.
(See, e.g.~\cite{Gonzales&Kamran&Olver1992a,heredero&olver1996} for other physical applications
of realizations of Lie algebras.)
Thus, without exaggeration this problem
has a major place in modern group analysis of differential equations.

In spite of its importance for applications, the problem of complete
description of reali\-zations has not been solved even for the cases
when either the dimension of algebras or the dimension of
realization space is a fixed small integer. An exception is Lie's
classification of all possible Lie groups of point
transformations acting on the two-dimensional complex or real space without
fixed points~\cite{lie1880,lie1893}, which is equivalent to classification
of all possible realizations of Lie algebras in vector fields on
the two-dimensional complex (real) space (see also \cite{Gonzales&Kamran&Olver1992a}).

In this paper  we construct a complete set of inequivalent realizations of real
Lie algebras of dimension no greater than four in vector fields on
a space of an arbitrary (finite) number of variables.
For solving this problem, we propose a new powerful technique based on the notion of
megaideal.

The plan of the paper is as follows.
Results on classifications of abstract Lie algebras are reviewed in Section~2.
In Section~3 we give necessary definitions and statements on megaideals and realizations of Lie algebras,
which form the theoretical basis of our technique.
Previous results on classifications of realizations are reviewed in Section~4.
In this section we also explain used notations, abbreviations and conventions
and describe the classification technique.
The results of our classification are formulated in the form of Tables~2--6.
An example of  classification of realizations for a four-dimensional algebra is discussed in detail in Section~5.
In Section~\ref{secWafo} we compare our results with those
of~\cite{soh&mahomed2001jphysa}.
Section~\ref{secConl} contains discussion.
In appendix~\ref{SectionOfResultsOnLowDimRealLieAlgebras}
different results on low-dimensional real Lie algebras (classifications, automorphisms, subalgebras etc.)
are collected.

\section{On classification of Lie algebras}

The necessary step to classify realizations of Lie algebras is
classification of these algebras, i.e.\ classification of possible
commutative relations between basis elements. By the Levi--Maltsev
theorem any finite-dimensional Lie algebra over a field of characteristic~0 is
a semi-direct sum (the Levi--Maltsev decomposition) of  the
radical (its maximal solvable ideal) and a semi-simple subalgebra
(called the Levi factor)~(see, e.g., \cite{Jacobson}). This result
reduces the task of classifying all Lie algebras to the following
problems:

1) classification of all semi-simple Lie algebras;

2) classification of all solvable Lie algebras;

3) classification of all algebras that are semi-direct sums
of semi-simple Lie algebras and solvable Lie algebras.

Of the problems listed above, only that of classifying all
semi-simple Lie algebras is completely solved in the well-known
Cartan theorem: any semi-simple complex or real Lie algebra can
be decomposed into a direct sum of ideals which are simple
subalgebras being mutually orthogonal with respect to the
Cartan--Killing form. Thus, the problem of classifying semi-simple
Lie algebras is equivalent to that of classifying all
non-isomorphic simple Lie algebras. This classification is known
(see, e.g., \cite{dynkin,barut&raczka1977}).

At the best of our knowledge,
the problem of classifying solvable Lie algebras is
completely solved only for Lie algebras of dimension up to
and including six (see, for example,
\cite{morozov1958,mubarakzyanov1963.1,mubarakzyanov1963.2,mubarakzyanov1963.3,turkowski1988,
turkowski1990}). Below we  shortly list some results on
classifying of low-dimensional Lie algebras.

All the possible complex Lie algebras of dimension $\leq 4$ were
listed by S.~Lie himself~\cite{lie1893}. In~1918
L.~Bianchi investigated three-dimensional real Lie
algebras~\cite{bianchi1918}.
Considerably later this problem was again considered by
H.C.~Lee~\cite{lee1947} and G.~Vranceanu~\cite{vranceanu1947}, and
their classifications are equivalent to Bianchi's one. Using Lie's
results on complex structures,
G.I.~Kruchkovich~\cite{kruchkovich1954,kruchkovich1957,kruchkovich1977}
classified four-dimensional real Lie algebras which do not contain
three-dimensional abelian subalgebras.

Complete, correct and easy to use classification of real Lie algebras
of dimension $\leq 4$ was first carried out by
G.M.~Mubarakzyanov~\cite{mubarakzyanov1963.1} (see also citation
of these results as well as description of subalgebras and
invariants of real low-dimensional Lie algebras
in~\cite{patera&sharp&winternitz1976,patera&winternitz1977}).
At the same year a variant of such classification was obtained by J.~Dozias~\cite{dozias1963}
and then adduced in~\cite{vergne1972}.
Analogous results are given in~\cite{petrov1966}. Namely, after
citing classifications of L.~Bianchi~\cite{bianchi1918} and
G.I.~Kruchkovich~\cite{kruchkovich1954}, A.Z.~Petrov classified
four-dimensional real Lie algebras containing three-dimensional
abelian ideals.

In \cite{MacCallum1999} M.A.H.~MacCallum proposed an alternative scheme of
classification and numeration for four-dimensional real Lie algebras.
He also reviewed and compared different approaches and results concerning this problem,
which were adduced in the previous papers of
G.M.~Mubarakzyanov~\cite{mubarakzyanov1963.1},
F.~Bratzlavsky~\cite{bratzlavsky1959},
G.I.~Kruchkovich and A.Z.~Petrov~\cite{kruchkovich1954,petrov1966},
Patera et al~\cite{patera&sharp&winternitz1976}.
In preprint~\cite{Andrada2004} authors reproduced the
classification of four-dimensional real Lie algebras and
compared their results with ones by G.M.~Mubarakzyanov~\cite{mubarakzyanov1963.1},
J.~Dozias~\cite{dozias1963} and Patera et al~\cite{patera&sharp&winternitz1976}.

In the series of papers
\cite{mubarakzyanov1963.2,mubarakzyanov1963.3,mubarakzyanov1966.4}
G.M.~Mubarakzyanov continued his investigations of low-dimensional Lie algebras.
He classified five-dimensional real Lie
algebras as well as six-dimensional solvable ones with one linearly
independent non-nilpotent element. Let us note that for
six-dimensional solvable real Lie algebras dimension $m$ of the
nilradical is greater than or equal to~$3$. In the case $m=3$
such algebras are decomposable. Classification of
six-dimensional nilpotent Lie algebras ($m=6$) was obtained by
K.A.~Umlauf~\cite{Umlauf1891} over complex field and generalized
by V.V.~Morozov~\cite{morozov1958} to the case of arbitrary field
of characteristic~0.

J.~Patera, R.T.~Sharp, P.~Winternitz and H.~Zassenhaus  in
\cite{patera&sharp&winternitz1976} revised the classification
of four- and five-dimensional real Lie algebras by G.M.~Mubarakzyanov
and nilpotent sixth-dimensional Lie algebras by V.V. Morozov.

Using the proposed in~\cite{patera&zassenhaus1990a} notion of nilpotent frame
of a solvable algebra, J.~Patera and H.~Zassenhaus~\cite{patera&zassenhaus1990b} classified
solvable Lie algebras of dimension $\leq 4$ over any perfect field.
These results were generalized by W.A. de Graaf~\cite{Graaf2004} to arbitrary fields using
the computer algebra system MAGMA.

In~\cite{turkowski1988,turkowski1992} P.~Turkowski classified all
real Lie algebras of dimension up to 9, which admit non-trivial
Levi decomposition. P.~Turkowski~\cite{turkowski1990} also
completed Mubarakzyanov's classification of six-dimensional
solvable Lie algebras over $\R$, by classifying real Lie
algebras of dimension 6 that contain four-dimensional nilradical
($m=4$).

The recent results and references on seven-dimensional nilpotent Lie
algebras can be found in \cite{seeley1993}.

In the case when the dimension of algebra is not fixed
sufficiently general results were obtained only in classification of
algebras with nilradicals having special structures (e.g., abelian
\cite{Ndogmo&Wintenitz1994a}, Heisenberg~\cite{Rubin&Winternitz1993}, or triangular algebras
\cite{Tremblay&Winternitz1998}) as their nilradical. Invariants of
these algebras, i.e. their generalized Casimir operators, were
investigated in
\cite{Ndogmo2000,Ndogmo&Wintenitz1994b,Tremblay&Winternitz1998,Tremblay&Winternitz2001}.

In Table 1 we summarize  known results about classification Lie algebras in form of the table
($A$~is Lie algebra over the field $P$, $\dim A$ is the dimension of $A$,
$\dim N$ is the dimension of the nilradical of~$A$).

\section{Megaideals and realizations of Lie algebras}

Now we define the notion of megaideal that is useful for constructing
realizations and proving their inequivalence in a simpler way.
Let $A$ be an $m$-dimensional (real or complex) Lie algebra ($m\in\N$) and
let ${\rm Aut}(A)$ and ${\rm Int}(A)$ denote the groups of all the automorphisms of $A$
and of its inner automorphisms respectively.
The Lie algebra of the group ${\rm Aut}(A)$ coincides with the Lie algebra ${\rm Der}(A)$
 of all the derivations
of the algebra $A.$ (A derivation $D$ of $A$ is called a linear mapping from $A$ into itself which
satisfy the condition $D[u,v]=[Du,v]+[u,Dv]$ for all $u,v\in A.$)
${\rm Der}(A)$ contains as an ideal the algebra ${\rm Ad}(A)$ of inner derivations of $A,$
which is the Lie algebra of ${\rm Int}(A).$
(The inner derivation corresponding to $u\in A$ is the mapping $\mathop{\rm ad}\nolimits u\,{:}\:v\rightarrow [v,u].$)
Fixing a basis $\{e_\mu, \mu=\overline{1,m}\}$ in $A,$ we associate an arbitrary linear mapping $l: A \rightarrow A$
(e.g., an automorphism or a derivation of $A$) with a~matrix $\alpha=(\alpha_{\nu\mu})_{\mu,\nu=1}^m$
by means of the expanding $l(e_\mu)=\alpha_{\nu\mu}e_\nu.$
Then each group of automorphisms of $A$ is associated with a subgroup
of the general linear group $GL(m)$ of all the non-degenerated $m\times m$ matrices (over $\R$ or $\C$)
as well as each algebra of derivations of $A$ is associated with a subalgebra
of the general linear algebra $gl(m)$ of all the $m\times m$ matrices.

\medskip

\noindent
{\bf Definition.} We call a vector subspace of $A,$ which is invariant under
any transformation from ${\rm Aut}(A),$ a {\it megaideal} of $A.$

\medskip

Since ${\rm Int}(A)$ is a normal subgroup of ${\rm Aut}(A)$,
it is clear that any megaideal of $A$ is a subalgebra and, moreover, an ideal in $A.$
But when ${\rm Int}(A)\not={\rm Aut}(A)$ (e.g., for nilpotent algebras)
there exist ideals in $A,$ which are not megaideals. Moreover,
any megaideal $I$ of $A$ is invariant with respect to all the derivations of $A$:
${\rm Der}(A)I\subset I,$ i.e. it is a characteristic subalgebra.
Characteristic subalgebras which are not megaideals can exist only if
${\rm Aut}(A)$ is a disconnected Lie group.

Both improper subsets of $A$ (the empty set and $A$ itself) are always megaideals in $A.$
The following lemmas are obvious.

\medskip

\noindent
{\bf Lemma 1.} If $I_1$ and $I_2$ are megaideals of $A$ then so are
$I_1+I_2,$ $I_1\cap I_2$ and $[I_1,I_2]$,
i.e.\ sums, intersections and Lie products of megaideals are also megaideals.

\medskip

\noindent
{\bf Corollary 1.} All the members of the commutator (derived) and
the lower central series of $A$,
i.e. all the derivatives $A^{(n)}$ and all the Lie powers $A^n$
($A^{(n)}=[A^{(n-1)},A^{(n-1)}]$, $A^n=[A,A^{n-1}]$, $A^{(0)}=A^0=A$)
are megaideals in $A.$

\medskip

This corollary follows from Lemma 1 by induction since $A$ is a megaideal in $A.$

\medskip

\noindent
{\bf Corollary 2.} The center~$A_{(1)}$ and all the other members of the upper central series~$\{A_{(n)},\, n\ge0\}$
of~$A$ are megaideals in $A.$

\medskip

Let us remind that $A_{(0)}=\{0\}$ and $A_{(n+1)}/A_{(n)}$ is the center of $A/A_{(n)}$.

\medskip

\noindent
{\bf Lemma 2.} The radical (i.e.\ the maximal solvable ideal) and the nil-radical (i.e.\ the maximal nilpotent ideal)
of $A$ are its megaideals.

\medskip

The above lemmas give a number of invariant subspaces of all the automorphisms in $A$ and,
therefore, simplify calculating ${\rm Aut}(A).$

\medskip

\newpage

\begin{center}
{\bf Table 1.} Short review of results on classification of low-dimensional algebras

\vspace{2mm}
\begin{tabular}{llcccl}
\hline
&&$P$&$\dim A$\ &$\dim N$&Remarks\\[1mm]
\hline
&&&&&\\[-2mm]
Lie~\cite{lie1893} & 1893 & $\C$ &  $\leq 4$ & &\\[1mm]
{\bfseries \itshape Bianchi}~\cite{bianchi1918} & 1918 & $\R$ & 3 && \\[1mm]
Lee~\cite{lee1947} & 1947 & $\R$ & 3 & & \parbox{4cm}{\small Bianchi results}\\[1mm]
Vranceanu~\cite{vranceanu1947} & 1947 & $\R$ & 3 & & \parbox{4cm}{\small Bianchi results} \\[5mm]
Dobresku~\cite{dobrescu1953} & 1953 & $\R$ & 4 && \parbox{4cm}{\small according to \cite{MacCallum1999}} \\[1mm]
Kruchkovich~\cite{kruchkovich1954,kruchkovich1957} & 1954, 1957& $\R$ & 4 &&
\parbox{4cm}{\small without 3-dim.\\ Abelian ideals}\\[5mm]
Bratzlavsky~\cite{bratzlavsky1959} & 1959 & $\R$ & 4 && \parbox{4cm}{\small according to \cite{MacCallum1999}} \\[1mm]
{\bfseries \itshape Mubarakzyanov}~\cite{mubarakzyanov1963.1} & 1963 & $\R,\C$ &4  && \\[1mm]
Dozias~\cite{dozias1963} & 1963 & $\R$ &4  && \parbox{4cm}{\small see \cite{vergne1972}} \\[1mm]
Petrov~\cite{petrov1966} & 1966 & $\R$ &$4$  &&
%\raisebox{-1.3ex}[0ex][0ex]{
\parbox{4cm}{\small completed\\ Kruchkovich's\\ classification}\\[5mm]
Ellis, Sciama~\cite{ellis&sciama1966} & 1966 & $\R$ & 4 && \parbox{4cm}{\small according to \cite{MacCallum1999}}\\[1mm]
MacCallum~\cite{MacCallum1999} & 1979, 1999  & $\R$ & 3, 4 && \\[1mm]
Patera, Zassenhaus~\cite{patera&zassenhaus1990b} & 1990  & \parbox{1.2cm}{\small perfect} & $\leq 4$ && \\[1mm]
Andrada et al~\cite{Andrada2004} & 2004  & $\R$ & 4 && \\[1mm]
de Graaf~\cite{Graaf2004} & 2004  & \parbox{1.4cm}{\small arbitrary}& 4 && \\[1mm]
{\bfseries \itshape Mubarakzyanov}~\cite{mubarakzyanov1963.2} & 1963 & $\R,\C$ & 5 && \\[1mm]
Umlauf~\cite{Umlauf1891} & 1891& $\C$& 6 & 6 & \parbox{4cm}{\small nilpotent}\\[1mm]
{\bfseries \itshape Morozov}~\cite{morozov1958} & 1958 &$\mathop{\rm char}\nolimits=0$  &6 & 6 &
\parbox{4cm}{\small nilpotent} \\[1mm]
{\bfseries \itshape Mubarakzyanov}~\cite{mubarakzyanov1963.3} & 1963 & $\R$ & 6 & 3 &
%\raisebox{0ex}[0ex][0ex]{
\parbox{4cm}{\small solvable\\ ($\Rightarrow$decomposable)}\\[5mm]
{\bfseries \itshape Mubarakzyanov}~\cite{mubarakzyanov1963.3} & 1963 & $\R$ & 6 & 5 &
\parbox{4cm}{\small ($\Rightarrow$solvable)}\\[1mm]
{\bfseries \itshape Turkowski}~\cite{turkowski1990} & 1990 & $\R$ & 6 & 4 &
\parbox{4cm}{\small ($\Rightarrow$solvable)}\\[1mm]
Turkowski~\cite{turkowski1988} & 1988 &  $\R$ & $\le 8$ &&
%\raisebox{-1.5ex}[0ex][0ex]{
\parbox{4cm}{\small admit non-trivial\\ Levi decomposition}\\[5mm]
Turkowski~\cite{turkowski1992} &  1992 &  $\R$ & $9$ && \parbox{4cm}{\small admit non-trivial\\ Levi decomposition}
\\[5mm]
%\raisebox{0ex}[0ex][0ex]{
\parbox{4cm}{Patera et al~\cite{patera&sharp&winternitz1976}} &  1976 &  $\R$ & $\le 6$ &&
%\raisebox{0ex}[0ex][0ex]{
\parbox{4cm}{\small only nilpotent\\ for $\dim A=6$}\\[5mm]
Safiullina~\cite{safiullina1964}  & 1964  & $\R$ & 7  & 7 & \parbox{4cm}{\small nilpotent,\\ according
to \cite{seeley1993}}\\[5mm]
Magnin~\cite{magnin1986}  & 1986 & $\R$ & $\leq 7$  &$\dim A$ & \parbox{4cm}{\small nilpotent,\\ according
to \cite{MacCallum1999}}\\[5mm]
Seeley~\cite{seeley1993}  & 1993& $\R$ & 7  & 7 & \parbox{4cm}{\small nilpotent}\\[1mm]
Tsagas~\cite{tsagas1999}  & 1999 & $\R$ & 8 & 8 & \parbox{4cm}{\small nilpotent}\\[1mm]
Tsagas et al~\cite{tsagas2000}  & 2000 & $\R$ & 9 &9 & \parbox{4cm}{\small nilpotent}\\[1mm]
\hline
\end{tabular}
\end{center}

%\bigskip
\noindent
{\bf Remark.} In the above table we emphasize consequent results
of Bianchi, Mubarakzyanov, Morozov and  Turkowski, which together
form classification of real Lie algebras of dimensions no greater than~6.

\newpage

\noindent {\bf Example 1.} Let $mA_1$ denote the $m$-dimensional
abelian algebra. ${\rm Aut}(mA_1)$ coincides with the group of all
the non-degenerated linear transformations of the $m$-dimensional
linear space (${}\sim GL(m)$) and ${\rm Int}(mA_1)$ contains only
the identical transformation. Any vector subspace in the abelian
algebra $mA_1$ is a subalgebra and an ideal in $mA_1$ and is not a
characteristic subalgebra or a megaideal. Therefore,  the abelian
algebra $mA_1$ do not contain proper megaideals.

\medskip

\noindent {\bf Example 2.} Let us fix the canonical basis
$\{e_1,e_2,e_3\}$ in the algebra $A=A_{2.1}\oplus
A_1$~\cite{mubarakzyanov1963.1}, in which only two first elements
has the non-zero commutator $[e_1,e_2]=e_1.$ In this basis
\[
{\rm Aut}(A) \sim
\left\{\left.\left( \!\begin{array}{ccc}
\alpha_{11}&\alpha_{12}&0\\ 0&1&0\\ 0&\alpha_{32}&\alpha_{33}
\end{array} \right)\;\right|\;\alpha_{11}\alpha_{33}\not=0\right\}, \quad
{\rm Int}(A) \sim \left\{\!\left.\left( \begin{array}{ccc}
e^{\varepsilon_1}&\varepsilon_2&0\\ 0&1&0\\ 0&0&1
\end{array} \right)\:\right|\;\varepsilon_1,\varepsilon_2\in\R\right\},
\]
\[
{\rm Der}(A) \sim \left\langle
\left( \begin{array}{ccc} 1&0&0\\ 0&0&0\\ 0&0&0\end{array} \right),\;
\left( \begin{array}{ccc} 0&1&0\\ 0&0&0\\ 0&0&0\end{array} \right),\;
\left( \begin{array}{ccc} 0&0&0\\ 0&0&0\\ 0&1&0\end{array} \right),\;
\left( \begin{array}{ccc} 0&0&0\\ 0&0&0\\ 0&0&1\end{array} \right)
\right\rangle,
\]
\[
{\rm Ad}(A) \sim \left\langle
\left( \begin{array}{ccc} 1&0&0\\ 0&0&0\\ 0&0&0\end{array} \right),\;
\left( \begin{array}{ccc} 0&1&0\\ 0&0&0\\ 0&0&0\end{array} \right)
\right\rangle.
\]
A complete set of ${\rm Int}(A)$-inequivalent proper subalgebras of $A$ is exhausted
by the following ones~\cite{patera&winternitz1977}:
\[\arraycolsep=0em\begin{array}{l}
\mbox{one-dimensional:}\;\: \langle pe_2+q e_3\rangle, \;
\langle e_1\pm e_3\rangle, \;  \langle e_1\rangle, \ p^2+q^2=1;
\vspace{1mm}\\
\mbox{two-dimensional:}\;\: \langle e_1, e_2+\varkappa e_3\rangle,
\;  \langle e_1, e_3\rangle, \;  \langle e_2, e_3\rangle.
\end{array} \]
Among them only $\langle e_1\rangle,$
$\langle e_3\rangle,$ $\langle e_1, e_3\rangle$ are megaideals,
$ \langle e_1,e_2+\varkappa e_3\rangle$ is an ideal
and is not a characteristic subalgebra (and, therefore, a megaideal).

\medskip

\noindent {\bf Example 3.} Consider the algebra $A=A_{3.4}^{-1}$
from the series $A_{3.4}^a$, $-1\le a<1,$ $a\not=0$~\cite{
mubarakzyanov1963.1}. The non-zero commutators of its canonical
basis elements are $[e_1,e_3]=e_1$ and $[e_2,e_3]=-e_2.$ ${\rm
Aut}(A)$ is not connected for this algebra:
\[
{\rm Aut}(A) \sim
\left\{\left.\left( \begin{array}{ccc}
\alpha_{11}&0&\alpha_{13}\\ 0&\alpha_{22}&\alpha_{23}\\ 0&0&1
\end{array} \right)\;\right|\;\alpha_{11}\alpha_{22}\not=0\right\}\bigcup
\left\{\left.\left( \begin{array}{ccc}
0&\alpha_{12}&\alpha_{13}\\ \alpha_{21}&0&\alpha_{23}\\ 0&0&-1
\end{array} \right)\;\right|\;\alpha_{12}\alpha_{21}\not=0\right\},
\]
\[
{\rm Int}(A) \sim \left\{\!\left.\left( \begin{array}{ccc}
e^{\varepsilon_1}&0&\varepsilon_2\\ 0&e^{-\varepsilon_1}&\varepsilon_3\\ 0&0&1
\end{array} \right)\:\right|\;\varepsilon_1,\varepsilon_2,\varepsilon_3\in\R\right\},
\]
\[
{\rm Der}(A) \sim \left\langle
\left( \begin{array}{ccc} 1&0&0\\ 0&0&0\\ 0&0&0\end{array} \right),\;
\left( \begin{array}{ccc} 0&0&0\\ 0&1&0\\ 0&0&0\end{array} \right),\;
\left( \begin{array}{ccc} 0&0&1\\ 0&0&0\\ 0&0&0\end{array} \right),\;
\left( \begin{array}{ccc} 0&0&0\\ 0&0&1\\ 0&0&0\end{array} \right)
\right\rangle,
\]
\[
{\rm Ad}(A) \sim \left\langle
\left( \begin{array}{ccc} 1&0&0\\ 0&-1&0\\ 0&0&0\end{array} \right),\;
\left( \begin{array}{ccc} 0&0&1\\ 0&0&0\\ 0&0&0\end{array} \right),\;
\left( \begin{array}{ccc} 0&0&0\\ 0&0&1\\ 0&0&0\end{array} \right)
\right\rangle.
\]
A complete set of ${\rm Int}(A)$-inequivalent proper subalgebras of $A$ is exhausted
by the following ones~\cite{patera&winternitz1977}:
\[\arraycolsep=0em\begin{array}{l}
\mbox{one-dimensional:}\;\: \langle e_1\rangle, \;  \langle e_2\rangle, \;  \langle e_3\rangle,
 \; \langle e_1\pm e_2\rangle;
\vspace{1mm}\\
\mbox{two-dimensional:}\;\: \langle e_1,e_2\rangle, \;  \langle e_1, e_3\rangle, \;  \langle e_2, e_3\rangle.
\end{array} \]

Among them only $\langle e_1, e_2\rangle$ is a megaideal, $\langle
e_1\rangle$ and $\langle e_2\rangle$ are characteristic
subalgebras and  are not megaideals.

\medskip

Let $M$ denote a $n$-dimensional smooth manifold and ${\rm
Vect}(M)$ denote the Lie algebra of smooth vector fields (i.e.\
first-order linear differential operators) on $M$ with the Lie
bracket of vector fields as a commutator. Here and below
smoothness means analyticity.

\medskip

\noindent {\bf Definition.} {\it A realization of a Lie algebra
$A$ in vector fields on $M$} is called a homomorphism $R{:}\; A
\rightarrow {\rm Vect}(M).$ The realization is said {\it faithful}
if $\mathop{\rm ker}\nolimits R=\{0\}$ and {\it unfaithful}
otherwise. Let $G$ be a subgroup of ${\rm Aut}(A).$ The
realizations $R_1{:}\; A \rightarrow {\rm Vect}(M_1)$ and
$R_2{:}\; A \rightarrow {\rm Vect}(M_2)$ are called {\it
$G$-equivalent} if there exist $\varphi\in G$ and a diffeomorphism
$f$ from $M_1$ to $M_2$ such that $R_2(v)=f_{*}\,R_1(\varphi(v))$
for all $v\in A.$ Here $f_{*}$ is the isomorphism from ${\rm
Vect}(M_1)$ to ${\rm Vect}(M_2)$ induced by $f.$ If $G$ contains
only the identical transformation, the realizations are called
{\it strongly equivalent.} The realizations are {\it weakly
equivalent} if $G={\rm Aut}(A).$ A restriction of the realization
$R$ on a subalgebra $A_0$ of the algebra $A$ is called {\it
a~realization induced by $R$} and is denoted as $R\bigr|_{A_0}.$

\medskip

Within  the framework of local approach that we use $M$ can be
considered as an open subset of $\R^n$
and all the diffeomorphisms are local.

Usually realizations of a Lie algebra have been classified with respect to the weak equivalence.
This it is reasonable
although the equivalence used in the representation theory is similar to the strong one.
The strong equivalence suits better for construction of
 realizations of algebras using realizations of their subalgebras
and is verified in a simpler way than the weak equivalence. It is
not specified in some papers what equivalence has been used, and
this results in classification mistakes.

To classify realizations of a $m$-dimensional Lie algebra $A$ in the most direct way,
we have to take $m$ linearly independent vector fields of the general form $e_i=\xi^{ia}(x)\p_a,$ where
$\p_a=\p/\p x_a,$ $x=(x_1,x_2,\ldots, x_n)\in M,$ and require
 them to satisfy the commutation relations of $A.$
As a result, we obtain a system of first-order PDEs for the
coefficients~$\xi^{ia}$ and integrate it, considering all the
possible cases. For each case we transform the solution into the
simplest form, using either local diffeomorphisms of the space of
$x$ and automorphisms of $A$ if the weak equivalence is meant or
only local diffeomorphisms of the space of $x$ for the strong
equivalence. A drawback of this method is the necessity to solve
a complicated nonlinear system of PDEs. Another way is
to classify sequentially realizations of a series of nested
subalgebras of $A,$ starting with a one-dimensional subalgebra and
ending up with $A.$

 Let $V$ be a subset of ${\rm Vect}(M)$ and
$r(x)=\dim \langle V(x)\rangle,$ $x\in M.$ $0\le r(x)\le n.$ The
general value of $r(x)$ on $M$ is called the {\it rank} of $V$ and
is denoted as $\rank V.$

\medskip

\noindent
{\bf Lemma 3.} Let $B$ be a subset and  $R_1$ and $R_2$ be realizations of the algebra $A.$
If $R_1(B)$ and $R_2(B)$ are inequivalent with respect to endomorphisms of ${\rm Vect}(M)$
generated by diffeomorphisms on $M$. Then $R_1$ and $R_2$ are strongly inequivalent.

\medskip

\noindent
{\bf Corollary 2.} If there exists a subset $B$ of $A$ such that $\rank R_1(B)\not=\rank R_2(B)$ then
the realizations $R_1$ and $R_2$ are strongly inequivalent.

\medskip

\noindent
{\bf Lemma 4.} Let $I$ be a megaideal and  $R_1$ and $R_2$ be realizations of the algebra $A.$
If $R_1\bigr|_{I}$ and $R_2\bigr|_{I}$ are ${\rm Aut}(A)|_{I}$-inequivalent then
$R_1$ and $R_2$ are weakly inequivalent.

\medskip

\noindent
{\bf Corollary 3.} If $R_1\bigr|_{I}$ and $R_2\bigr|_{I}$ are weakly inequivalent then
$R_1$ and $R_2$ also are weakly inequivalent.

\medskip

\noindent
{\bf Corollary 4.} If there exists a megaideal $I$ of $A$ such that $\rank R_1(I)\not=\rank R_2(I)$ then
the realizations $R_1$ and $R_2$ are weakly inequivalent.

\medskip

\noindent {\bf Remark.} In this paper we consider faithful
realizations only. If the faithful realizations of Lie algebras of
dimensions less than $m$ are known, the unfaithful realizations of
$m$-dimensional algebras can be constructed in an easy way.
Indeed, each unfaithful realization of  $m$-dimensional algebra
$A$, having the kernel $I$, yields a faithful realization of the
quotient algebra $A/I$ of dimension less than $m.$ This
correspondence is well-defined since the kernel of any
homomorphism from an algebra $A$ to an algebra~$A'$ is an ideal in
$A.$

\section{Realizations of low-dimensional real Lie algebras}

The most important and elegant results on realizations of Lie
algebras were obtained by S.~Lie himself. He classified
non-singular Lie algebras of vector fields in one real variable,
one complex variable and two complex
variables~\cite{lie1924,lie1880}. Using an ingenious geometric
argument, Lie
 also listed the Lie algebras of vector fields in two
real variables~\cite[Vol.3]{lie1893} (a more complete discussion can be found in~\cite{Gonzales&Kamran&Olver1992a}).
Finally, in \cite[Vol.3]{lie1893} he claimed to have completely classified
all Lie algebras of vector field in three complex variables
(in fact he gives details in the case of primitives algebras, and divides the imprimitive cases into
three classes, of which only the first two are treated~\cite{Gonzales&Kamran&Olver1992a}).

Using Lie's classification of Lie algebras of vector fields in two complex variables,
 A.~Gonz\'alez-L\'opez, N.~Kamran and P.~Olver~\cite{Gonzales&Kamran&Olver1992b} studied finite-dimensional
Lie algebras of  first order differential operators $Q=\xi^i(x)\p_{x_i} +f(x)$ and classified all of such
algebras with two complex variables.

In \cite{mahomed&leach1989} F.M. Mahomed and P.G.L. Leach investigated
realizations of three-dimensional real Lie algebras in terms of
Lie vector fields in two variables and used them for treating
third order ODEs. Analogous realizations for four-dimensional real
Lie algebras without commutative three-dimensional subalgebras
were considered by A.~Schmucker and
G.~Czichowski~\cite{Schmucker&Czichowski1998}.

All the possible realizations of algebra $so(3)$ in the real vectors
fields were first classified in \cite{lahno&zhdanov2000,zhdanov&lahno&fushchych2000}.
Covariant realizations of physical algebras (Galilei, Poincar\'e and Euclid ones)
were constructed in
\cite{fushchych&lahno&zhdanov1993,fushchych&tsyfra&boyko,fushchych&zhdanov1997,fushchych&zhdanov&lahno1994,
lahno&zhdanov2000,max2002,max2000,
yehorchenko1992,zhdanov&lahno2000,zhdanov&lahno&fushchych2000}.
Complete description of
realizations of the Galilei algebra  in the space of two
dependent and two independent variables was obtained in
\cite{rideau&winternitz1993,zhdanov&fushchych1997}.
In \cite{Kantor&Patera1994}
I.L.~Kantor and J.~Patera described
finite-dimensional Lie algebras of polynomial (degree $\le 3$)
vector fields in $n$ real variables
that contain the vector fields $\partial_{x_i}$ $(i=\overline{1,n})$.
In \cite{post1994,post2001,post2002} G.~Post studied finite-dimensional Lie algebras
of polynomial vector fields of $n$ variables that contain
the vector fields $\partial_{x_i}$ $(i=\overline{1,n})$ and $x_i \partial_{x_i}$.

C.~Wafo Soh and F.M.~Mahomed~\cite{soh&mahomed2001jphysa} used
Mubarakzyanov's results~\cite{mubarakzyanov1963.1} in order to classify
realizations of three- and four-dimensional real Lie algebras in
the space of three variables and to describe systems of two
second-order ODEs admitting real four-dimensional symmetry
Lie algebras, but unfortunately their paper contains some
misprints and incorrect statements (see Section~\ref{secWafo} of our paper).
Therefore, this classification cannot be regarded as complete.
The results of~\cite{soh&mahomed2001jphysa} are used
in~\cite{soh&mahomed2001jnlm} to solve the problem of
linearization of systems of second-order ordinary differential
equations, so some results from~\cite{soh&mahomed2001jnlm} also
are not completely correct.

A preliminary classification of realizations of solvable
three-dimensional Lie algebras in the space of any (finite) number
of variables was given in~\cite{lutfullin2000}. Analogous results
on a complete set of inequivalent realizations for real
four-dimensional solvable Lie algebras were announced at the Fourth
International Conference ``Symmetry in Nonlinear Mathematical
Physics'' (9--15 July, 2001, Kyiv) and were published in the
proceedings of this
conference~\cite{lutfullin&popovych2002,nesterenko&boyko2002}.

In this paper we present final results of our classifications of
realizations of all the Lie algebras of dimension up to 4. On
account of them being cumbersome we adduce only classification of
realizations with respect to weak equivalence because it is more
complicated to obtain, is more suitable for applications and can
be presented in a more compact form. The results are formulated in
the form of Tables~2--6. Below equivalence indicates weak
equivalence.

\medskip

\noindent {\bf Remarks for Tables 2--6.} We use the following
notation, contractions and agreements.
\begin{itemize}\renewcommand{\itemsep}{-1pt}
\item
We treat Mu\-ba\-rak\-zya\-nov's classification of abstract
Lie algebras and follow, in general, his numeration of Lie algebras.
For each algebra we write down only non-zero commutators between the basis elements.
$\p_i$ is a shorthand for $\p/\p x_i.$
$R(A,N)$ denotes the $N$-th realization of the algebra $A$ corresponding
to position in the table,
and the algebra symbol $A$ can be omitted if is clear what algebra is meant.
If it is necessary we also point out parameter symbol
$\alpha_1,\ldots,\alpha_k$ in the designation
$R(A,N,(\alpha_1,\ldots,\alpha_k))$ of series of realizations.
\item
The constant parameters of series of solvable Lie algebras (e.g.,
$A_{4.2}^b$) are denoted as $a,$ $b$ or $c.$ All the other constants as
well as the functions in Tables~2--6 are parameters of realization
series. The functions are arbitrary differentiable real-valued functions of
their arguments, satisfying only the conditions given in remarks after
the respective table. The presence of such remark for a realization
is marked in the last column of the table.
All the constants are real.
The constant $\varepsilon$ takes only two values 0 or 1, i.e.
$\varepsilon\in\{0;1\}.$ The conditions for the other constant
parameters of realization series are given in remarks after
the corresponding table.
\item
For each series of solvable Lie
algebras we list, at first, the ``common'' inequivalent
realizations (more precisely, the inequivalent realizations series
parametrized with the parameters of algebra series) existing for
all the allowed values of the parameters of algebra series.
Then, we list the ``specific'' realizations which exist or are
inequivalent to ``common'' realizations only for some ``specific''
sets of values of the parameters. Numeration of ``specific''
realizations for each ``specific'' set  of values of the
parameters is continuation of that for ``common'' realizations.
\item
In all the conditions of algebra equivalence, which are
given in remarks after tables, $(\alpha_{\mu\nu})$ is a
non-degenerate $(r\times r)$-matrix, where $r$ is the dimension of the
algebra under consideration.
\item
The summation over repeated indices is implied unless stated otherwise.
\end{itemize}

\noindent
{\bf Remarks on the series $\boldsymbol{A_{4,5}}$ and $\boldsymbol{A_{4,6}}$.}
Consider the algebra series $\{A_{4,5}^{a_1,a_2,a_3}\:|\: a_1a_2a_3\not=0\}$ generated by the
algebras for which the non-zero commutation relations  have the form
$[e_1, e_4]=a_1e_1,$ $[e_2, e_4]=a_2e_2,$ $[e_3, e_4]=a_3e_3.$
Two  algebras from this series,
with the parameters $(a_1,a_2,a_3)$ and $(\tilde a_1,\tilde a_2,\tilde a_3)$ are
equivalent iff there exist a real $\lambda\not=0$ and a permutation $(j_1,j_2,j_3)$ of the set $\{1;2;3\}$
such that the condition $\tilde a_i=\lambda a_{j_i}$ ($i=\overline{1,3}$) is satisfied.
For the algebras under consideration to be inequivalent, one has to constrain the set of
parameter values.
There are different ways to do this. A traditional way~\cite{ Cerquetelli&Ciccoli&Nucci2002,mubarakzyanov1963.1,
patera&sharp&winternitz1976,patera&winternitz1977,soh&mahomed2001jphysa}
is to apply the condition $-1\le a_2\le a_3\le a_1=1.$ But this condition is not sufficient
to select inequivalent algebras since the algebras $A_{4,5}^{1,-1,b}$ and $A_{4,5}^{1,-1,-b}$
are equivalent in spite of their parameters satisfying the above constraining condition if $|b|\le 1.$
The additional condition $a_3\ge0$ if $a_2=-1$ guarantees for the algebras with constrained parameters
to be inequivalent.

Moreover, it is convenient for us to break the parameter set into
three disjoint subsets depending on the number of equal
parameters. Each from these subsets is normalized individually. As
a result we obtain three inequivalent cases:
\[
a_1=a_2=a_3=1; \quad a_1=a_2=1,\; a_3\not=1,0; \quad -1\le a_1<a_2<a_3=1, \; a_2>0\; \mbox{if}\; a_1=-1.
\]

An analogous remark is true also for the algebra series $\{A_{4,6}^{a,b}\:|\: a\not=0\}$
generated by the algebras for which the non-zero commutation relations  have the form
$[e_1, e_4]=ae_1,$ $[e_2, e_4]=be_2-e_3,$ $[e_3, e_4]=e_2+be_3.$
Two  algebras from this series
with the different parameters $(a,b)$ and $(\tilde a,\tilde b)$ are equivalent iff
$\tilde a=-a,$ $\tilde b=-b.$ A traditional way of constraining the set of parameter values is
to apply the condition $b \ge 0$ that does not exclude the equivalent algebras of the form
$A_{4,6}^{a,0}$ and $A_{4,6}^{-a,0}$ from consideration.
That is why it is more correct to use the condition $a>0$ as a constraining relation for the
parameters of this series.

\medskip

The technique of classification is the following.
\begin{itemize}
\item
For each algebra $A$ from Mu\-ba\-rak\-zya\-nov's classification~\cite{ mubarakzyanov1963.1}
of abstract Lie algebras of dimension $m\le4$ we find the automorphism group ${\rm Aut}(A)$
and the set of megaideals of $A.$
(Our notions of low-dimensional algebras, choice of their basis elements, and, consequently,
the form of commutative relations coincide with Mu\-ba\-rak\-zya\-nov's ones.)
Calculations of this step is quite simple due to low dimensions and simplicity of the canonical
commutation relations.
Lemmas~1 and~2, Corollary~1 and other similar statements are useful for such calculations.
See also the remarks below.

\item
We choose a maximal proper subalgebra $B$ of $A.$ As rule, dimension of $B$ is equal to
$m-1.$
So, if $A$ is solvable, it necessarily contains a $(m-1)$-dimensional ideal.
The simple algebra $sl(2,\R)$ has a two-dimensional subalgebra.
The Levi factors of unsolvable four-dimensional algebras ($sl(2,\R)\oplus A_1$ and $so(3)\oplus A_1$)
are three-dimensional ideals of these algebras.
Only $so(3)$ does not contain a subalgebra of dimension $m-1=2$ that is a reason of
difficulties in constructing realizations for this algebra.
Moreover, the algebras $sl(2,\R),$ $so(3),$ $mA_1,$ $ A_{3.1},$ $A_{3.1} \oplus A_1$ and $2A_{2.1}$
exhaust the list of algebras under consideration that do not contain
$(m-1)$--dimensional megaideals.
\item
Let us suppose that a complete list of strongly inequivalent realizations of $B$ has been already constructed.
(If $B$ is a megaideal of $A$ and realizations of $A$ are classified only with respect to the weak equivalence,
it is sufficient to use only ${\rm Aut}(A)|_B$-inequivalent realizations of $B.$)
For each realization $R(B)$ from this list we make the following procedure.
We find the set ${\rm Diff}^{R(B)}$ of local diffeomorphisms of the space of $x,$ which preserve $R(B).$
Then, we realize the basis vector $e_i$ (or the basis vectors in the case of $so(3)$) from $A\backslash B$
in the most general form $e_i=\xi^{ia}(x)\p_a,$ where $\p_a=\p/\p x_a,$
and require that it satisfied the commutation relations of $A$ with the basis vectors from $R(B).$
As a result, we obtain a system of first-order PDEs for the coefficients~$\xi^{ia}$
and integrate it, considering all possible cases.
For each case we reduce the found solution to the simplest form,
using either diffeomorphisms from ${\rm Diff}^{R(B)}$ and automorphisms of $A$ if
the weak equivalence is meant or only diffeomorphisms from ${\rm Diff}^{R(B)}$
 for the strong equivalence.
\item
The last step is to test inequivalence of the constructed
realizations. We associate the $N$-th one of them with the ordered
collection of integers $(r_{Nj}),$ where $r_{Nj}$ is equal to the
rank of the elements of $S_j$ in the realization $R(A,N).$ Here
$S_j$ is either the $j$-th subset of basis of $A$ with $|S_j|>1$
in the case of strong equivalence or the basis of the $j$-th
megaideals $I_j$ of $A$ with $\dim I_j>1$ in the case of weak
equivalence. Inequivalence of realizations with different
associated collection of integers immediately follows from
Corollary~2 or Corollary~4 respectively. Inequivalence of
realizations in the pairs with identical collections of ranks is
proved using another method, e.g. Casimir operators (for simple algebras),
Lemmas~2 and~3, Corollary~3 and
the rule of constraries (see the following section).
\end{itemize}
We rigorously proved inequivalence of all the constructed
realizations. Moreover, we compared our classification with
results of the papers, cited in the beginning of the section (see
Section~\ref{secWafo} for details of comparison with results of
one of them).

\medskip

\noindent
{\bf Remark.} Another interesting method to construct realizations of Lie algebras
in vector fields was proposed by
I.~Shirokov~\cite{shirokov1997,shirokov2003,baranovskii&shirokov2003}.
This method is also simple to use and based on classification of subalgebras of Lie algebras.

\medskip

\noindent
{\bf Remark.} The automorphisms of semi-simple algebras are well-known~\cite{Jacobson}.
The automorphisms of three-dimensional algebras were considered in \cite{Harvey1979}.
Let us note that in this paper only connected components of the unity
of the automorphism groups were constructed really, i.e. the discrete transformations were
missed.
The automorphism groups of four-dimensional algebras
were published in~\cite{Christodoulakis&Papadopoulos&Dimakis2003} (with a few misprints, which were
corrected in Corrigendum).
Namely, for the algebra $A^\alpha_{4.2}$ the automorphism groups
are to have the following form (we preserve the notations
of~\cite{Christodoulakis&Papadopoulos&Dimakis2003}):
\begin{gather*}
{\rm Aut}(A_{4.2}^\alpha)=\left\{\left.\left(\begin{array}{cccc}
a_1 & 0 & 0 & a_4\\
0 & a_6 & a_7 & a_8\\
0 & 0 & a_6 & a_{12}\\
0 & 0 & 0 & 1\end{array}\right) \;\right| \ a_1a_6\not=0\right\}, \quad \alpha\not=0,1;\\
{\rm Aut}(A_{4.2}^1)=\left\{\left.\left(\begin{array}{cccc}
a_1 & 0 & a_3 & a_4\\
a_5 & a_6 & a_7 & a_8\\
0 & 0 & a_6 & a_{12}\\
0 & 0 & 0 & 1\end{array}\right) \;\right| \ a_1a_6\not=0\right\}.
\end{gather*}

\medskip

\noindent
{\bf Remark.}
As for any classification which are performed up to an equivalence relation,
there exist the problem of choice of canonical forms of realizations for an
arbitrary fixed Lie algebra.
Such choice can be made in a number of ways and depends, in particular,
on the choice of canonical form of commutation relations (i.e. structure constants) of the algebra
and the dimension of realization manifold (see e.g. Remark after Table 6).

The closed problem on reduction of realizations to the linear form (i.e. on search of realizations in
vector fields having linear coefficients with respect to $x$), including questions of 
existence of such forms and a minimum of possible dimensions of the realization manifold, is also opened.
This problem is equivalent to construction of finite-dimensional representations of the algebras under consideration.

\section{Example: realizations of $\boldsymbol{A_{4.10}}$}

We consider in detail constructing of a list of inequivalent realizations for the algebra $A_{4.10}$.
The non-zero commutators between the basis elements of $A_{4.10}$ are as follows:
\[
[e_1,e_3]=e_1, \quad [e_2,e_3]=e_2, \quad [e_1,e_4]=-e_2, \quad [e_2,e_4]=e_1.
\]
The automorphism group ${\rm Aut}(A_{4.10})$ is generated by the basis transformations
of the form $\tilde e_\mu=\alpha_{\nu\mu}e_\nu,$ where $\mu,\nu=\overline{1,4},$
\begin{equation}\label{gen.form.matrix.aut.A4.10}
(\alpha_{\nu\mu})=
\left(\begin{array}{cccc}
\pm \alpha_{22}& \alpha_{12} & \alpha_{13} & \pm\alpha_{23} \\
\mp \alpha_{12} & \alpha_{22} & \alpha_{23} & \mp \alpha_{13}\\
0 & 0 & 1 & 0\\
0 & 0 & 0 & \pm 1
\end{array}\right).
\end{equation}

The algebra $A_{4.10}$ contains four non-zero megaideals:
\[\arraycolsep=0em\begin{array}{l}
I_1=\langle e_1,e_2\rangle \sim 2A_1, \quad
I_2=\langle e_1,e_2,e_3\rangle \sim A_{3.3}, \quad
I_3=\langle e_1,e_2,e_4\rangle \sim A_{3.5}^0, \\[1ex]
I_4=\langle e_1,e_2,e_3,e_4\rangle \sim A_{4.10}. \quad
\end{array}\]
Realizations of two three-dimensional megaideals $I_2$ and $I_3$ can be extended
by means of one additional basis element to realizations of $A_{4.10}.$
To this end we use $I_2$. This megaideal has four inequivalent realizations
$R(A_{3.3},N)$ $(N=\overline{1,4})$ in Lie vector fields (see Table~3).
Let us emphasize that it is inessential for the algebra $A_{3.3}$ which equivalence (strong or weak)
has been used for classifying realizations.
For each of these realizations we perform the following procedure.
Presenting the fourth basis element in the most general form $e_4=\xi^a(x)\p_a$
and commuting $e_4$ with the other basis elements, we obtain a linear overdetermined
system of first-order PDEs for the functions $\xi^a.$ Then we solve this system and simplify
its general solution by means of transformations $\tilde x_a=f^a(x)$ ($a=\overline{1,n}$)
which preserve the forms of $e_1,$ $e_2,$ and $e_3$ in the considered realization of $A_{3.3}.$
To find the appropriate functions $f^a(x)$, we are to solve one more system of PDEs which results from
the conditions $e_i\bigl|_{x_a\rightarrow \tilde x_a}=(e_i f^a)(x)\p_{\tilde x_a}$ if $\tilde x_a=f^a(x),$ $i=\overline{1,3}.$
The last step is to prove inequivalence of all the constructed realizations.

So, for the realization $R(A_{3.3},1)$ we have $e_1=\p_1,$ $e_2=\p_2,$ $e_3=x_1\p_1+x_2\p_2+\p_3,$
and the commutation relations imply the following system on the functions $\xi^a$:
\[\arraycolsep=0em\begin{array}{llll}
[e_1,e_4]=-e_2\quad &\Rightarrow\quad \xi^1_1=0, \quad &\xi^2_1=-1, \quad &\xi^k_1=0,\\[1ex]
[e_2,e_4]=e_1 \quad &\Rightarrow\quad \xi^1_2=1, \quad &\xi^2_2=0, \quad &\xi^k_2=0, \qquad k=\overline{3,n},\\[1ex]
[e_3,e_4]=0     \quad &\Rightarrow\quad \xi^1_3=\xi^1-x_2, \quad& \xi^2_3=\xi^2+x_1, \quad &\xi^k_3=0,
\end{array}\]
the general solution of which can be easy found:
\[
\xi^1=x_2+\theta^1(\hat x)e^{x_3}, \quad \xi^2=-x_1+\theta^2(\hat x)e^{x_3}, \quad
\xi^k=\theta^k(\hat x),  \quad k=\overline{3,n},
\]
where $\theta^a$ ($a=\overline{1,n}$) are arbitrary smooth functions of $\hat x=(x_4,\ldots,x_n).$
The form of $e_1,$ $e_2,$ and $e_3$ are preserved only by the transformations
\[
\tilde x_1=x_1+f^1(\hat x)e^{x_3}, \quad
\tilde x_2=x_2+f^2(\hat x)e^{x_3}, \quad
\tilde x_3=x_3+f^3(\hat x), \quad
\tilde x_\alpha=f^\alpha(\hat x), \quad \alpha=\overline{4,n},
\]
where $f^a$ ($a=\overline{1,n}$) are arbitrary
smooth functions of $\hat x,$  and  $f^\alpha$ ($\alpha=\overline{4,n}$) are functionally independent.
Depending on values of the parameter-functions $\theta^k$ ($k=\overline{3,n}$)
there exist three cases of possible reduction of $e_4$ to canonical forms by means of these transformations, namely,
\[\arraycolsep=0em\begin{array}{lll}
\exists\:\alpha\,{:}\; \theta^\alpha\not=0 \quad & \Rightarrow\quad
e_4=x_2\p_1-x_1\p_2+\p_4 \quad & (\mbox{the realization}\: R(A_{4.10},1) ); \\[1ex]
\theta^\alpha=0, \; \theta^3\not=\mathop{\rm const}\nolimits\quad & \Rightarrow\quad
e_4=x_2\p_1-x_1\p_2+x_4\p_3 \quad & (\mbox{the realization}\: R(A_{4.10},2) ); \\[1ex]
\theta^\alpha=0, \; \theta^3=\mathop{\rm const}\nolimits\quad & \Rightarrow\quad
e_4=x_2\p_1-x_1\p_2+C\p_3 \quad & (\mbox{the realization}\: R(A_{4.10},3,C) ).
\end{array}\]
Here $C$ ia an arbitrary constant.

The calculations for other realizations of $A_{3.3}$ are easier than for the first one.
Below for each from these realizations we adduce brief only
the general solution of the system of PDEs for the coefficients $\xi^a,$ the transformations
which preserve the forms of $e_1,$ $e_2,$ and $e_3$ in the considered realization of $A_{3.3}$,
and the respective realizations of $A_{4.10}$.
\[\arraycolsep=0em\begin{array}{ll}
R(A_{3.3},2){:}\quad
&\xi^1=x_2, \; \xi^2=-x_1, \; \xi^k=\theta^k(\bar x), \quad k=\overline{3,n}, \quad \bar x=(x_3, \ldots, x_n); \\[1ex]
&\tilde x_1=x_1, \quad \tilde x_2=x_2, \quad \tilde x_k=f^k(\bar x); \\[1ex]
&R(A_{4.10},5)\quad \mbox{if}\quad\exists \: k{:}\; \theta^k\not=0\quad \mbox{and} \quad
R(A_{4.10},6)\quad \mbox{if}\quad \theta^k=0.
\\[2ex]
R(A_{3.3},3){:}\quad
&\xi^1=-x_1x_2+\theta^1(x')e^{x_3}, \; \xi^2=-(1+x_2^2), \; \xi^k=\theta^k(x'); \\[1ex]
&\tilde x_1=x_1+f(x')e^{x_3}, \quad \tilde x_2=x_2, \quad \tilde x_3=x_3+f^3(x'),
\quad \tilde x_\alpha=f^\alpha(x'); \\[1ex]
&R(A_{4.10},4); \quad k=\overline{3,n}, \quad \alpha=\overline{4,n}, \quad x'=(x_2,x_4,x_5, \ldots, x_n).
\\[2ex]
R(A_{3.3},4){:}\quad
&\xi^1=-x_1x_2, \; \xi^2=-(1+x_2^2), \; \xi^k=\theta^k(\check x), \quad
k=\overline{3,n}, \quad \check x=(x_2, \ldots, x_n);\\[1ex]
&\tilde x_1=x_1, \quad \tilde x_2=x_2, \quad \tilde x_k=f^k(\check x); \\[1ex]
&R(A_{4.10},7).
\end{array}\]
Here $\theta^a$ ($a=\overline{1,n}$) are arbitrary smooth functions of their arguments,
and $f^a$ ($a=\overline{1,n}$) are such smooth functions of their arguments that the respective
transformation of $x$ is not singular.

To prove inequivalence of the constructed realizations, we associate the $N$-th of them
with the ordered collection of integers $(r_{N1},r_{N2},r_{N3},r_{N4}),$
where $r_{Nj}=\mathop{\rm rank}\nolimits R(A_{4.10},N)\bigr|_{I_j},$ i.e.
$r_{Nj}$ is equal to the rank of basis elements of the megaideals $I_j$ in the realization $R(A_{4.10},N),$
($N=\overline{1,7},$ $j=\overline{1,4}$):
\[\arraycolsep=0em\begin{array}{l}
R(A_{4.10},1) \longrightarrow (2,3,3,4); \quad
R(A_{4.10},2) \longrightarrow (2,3,3,3);  \\[1ex]
R(A_{4.10},3,C) \longrightarrow (2,3,3,3) \quad\mbox{if}\quad C\not=0 \quad \mbox{and}\quad
R(A_{4.10},3,0) \longrightarrow (2,3,2,3); \\[1ex]
R(A_{4.10},4) \longrightarrow (1,2,2,3); \quad
R(A_{4.10},5) \longrightarrow (2,2,3,3);  \\[1ex]
R(A_{4.10},6) \longrightarrow (2,2,2,2);  \quad
R(A_{4.10},7) \longrightarrow (1,1,2,2).
\end{array}\]
Inequivalence of realizations with different associated collections of integers follows immediately from Corollary~4.
The collections of ranks of megaideals coincide only for the pairs of realizations of two forms
\[
\{R(A_{4.10},2),R(A_{4.10},3,C)\}\quad \mbox{and}\quad \{R(A_{4.10},3,C),R(A_{4.10},3,\tilde C)\},
\]
where $C,\tilde C\not=0.$ Inequivalence of realizations in these
pairs is to be proved using another method, e.g. the rule of
contraries.

Let us suppose that the realizations $R(A_{4.10},2)$ and $R(A_{4.10},3,C)$ are equivalent
and let us fix their bases given in Table~5.
Then, by the definition of equivalence there exists an automorphism of $A_{4.10}$
$\tilde e_\mu=\alpha_{\nu\mu}e_\nu$ and a change of variables $\tilde x_a=g^a(x)$
which transform the basis of $R(A_{4.10},2)$ into the basis of $R(A_{4.10},3,C).$
(Here $\mu,\nu=\overline{1,4},$ $a=\overline{1,n},$ and the matrix $(\alpha_{\nu\mu})$
has the form~(\ref{gen.form.matrix.aut.A4.10}).)
For this condition to hold true, the function $g^3$ is to satisfy the following system of
PDEs:
\[
g^3_1=0, \quad g^3_2=0, \quad g^3_3=1, \quad x_4g^3_3=C
\]
which implies the contradictory equality $x_4=C.$
Therefore, the considered realizations are inequivalent.

In an analogous way we obtain that the realizations $R(A_{4.10},3,C)$ and $R(A_{4.10},3,\tilde C)$
are equivalent iff $C=\tilde C.$

\newpage

\centerline{\small {\bf Table 2.} Realizations of one and two-dimensional real Lie algebras}

\vspace{-2mm}

\begin{center}\small
\begin{tabular}{|p{3cm}|r|p{11.0cm}|p{0.5cm}|}
\hline\vspacebefore
\hfil Algebra & $N$ &\hfil  Realization&\hfil  $(*)$
\\ \hline\vspacebefore
$A_1$&$1$&$\p_1$&
\myhline
$2A_1$&1&$\p_1$, $\p_2$ &\\
&2 &$\p_1$, $x_2\p_1$&
\myhline
$A_{2.1}$ &1 &$\p_1$, $x_1\p_1+\p_2$ &\\
$[e_1,e_2]=e_1$&2 & $\p_1$, $x_1\p_1$&
\vspaceafter\\\hline
\end{tabular}
\end{center}

\centerline{\small {\bf Table 3.} Realizations of three-dimensional solvable Lie algebras}

\vspace{-2mm}

\begin{center}\small
\begin{tabular}{|p{3cm}|r|p{11cm}|p{0.5cm}|}
\hline \vspacebefore
\hfil Algebra & $N$ &\hfil  Realization&\hfil  $(*)$
\\ \hline\vspacebefore
$3A_1$
&1&$\p_1$, $\p_2$, $\p_3$&\\
&2&$\p_1$, $\p_2$, $x_3\p_1+x_4\p_2$&\\
&3&$\p_1$, $\p_2$, $x_3\p_1+\varphi(x_3)\p_2$&\hfil  $(*)$\\
&4&$\p_1$, $x_2\p_1$, $x_3\p_1$&\\
&5&$\p_1$, $x_2\p_1$, $\varphi \left(x_2\right)\p_1$ &\hfil  $(*)$
\myhline
$A_{2.1}\oplus A_1$
&1&$\p_1$, $x_1\p_1+\p_3$, $\p_2$&\\
$[e_1,e_2]=e_1$
&2&$\p_1$, $x_1\p_1+x_3\p_2$, $\p_2$&\\
&3&$\p_1,x_1\p_1$, $\p_2$&\\
&4&$\p_1$, $x_1\p_1+x_2\p_2$, $x_2\p_1$&
\myhline
$A_{3.1}$
&1&$\p_1$, $\p_2$, $x_2\p_1+\p_3$&\\
$[e_2,e_3]=e_1$
&2&$\p_1$, $\p_2$, $x_2\p_1+x_3\p_2$&\\
&3&$\p_1$, $\p_2$, $x_2\p_1$&
\myhline
$A_{3.2}$
&1&$\p_1$, $\p_2$, $\left(x_1+x_2\right)\p_1+x_2\p_2+\p_3$&\\
$[e_1,e_3]=e_1$
&2&$\p_1$, $\p_2$,  $\left(x_1+x_2\right)\p_1+x_2\p_2$&\\
$[e_2,e_3]=e_1+e_2$
&3&$\p_1$, $x_2\p_1$, $x_1\p_1-\p_2$&
\myhline
$A_{3.3}$
&1&$\p_1$, $\p_2$, $x_1\p_1+x_2\p_2+\p_3$&\\
$[e_1,e_3]=e_1$
&2&$\p_1$, $\p_2$, $x_1\p_1+x_2\p_2$&\\
$[e_2,e_3]=e_2$
&3&$\p_1$, $ x_2\p_1$, $x_1\p_1+\p_3$&\\
&4&$\p_1$, $ x_2\p_1$, $x_1\p_1$&
\myhline
$A_{3.4}^a,$  $|a|\!\leq\! 1,$ $a\!\not=\!0,1$
&1&$\p_1$, $\p_2$, $x_1\p_1+ax_2\p_2+\p_3$&\\
$[e_1,e_3]=e_1$
&2&$\p_1$, $\p_2$, $x_1\p_1+ax_2\p_2$&\\
$[e_2,e_3]=ae_2$
&3&$\p_1$, $x_2\p_1$, $x_1\p_1+(1-a)x_2\p_2$&
\myhline
$A_{3.5}^b,$ $\;b\geq 0$
&1&$\p_1$, $\p_2$, $(bx_1+x_2)\p_1+(-x_1+bx_2)\p_2 +\p_3$&\\
 $[e_1,e_3]=be_1-e_2$
&2&$\p_1$, $\p_2$, $(bx_1+x_2)\p_1+(-x_1+bx_2)\p_2$&\\
$[e_2,e_3]=e_1+be_2$
&3&$\p_1$, $x_2\p_1$, $(b-x_2)x_1\p_1-(1+x_2^2)\p_2$&
\vspaceafter\\ \hline
\end{tabular}
\end{center}

\noindent
{\bf Remarks for Table 3.}
\begin{description}\renewcommand{\itemsep}{0ex}\vspace{-1ex}
\item[$R(3A_1,3,\varphi).$] $\varphi=\varphi(x_3)$.
The realizations $R(3A_1,3,\varphi)$ and $R(3A_1,3,\tilde\varphi)$
are equivalent iff
\begin{equation}\label{equiv.cond.r3a1.3}\arraycolsep=0em\begin{array}{l}
\tilde x_3=-(\alpha_{11}x_3+\alpha_{12}\varphi(x_3)-\alpha_{13})/
(\alpha_{31}x_3+\alpha_{32}\varphi(x_3)-\alpha_{33}),\\
\lefteqn{\tilde \varphi}\phantom{x_3}=-(\alpha_{21}x_3+\alpha_{22}\varphi(x_3)-\alpha_{23})/
(\alpha_{31}x_3+\alpha_{32}\varphi(x_3)-\alpha_{33}).
\end{array}\end{equation}
\item[$R(3A_1,5,\varphi).$] $\varphi=\varphi(x_2)$, $\varphi''\not=0.$
The realizations $R(3A_1,5,\varphi)$ and $R(3A_1,5,\tilde\varphi)$
are equivalent iff
\begin{equation}\label{equiv.cond.r3a1.5}\arraycolsep=0em\begin{array}{l}
\tilde x_2=-(\alpha_{21}x_2+\alpha_{22}\varphi(x_2)-\alpha_{23})/
(\alpha_{11}x_2+\alpha_{12}\varphi(x_2)-\alpha_{13}),\\
\lefteqn{\tilde \varphi}\phantom{x_2}=-(\alpha_{31}x_2+\alpha_{32}\varphi(x_2)-\alpha_{33})/
(\alpha_{11}x_2+\alpha_{12}\varphi(x_2)-\alpha_{13}).
\end{array}\end{equation}
\end{description}

\newpage

\centerline{\small {\bf Table 4.} Realizations of real decomposable solvable four-dimensional Lie algebras}

\vspace{-2mm}

\begin{center}\small
\begin{tabular}{|p{3cm}|r|p{11cm}|p{0.5cm}|}
\hline \vspacebefore
\hfil Algebra & $N$ &\hfil  Realization&\hfil  $(*)$
\\ \hline\vspacebefore
$ 4A_1$
& 1 & $\p_1,$ $\p_2,$ $\p_3,$ $\p_4$ &\\
& 2 & $\p_1,$ $\p_2,$ $\p_3,$ $x_4\p_1+x_5\p_2+x_6\p_3$  &\\
& 3 & $\p_1,$ $\p_2,$ $\p_3,$ $x_4\p_1+x_5\p_2+\theta(x_4,x_5)\p_3$  &\hfil  $(*)$\\
& 4 & $\p_1,$ $\p_2,$ $\p_3,$ $x_4\p_1+\varphi(x_4)\p_2+\psi(x_4)\p_3$  &\hfil  $(*)$\\
& 5 & $\p_1,$ $\p_2,$ $x_3\p_1+x_4\p_2,$ $x_5\p_1+x_6\p_2$ &\\
& 6 & $\p_1,$ $\p_2,$ $x_3\p_1+x_4\p_2,$ $x_5\p_1+\theta(x_3,x_4,x_5)\p_2$ &\hfil  $(*)$\\
& 7 & $\p_1,$ $\p_2,$ $x_3\p_1+\varphi(x_3,x_4)\p_2,$ $x_4\p_1+\psi(x_3,x_4)\p_2$ &\hfil  $(*)$\\
& 8 & $\p_1,$ $\p_2,$ $x_3\p_1+\varphi(x_3)\p_2,$ $\theta(x_3)\p_1+\psi(x_3)\p_2$ &\hfil  $(*)$\\
& 9 & $\p_1,$ $x_2\p_1,$ $x_3\p_1,$ $x_4\p_1$ &\\
& 10 & $\p_1,$ $x_2\p_1,$ $x_3\p_1,$ $\theta(x_2,x_3)\p_1$ &\hfil  $(*)$\\
& 11 & $\p_1,$ $x_2\p_1,$ $\varphi(x_2)\p_1,$ $\psi(x_2)\p_1$ &\hfil  $(*)$
\myhline
$ A_{2.1} \oplus 2 A_1$
& 1 & $\p_1,$ $x_1\p_1+\p_4,$ $\p_2,$ $\p_3$ &\\
$[e_1,e_2]=e_1$
& 2 & $\p_1,$ $x_1\p_1+x_4\p_2+x_5\p_3,$ $\p_2,$ $\p_3$ &\\
& 3 & $\p_1,$ $x_1\p_1+x_4\p_2+\varphi(x_4)\p_3,$ $\p_2,$ $\p_3$ &\hfil  $(*)$\\
& 4 & $\p_1,$ $x_1\p_1,$ $\p_2,$ $\p_3$ &\\
& 5 & $\p_1,$ $x_1\p_1+x_3\p_3,$ $\p_2,$ $x_3\p_1+x_4\p_2$ &\\
& 6 & $\p_1,$ $x_1\p_1+x_3\p_3,$ $\p_2,$ $x_3\p_1$ &\\
& 7 & $\p_1,$ $x_1\p_1+\p_4,$ $\p_2,$ $x_3\p_2$ &\\
& 8 & $\p_1,$ $x_1\p_1+x_4\p_2,$ $\p_2,$ $x_3\p_2$ &\\
& 9 & $\p_1,$ $x_1\p_1+\varphi(x_3)\p_2,$ $\p_2,$ $x_3\p_2$ &\hfil  $(*)$\\
& 10 & $\p_1,$ $x_1\p_1+x_2\p_2+x_3\p_3,$ $x_2\p_1,$ $x_3\p_1$ &
\myhline
$ 2A_{2.1}$
& 1 & $\p_1,$ $x_1\p_1+\p_3,$ $\p_2,$ $x_2\p_2+\p_4$ &\\
$[e_1,e_2]=e_1$
& 2 & $\p_1,$ $x_1\p_1+\p_3,$ $\p_2,$ $x_2\p_2+x_4\p_3$  &\\
$[e_3,e_4]=e_3$
& 3 & $\p_1,$ $x_1\p_1+\p_3,$ $\p_2,$ $x_2\p_2+C\p_3$ &\hfil  $(*)$\\
& 4 & $\p_1,$ $x_1\p_1+x_3\p_2,$ $\p_2,$ $x_2\p_2+x_3\p_3$ &\\
& 5 & $\p_1,$ $x_1\p_1,$ $\p_2,$ $x_2\p_2$ &\\
& 6 & $\p_1,$ $x_1\p_1+x_2\p_2,$ $x_2\p_1,$ $-x_2\p_2+\p_3$ &\\
& 7 & $\p_1,$ $x_1\p_1+x_2\p_2,$ $x_2\p_1,$ $-x_2\p_2$ &
\myhline
$ A_{3.1} \oplus A_1$
& 1 & $\p_1,$ $\p_3,$ $x_3\p_1+\p_4,$ $\p_2$ &\\
$[e_2,e_3]=e_1$
& 2 & $\p_1,$ $\p_3,$ $x_3\p_1+x_4\p_2+x_5\p_3,$ $\p_2$ &\\
& 3 % 3a
& $\p_1,$ $\p_3,$ $x_3\p_1+\varphi(x_4)\p_2+x_4\p_3,$ $\p_2$ &\hfil  $(*)$\\
& 4 % 3b
& $\p_1,$ $\p_3,$ $x_3\p_1+x_4\p_2,$ $\p_2$ &\\
& 5 % 3c
& $\p_1,$ $\p_3,$ $x_3\p_1,$ $\p_2$ &\\
& 6 %4
& $\p_1,$ $\p_3,$ $x_3\p_1+\p_4,$ $x_2\p_1$ &\\
& 7 %5
& $\p_1,$ $\p_3,$ $x_3\p_1+x_4\p_3,$ $x_2\p_1$ &\\
& 8 %6
& $\p_1,$ $\p_3,$ $x_3\p_1+\varphi(x_2)\p_3,$ $x_2\p_1$ &\hfil  $(*)$
\myhline
$ A_{3.2} \oplus A_1$
& 1 & $\p_1,$ $\p_2,$ $(x_1+x_2)\p_1+x_2\p_2+\p_3,$ $\p_4$ &\\
$[e_1,e_3]=e_1$
& 2 & $\p_1,$ $\p_2,$ $(x_1+x_2)\p_1+x_2\p_2+\p_3,$ $x_4\p_3$ &\\
$[e_2,e_3]=e_1+e_2$
& 3 & $\p_1,$ $\p_2,$ $(x_1+x_2)\p_1+x_2\p_2,$ $\p_3$ &\\
& 4 & $\p_1,$ $\p_2,$ $(x_1+x_2)\p_1+x_2\p_2+\p_3,$ $x_4e^{x_3}(x_3\p_1+\p_2)$ &\\
& 5 & $\p_1,$ $\p_2,$ $(x_1+x_2)\p_1+x_2\p_2+\p_3,$ $e^{x_3}(x_3\p_1+\p_2)$ &\\
& 6 & $\p_1,$ $\p_2,$ $(x_1+x_2)\p_1+x_2\p_2+\p_3,$ $e^{x_3}\p_1$ &\\
& 7 & $\p_1,$ $x_2\p_1,$ $x_1\p_1-\p_2,$ $\p_3$ &\\
& 8 & $\p_1,$ $x_2\p_1,$ $x_1\p_1-\p_2,$ $x_3e^{-x_2}\p_1$ &\\
& 9 & $\p_1,$ $x_2\p_1,$ $x_1\p_1-\p_2,$ $e^{-x_2}\p_1$ &
\myhline
$ A_{3.3} \oplus A_1$
& 1 & $\p_1,$ $\p_2,$ $x_1\p_1+x_2\p_2+\p_3,$ $\p_4$ &\\
$[e_1,e_3]=e_1$
& 2 & $\p_1,$ $\p_2,$ $x_1\p_1+x_2\p_2+\p_3,$ $x_4\p_3$ &\\
$[e_2,e_3]=e_2$
& 3 & $\p_1,$ $\p_2,$ $x_1\p_1+x_2\p_2,$ $\p_3$ &\\
& 4 & $\p_1,$ $\p_2,$ $x_1\p_1+x_2\p_2+\p_3,$ $e^{x_3}(\p_1+x_4\p_2)$ &\\
& 5 & $\p_1,$ $\p_2,$ $x_1\p_1+x_2\p_2+\p_3,$ $e^{x_3}\p_1$ &\\
& 6 & $\p_1,$ $x_2\p_1,$ $x_1\p_1+\p_3,$ $\p_4$ &\\
& 7 & $\p_1,$ $x_2\p_1,$ $x_1\p_1+\p_3,$ $x_4\p_3$ &\\
& 8 & $\p_1,$ $x_2\p_1,$ $x_1\p_1+\p_3,$ $\varphi(x_2)\p_3$ &\hfil  $(*)$\\
& 9 & $\p_1,$ $x_2\p_1,$ $x_1\p_1+\p_3,$ $e^{x_3}\p_1$ &
\vspaceafter\\ \hline
\end{tabular}\end{center}

\newpage

\centerline{\small {\bf Continuation of Table 4.}}

\vspace{-2mm}

\begin{center}\small
\begin{tabular}{|p{3cm}|r|p{11cm}|p{0.5cm}|}
\hline \vspacebefore
\hfil Algebra & $N$ &\hfil  Realization&\hfil  $(*)$
\\ \hline\vspacebefore
$ A_{3.4}^a \oplus A_1$
& 1 & $\p_1,$ $\p_2,$ $x_1\p_1+ax_2\p_2+\p_3,$ $\p_4$ &\\
$|a|\le 1,$ $\;a\not=0,1$
& 2 & $\p_1,$ $\p_2,$ $x_1\p_1+ax_2\p_2+\p_3,$ $x_4\p_3$ &\\
$[e_1,e_3]=e_1$
& 3 & $\p_1,$ $\p_2,$ $x_1\p_1+ax_2\p_2,$ $\p_3$ &\\
$[e_2,e_3]=ae_2$
& 4 & $\p_1,$ $\p_2,$ $x_1\p_1+ax_2\p_2+\p_3,$ $e^{x_3}\p_1+x_4e^{ax_3}\p_2$ &\\
& 5 & $\p_1,$ $\p_2,$ $x_1\p_1+ax_2\p_2+\p_3,$ $e^{x_3}\p_1+e^{ax_3}\p_2$ &\\
& 6 & $\p_1,$ $\p_2,$ $x_1\p_1+ax_2\p_2+\p_3,$ $e^{x_3}\p_1$ &\\
& 7 %8
& $\p_1,$ $x_2\p_1,$ $x_1\p_1+(1-a)x_2\p_2,$ $\p_3$ &\\
& 8 %9
& $\p_1,$ $x_2\p_1,$ $x_1\p_1+(1-a)x_2\p_2,$ $x_3|x_2|^{\frac{1}{1-a}}\p_1$ &\\
& 9 %10
& $\p_1,$ $x_2\p_1,$ $x_1\p_1+(1-a)x_2\p_2,$ $|x_2|^{\frac{1}{1-a}}\p_1$ &
\dotline
$a\not=-1$
& 10 %7
& $\p_1,$ $\p_2,$ $x_1\p_1+ax_2\p_2+\p_3,$ $e^{ax_3}\p_2$ &
\myhline
$A_{3.5}^b \oplus A_1$, $\;b\ge0$
& 1 & $\p_1,$ $\p_2,$ $(b x_1+x_2)\p_1+(-x_1+b x_2)\p_2+\p_3,$ $\p_4$ &\\
$[e_1,e_3]=b e_1-e_2$
& 2 & $\p_1,$ $\p_2,$ $(b x_1+x_2)\p_1+(-x_1+b x_2)\p_2+\p_3,$ $x_4\p_3$ &\\
$[e_2,e_3]=e_1+be_2$
& 3 & $\p_1,$ $\p_2,$ $(b x_1+x_2)\p_1+(-x_1+b x_2)\p_2,$ $\p_3$ &\\
& 4 & $\p_1,$ $\p_2,$ $(b x_1+x_2)\p_1+(-x_1+b x_2)\p_2+\p_3,$ $x_4e^{bx_3}(\cos x_3\p_1-\sin x_3\p_2)$ &\\
& 5 & $\p_1,$ $\p_2,$ $(b x_1+x_2)\p_1+(-x_1+b x_2)\p_2+\p_3,$ $e^{bx_3}(\cos x_3\p_1-\sin x_3\p_2)$ &\\
& 6 & $\p_1,$ $x_2\p_1,$ $(b-x_2)x_1\p_1-(1+x_2^2)\p_2,$ $\p_3$ &\\
& 7 & $\p_1,$ $x_2\p_1,$ $(b-x_2)x_1\p_1-(1+x_2^2)\p_2,$ $x_3\sqrt{1+x_2^2}\,e^{-b\arctan x_2}\p_1$ &\\
& 8 & $\p_1,$ $x_2\p_1,$ $(b-x_2)x_1\p_1-(1+x_2^2)\p_2,$ $\sqrt{1+x_2^2}\,e^{-b\arctan x_2}\p_1$ &
\vspaceafter\\\hline
\end{tabular}\end{center}

\noindent
{\bf Remarks for Table 4.}
\begin{description}\renewcommand{\itemsep}{0ex}\vspace{-1ex}

\item[$R(4A_1,3,\theta).$] $\theta=\theta(x_4,x_5).$
The realizations $R(4A_1,3,\theta)$ and $R(4A_1,3,\tilde\theta)$
are equivalent iff
\begin{equation}\label{equiv.cond.r4a1.3-4}
\tilde\xi^a=-(\xi^b\alpha_{ba}-\alpha_{4a})/(\xi^c\alpha_{c4}-\alpha_{44}),
\end{equation}
where
$\xi^1=x_4,$ $\xi^2= x_5,$ $\xi^3=\theta(x_4,x_5),$
$\tilde\xi^1=\tilde x_4,$ $\tilde\xi^2=\tilde x_5,$ $\tilde\xi^3=\tilde\theta(\tilde x_4,\tilde x_5),$
$a,b,c=\overline{1,3}.$

\item[$R(4A_1,4,(\varphi,\psi)).$] $\varphi=\varphi(x_4),$ $\psi=\psi(x_4).$
The realizations $R(4A_1,4,(\varphi,\psi))$ and $R(4A_1,4,(\tilde\varphi,\tilde\psi))$
are equivalent iff condition~(\ref{equiv.cond.r4a1.3-4}) is satisfied,
where
$\xi^1=x_4,$ $\xi^2=\varphi(x_4),$ $\xi^3=\psi(x_4),$
$\tilde\xi^1=\tilde x_4,$ $\tilde\xi^2=\tilde \varphi(\tilde x_4),$ $\tilde\xi^3=\tilde\psi(\tilde x_4).$

\item[$R(4A_1,6,\theta).$] $\theta=\theta(x_3,x_4,x_5).$
The realizations $R(4A_1,3,\theta)$ and $R(4A_1,3,\tilde\theta)$
are equivalent iff
\begin{equation}\label{equiv.cond.r4a1.6-8}
(\xi^{ik}\alpha_{k,2+j}-\alpha_{2+i,2+j})\tilde\xi^{jl}=-(\xi^{ik}\alpha_{kl}-\alpha_{2+i,l}),
\end{equation}
where
$\xi^{11}=x_3,$ $\xi^{12}=x_4,$ $\xi^{21}=x_5,$ $\xi^{22}=\theta(x_3,x_4,x_5),$
$\tilde\xi^{11}=\tilde x_3,$ $\tilde\xi^{12}=\tilde x_4,$
$\tilde\xi^{21}=\tilde x_5,$ $\tilde\xi^{22}=\tilde\theta(\tilde x_3,\tilde x_4,\tilde x_5),$
$i,j,k,l=1,2.$

\item[$R(4A_1,7,(\varphi,\psi)).$] $\varphi\!=\!\varphi(x_3,x_4),$ $\psi\!=\!\psi(x_3,x_4).$
The realizations $R(4A_1,7,(\varphi,\psi))$ and $R(4A_1,7,(\tilde\varphi,\tilde\psi))$
are equivalent iff condition~(\ref{equiv.cond.r4a1.6-8}) is satisfied,
where
$\xi^{11}=x_3,$ $\xi^{12}=\varphi(x_3,x_4),$ $\xi^{21}=x_4,$ $\xi^{22}=\psi(x_3,x_4),$
$\tilde\xi^{11}=\tilde x_3,$ $\tilde\xi^{12}=\tilde\varphi(\tilde x_3,\tilde x_4),$
$\tilde\xi^{21}=\tilde x_4,$ $\tilde\xi^{22}=\tilde\psi(\tilde x_3,\tilde x_4).$

\item[$R(4A_1,8,(\varphi,\psi,\theta)).$] $\varphi=\varphi(x_3),$ $\psi=\psi(x_3),$ $\theta=\theta(x_3),$
and the vector-functions $(x_3,\varphi)$ and  $(\theta,\psi)$ are linearly independent.
The realizations $R(4A_1,8,(\varphi,\psi,\theta))$ and $R(4A_1,8,(\tilde\varphi,\tilde\psi,\tilde\theta))$
are equivalent iff condition~(\ref{equiv.cond.r4a1.6-8}) is satisfied,
where
$\xi^{11}=x_3,$ $\xi^{12}=\varphi(x_3),$ $\xi^{21}=\theta(x_3),$ $\xi^{22}=\psi(x_3),$
$\tilde\xi^{11}=\tilde x_3,$ $\tilde\xi^{12}=\tilde\varphi(\tilde x_3),$
$\tilde\xi^{21}=\tilde\theta(\tilde x_3),$ $\tilde\xi^{22}=\tilde\psi(\tilde x_3).$

\item[$R(4A_1,10,\theta).$] $\theta=\theta(x_2,x_3),$
and the function $\theta$ is nonlinear with respect to $(x_2,x_3).$
The realizations $R(4A_1,10,\theta)$ and $R(4A_1,10,\tilde\theta)$
are equivalent iff
\begin{equation}\label{equiv.cond.r4a1.10-11}
(\xi^a\alpha_{1,b+1}-\alpha_{a+1,b+1})\tilde\xi^b=-(\xi^a\alpha_{11}-\alpha_{a1}),
\end{equation}
where
$\xi^1=x_2,$ $\xi^2=x_3,$ $\xi^3=\theta(x_2,x_3),$
$\xi^1=\tilde x_2,$ $\xi^2=\tilde x_3,$ $\xi^3=\tilde\theta(\tilde x_2,\tilde x_3),$
$a,b=\overline{1,3}.$

\item[$R(4A_1,11,(\varphi,\theta)).$] $\varphi=\varphi(x_2),$ $\psi=\psi(x_2),$
and the functions $1,$ $x_2,$ $\varphi$ and  $\psi$ are linearly independent.
The realizations $R(4A_1,11,(\varphi,\theta))$ and $R(4A_1,11,(\tilde\varphi,\tilde\theta))$
are equivalent iff condition~(\ref{equiv.cond.r4a1.10-11}) is satisfied,
where
$\xi^1=x_2,$ $\xi^2=\varphi(x_2),$ $\xi^3=\psi(x_2),$
$\xi^1=\tilde x_2,$ $\xi^2=\tilde \varphi(\tilde x_2),$ $\xi^3=\tilde \psi(\tilde x_2).$

\item[$R(A_{2.1}\oplus 2A_1,3,\varphi).$] $\varphi=\varphi(x_4).$
The realizations $R(A_{2.1}\oplus 2A_1,3,\varphi)$ and $R(A_{2.1}\oplus 2A_1,3,\tilde\varphi)$
are equivalent~iff \\
$\tilde x_4=-\alpha_{23}+\alpha_{33}x_4+\alpha_{43}\varphi,$\quad
$\tilde \varphi=-\alpha_{24}+\alpha_{34}x_4+\alpha_{44}\varphi$\\
($\tilde \varphi=\tilde \varphi(\tilde x_4),$ $\alpha_{22}=1,$
$\alpha_{12}=\alpha_{13}=\alpha_{14}=\alpha_{31}=\alpha_{32}=\alpha_{41}=\alpha_{42}=0$).

\item[$R(A_{2.1}\oplus 2A_1,9,\varphi).$] $\varphi=\varphi(x_3).$
The realizations $R(A_{2.1}\oplus 2A_1,9,\varphi)$ and $R(A_{2.1}\oplus 2A_1,9,\tilde\varphi)$
are equivalent~iff \\
$\tilde x_3=-(\alpha_{33}x_3-\alpha_{43})/(\alpha_{34}x_3-\alpha_{44}),$\quad
$\tilde \varphi=(\alpha_{33}+\alpha_{34}\tilde x_3)\varphi-(\alpha_{23}+\alpha_{24}\tilde x_3)$\\
($\tilde \varphi=\tilde \varphi(\tilde x_3),$ $\alpha_{22}=1,$
$\alpha_{12}=\alpha_{13}=\alpha_{14}=\alpha_{31}=\alpha_{32}=\alpha_{41}=\alpha_{42}=0$).

\item[$R(2A_{2.1},3,C).$] $|C|\le1.$
If $C\not=\tilde C$ ($|C|\le1,$ $|\tilde C|\le1$),
the realizations $R(2A_{2.1},3,C)$ and $R(2A_{2.1},3,\tilde C)$ are inequivalent.

\item[$R(A_{3.1}\oplus A_1,3,\varphi).$] $\varphi=\varphi(x_4).$
The realizations $R(A_{3.1}\oplus A_1,3,\varphi)$ and $R(A_{3.1}\oplus A_1,3,\tilde\varphi)$
are equivalent~iff \\
$\tilde x_4=-(\alpha_{22}x_4-\alpha_{32})/(\alpha_{23}x_4-\alpha_{33}),$\quad
$\tilde \varphi=-(\alpha_{44}\varphi+\alpha_{24}x_4-\alpha_{34})/(\alpha_{23}x_4-\alpha_{33})$\\
($\tilde \varphi=\tilde \varphi(\tilde x_4),$ $\alpha_{11}=\alpha_{22}\alpha_{33}-\alpha_{23}\alpha_{32},$
$\alpha_{12}=\alpha_{13}=\alpha_{14}=\alpha_{42}=\alpha_{43}=0$).

\item[$R(A_{3.1}\oplus A_1,8,\varphi).$] $\varphi=\varphi(x_2).$
The realizations $R(A_{3.1}\oplus A_1,8,\varphi)$ and $R(A_{3.1}\oplus A_1,8,\tilde\varphi)$
are equivalent~iff \\
$\tilde x_2=(\alpha_{11}x_2-\alpha_{41})/\alpha_{44},$\quad
$\tilde \varphi=-(\alpha_{22}\varphi-\alpha_{32})/(\alpha_{23}\varphi-\alpha_{33})$
\\
($\tilde \varphi=\tilde \varphi(\tilde x_2),$ $\alpha_{11}=\alpha_{22}\alpha_{33}-\alpha_{23}\alpha_{32},$
$\alpha_{12}=\alpha_{13}=\alpha_{14}=\alpha_{42}=\alpha_{43}=0$).

\item[$R(A_{3.3}\oplus A_1,8,\varphi).$] $\varphi=\varphi(x_2)\not=0.$
The realizations $R(A_{3.3}\oplus A_1,8,\varphi)$ and $R(A_{3.3}\oplus A_1,8,\tilde\varphi)$
are equivalent~iff \\
$\tilde x_2=-(\alpha_{11}x_2-\alpha_{21})/(\alpha_{12}x_2-\alpha_{22}),$\quad
$\tilde \varphi=-\varphi/(\alpha_{34}\varphi-\alpha_{44})$
\\
($\tilde \varphi=\tilde \varphi(\tilde x_2),$
$\alpha_{13}=\alpha_{14}=\alpha_{23}=\alpha_{24}=\alpha_{41}=\alpha_{42}=\alpha_{43}=0,$ $\alpha_{33}=1$).

\end{description}

\newpage

\centerline{\small{\bf Table 5.} Realizations of real indecomposable solvable four-dimensional Lie algebras}

\vspace{-2mm}

\begin{center}\small %\renewcommand{\arraystretch}{1.1}
\begin{tabular}{|p{3cm}|r|p{11cm}|p{0.5cm}|}
\hline\vspacebefore
\hfil Algebra & $N$ &\hfil  Realization&\hfil  $(*)$\\
\hline\vspacebefore
$A_{4.1}$
& 1 &$\p_1$, $\p_2$, $\p_3$, $x_2\p_1+x_3\p_2+\p_4$ &\\
$[e_2,e_4]=e_1$
& 2 &$\p_1$, $\p_2$, $\p_3$, $x_2\p_1+x_3\p_2+x_4\p_3$&\\
$[e_3,e_4]=e_2$
& 3 &$\p_1$, $\p_2$, $\p_3$, $x_2\p_1+x_3\p_2$&\\
& 4 &$\p_1$, $\p_2$, $x_3\p_1+x_4\p_2$, $x_2\p_1+x_4\p_3-\p_4$& \\
& 5 &$\p_1$, $\p_2$, $-\frac 12 x_3^2 \p_1+x_3\p_2$, $x_2\p_1-\p_3$& \\
& 6 &$\p_1$, $x_2\p_1$, $\p_3$, $x_2x_3\p_1-\p_2$& \\
& 7 &$\p_1$, $x_2\p_1$, $x_3\p_1$, $-\p_2-x_2\p_3$ & \\
& 8 &$\p_1$, $x_2\p_1$, $\frac 12 x_2^2 \p_1$,
 $-\p_2$ &
\myhline
 $A_{4.2}^b$, \ $b\not=0$
& 1 %1
&$\p_1$, $\p_2$, $\p_3$, $bx_1\p_1+(x_2+x_3)\p_2+x_3\p_3+\p_4$ & \\
$[e_1,e_4]=be_1$
& 2 %2
&$\p_1$, $\p_2$, $\p_3$, $bx_1\p_1+(x_2+x_3)\p_2+x_3\p_3$ & \\
$[e_2,e_4]=e_2$
& 3 %3
&$\p_1$, $\p_2$, $x_4\p_1+x_3\p_2$, $bx_1\p_1+x_2\p_2+(b-1)x_4\p_4-\p_3$ & \\
$[e_3,e_4]=e_2+e_3$
& 4 %4, \varepsilon=0
&$\p_1$, $\p_2$, $x_3\p_2$, $bx_1\p_1+x_2\p_2-\p_3$ & \\
& 5 %5
&$\p_1$, $x_2\p_1$, $\p_3$, $(bx_1+x_2x_3)\p_1+(b-1)x_2\p_2+x_3\p_3$& \\
& 6 %7
&$\p_1$, $x_2\p_1$, $x_3\p_1$, $bx_1\p_1+(b-1)x_2 \p_2+((b-1)x_3-x_2)\p_3$ &
\dotline
$b\ne1$
& 7 %4, \varepsilon=0, b\not=1
&$\p_1$, $\p_2$, $e^{(1-b)x_3}\p_1+x_3\p_2$, $bx_1\p_1+x_2\p_2-\p_3$ & \\
& 8 %8, b\not=1
&$\p_1$, $x_2\p_1$, $\frac{x_2}{1-b}\ln{|x_2|}\p_1$, $bx_1\p_1+(b-1)x_2\p_2$&
\dotline
$b=1$
& 7 %6,b=1
&$\p_1$, $x_2\p_1$, $\p_3$, $(x_1+x_2x_3)\p_1+x_3\p_3+\p_4$ &
\myhline
$A_{4.3}$
& 1 &$\p_1$, $\p_2$, $\p_3$, $x_1\p_1+x_3\p_2+\p_4$ & \\
$[e_1,e_4]=e_1$
& 2 &$\p_1$, $\p_2$, $\p_3$, $x_1\p_1+x_3\p_2+x_4\p_3$ & \\
$[e_3,e_4]=e_2$
& 3 &$\p_1$, $\p_2$, $\p_3$, $x_1\p_1+x_3\p_2$& \\
& 4 &$\p_1$,$\p_2$, $x_3\p_1+x_4\p_2$, $x_1\p_1+x_3\p_3-\p_4$ & \\
& 5 &$\p_1$, $\p_2$, $\varepsilon e^{-x_3}\p_1+x_3\p_2$, $x_1\p_1-\p_3$ & \\
& 6 & $\p_1$, $x_2\p_1$, $\p_3$, $(x_1+x_2x_3)\p_1+x_2\p_2$ & \\
& 7 &$\p_1$, $x_2\p_1$, $x_3\p_1$, $x_1\p_1+x_2\p_2+(x_3-x_2)\p_3$ & \\
& 8 &$\p_1$, $x_2\p_1$, $-x_2\ln{|x_2|}\p_1$, $x_1\p_1+x_2\p_2$ &
\myhline
$A_{4.4}$
& 1 &$\p_1$, $\p_2$, $\p_3$, $(x_1+x_2)\p_1+(x_2+x_3)\p_2+x_3\p_3+\p_4$ & \\
$[e_1,e_4]=e_1$
& 2 &$\p_1$, $\p_2$, $\p_3$, $(x_1+x_2)\p_1+(x_2+x_3)\p_2+x_3\p_3$ & \\
$[e_2,e_4]=e_1+e_2$
& 3 &$\p_1$, $\p_2$, $x_3\p_1+x_4\p_2$, $(x_1+x_2)\p_1+x_2\p_2+x_4\p_3-\p_4$ & \\
$[e_3,e_4]=e_2+e_3$
& 4 &$\p_1$, $\p_2$, $-\frac 12 x_3^2 \p_1+x_3\p_2$, $(x_1+x_2)\p_1+x_2\p_2-\p_3$ & \\
& 5 &$\p_1$, $x_2\p_1$, $\p_3$, $(x_1+x_2x_3)\p_1-\p_2+x_3\p_3$ & \\
& 6 &$\p_1$, $x_2\p_1$, $x_3\p_1$, $x_1\p_1-\p_2-x_2\p_3$ & \\
& 7 &$\p_1$, $x_2\p_1$, $\frac 12 x_2^2 \p_1$,  $x_1\p_1-\p_2$ &
\myhline
$A_{4.5}^{a,b,c},$ $abc\not=0$
& 1 &$\p_1$, $\p_2$, $\p_3$, $ax_1\p_1+bx_2\p_2+cx_3\p_3+\p_4$ & \\
$[e_1,e_4]=ae_1$
& 2 &$\p_1$, $\p_2$, $\p_3$, $ax_1\p_1+bx_2\p_2+cx_3\p_3$ & \\
$[e_2,e_4]=be_2$
& 3 &$\p_1$, $\p_2$, $x_3\p_1+x_4\p_2$, $ax_1\p_1+bx_2\p_2+(a-c)x_3\p_3+(b-c)x_4\p_4$ & \\
$[e_3,e_4]=ce_3$
& 4 %12
&$\p_1$, $x_2\p_1$, $x_3\p_1$, $ax_1\p_1+(a-b)x_2\p_2+(a-c)x_3\p_3$ &
\dotline
$a=b=c=1$
& 5 %17
& $\p_1$, $\p_2$, $x_3\p_1+x_4\p_2$, $x_1\p_1+x_2\p_2+\p_5$ & \\
& 6 %8
&$\p_1$, $\p_2$, $x_3\p_1+\varphi(x_3)\p_2$, $x_1\p_1+x_2\p_2+\p_4$ & \hfil  $(*)$\\
& 7 %9
&$\p_1$, $\p_2$, $x_3\p_1+\varphi(x_3)\p_2$, $x_1\p_1+x_2\p_2$ & \hfil  $(*)$\\
& 8 %13
&$\p_1$, $x_2\p_1$, $x_3\p_1,x_1\p_1+\p_4$ & \\
& 9% 14
&$\p_1$, $x_2\p_1$, $\varphi(x_2)\p_1$, $x_1\p_1+\p_3$ & \hfil  $(*)$\\
& 10 %15
&$\p_1$, $x_2\p_1$, $\varphi(x_2)\p_1$, $x_1\p_1$ &  \hfil  $(*)$%
\dotline
$a=b=1,$ $c\not=1$
& 5 %4
&$\p_1$, $x_2\p_1$, $\p_3$, $x_1\p_1+cx_3\p_3+\p_4$ & \\
& 6% 5
&$\p_1$, $x_2\p_1$, $\p_3$, $x_1\p_1+cx_3\p_3$  & \\
& 7 %10
&$\p_1$, $\p_2$, $e^{(1-c)x_3}\p_1$, $x_1\p_1+x_2\p_2+\p_3$ &
\dotline
$-1\le a<b<c=1$
& 5 %10
&$\p_1$, $\p_2$, $\varepsilon_1e^{(a-1)x_3}\p_1+\varepsilon_2e^{(b-1)x_3}\p_2$, $ax_1\p_1+bx_2\p_2+\p_3$ &\hfil  $(*)$ \\
$b>0$ if $a=-1$
& 6 %11
&$\p_1$, $x_2\p_1$, $\p_3$, $ax_1\p_1+(a-b)x_2\p_2+x_3\p_3$ &\\
& 7 %16
&$\p_1$, $e^{(a-b)x_2}\p_1$, $e^{(a-1)x_2}\p_1$, $ax_1\p_1+\p_2$&\vspaceafter\\
\hline
\end{tabular}
\end{center}

\newpage

\centerline{\small {\bf Continuation of Table 5.}}

\vspace{-2mm}

\begin{center}\small
\begin{tabular}{|p{3cm}|r|p{11cm}|p{0.5cm}|}
\hline\vspacebefore
\hfil Algebra & $N$ &\hfil  Realization&\hfil  $(*)$\\
\hline\vspacebefore
$A_{4.6}^{a,b},$ $a>0$
& 1 &$\p_1$, $\p_2$, $\p_3$, $ax_1\p_1+(bx_2+x_3)\p_2+(-x_2+bx_3)\p_3 +\p_4$ & \\
$[e_1,e_4]=ae_1$
& 2 &$\p_1$, $\p_2$, $\p_3$, $ax_1\p_1+(bx_2+x_3)\p_2+(-x_2+bx_3)\p_3$ & \\
$[e_2,e_4]=be_2-e_3$
& 3 &$\p_1$, $\p_2$, $x_3\p_1+x_4\p_2$, $(ax_1\!-\!x_2x_3)\p_1+(b\!-\!x_4)x_2\p_2+(a\!-\!b\!-\!x_4)x_3\p_3-(1\!+\!x_4^2)\p_4$ & \\
$[e_3,e_4]=e_2+be_3$
& 4 &$\p_1$, $\p_2$, $\varepsilon e^{(b-a)\arctan{x_3}}\sqrt{1+x_3^2}\p_1+x_3\p_2$,& \\
& &$(ax_1-\varepsilon x_2e^{(b-a)\arctan{x_3}}\sqrt{1+x_3^2}\,)\p_1+(b-x_3)x_2\p_2-(1+x_3^2)\p_3$& \\
& 5 &$\p_1$, $x_2\p_1$, $x_3\p_1$, $ax_1\p_1+((a-b)x_2+x_3)\p_2+(-x_2+(a-b)x_3)\p_3$ & \\
& 6 & $\p_1$, $e^{(a-b)x_2}\cos{x_2}\p_1$, $-e^{(a-b)x_2}\sin{x_2}\p_1$, $ax_1\p_1+\p_2$ &
\myhline
$A_{4.7}$
& 1 &$\p_1$, $\p_2$, $x_2\p_1+\p_3$, $(2x_1+\frac12 x_3^2)\p_1+(x_2+x_3)\p_2+x_3\p_3+\p_4$ & \\[0.2ex]
$[e_2,e_3]=e_1$
& 2 &$\p_1$, $\p_2$, $x_2\p_1+\p_3$, $(2x_1+\frac12 x_3^2)\p_1+(x_2+x_3)\p_2+x_3\p_3$ & \\[0.2ex]
$[e_1,e_4]=2e_1$
& 3 &$\p_1$, $\p_2$, $x_2\p_1+x_3\p_2$, $2x_1\p_1+x_2\p_2-\p_3$ & \\[0.2ex]
$[e_2,e_4]=e_2$
& 4 &$\p_1$, $x_2\p_1$, $-\p_2$, $(2x_1-\frac12 x_2^2)\p_1+x_2\p_2+\p_3$ & \\[0.2ex]
$[e_3,e_4]=e_2+e_3$
& 5 &$\p_1$, $x_2\p_1$, $-\p_2$, $(2x_1-\frac12 x_2^2)\p_1+x_2\p_2$ &
\myhline
$A_{4.8}^b,$ $|b|\leq 1$
& 1 &$\p_1$, $\p_2$, $x_2\p_1+\p_3$, $(1+b)x_1\p_1+x_2\p_2+bx_3\p_3+\p_4$ & \\
$[e_2,e_3]=e_1$
& 2 &$\p_1$, $\p_2$, $x_2\p_1+\p_3$, $(1+b)x_1\p_1+x_2\p_2+bx_3\p_3$ & \\
$[e_1,e_4]=(1+b)e_1$
& 3 %7
&$\p_1$, $\p_2$, $x_2\p_1+x_3\p_2$, $(1+b)x_1\p_1+x_2\p_2+(1-b)x_3\p_3$ & \\
$[e_2,e_4]=e_2$
& 4 %8
&$\p_1$, $\p_2$, $x_2\p_1$, $(1+b)x_1\p_1+x_2\p_2+\p_3$ & \\
$[e_3,e_4]=be_3$
& 5 %9
&$\p_1$, $\p_2$, $x_2\p_1$, $(1+b)x_1\p_1+x_2\p_2$ &
\dotline
$b=1$
& 6 &$\p_1$, $\p_2$, $x_2\p_1+x_3\p_2$, $2x_1\p_1+x_2\p_2+\p_4$ &
\dotline
$b=-1$
& 6 %4
&$\p_1$, $\p_2$, $x_2\p_1+\p_3$, $x_4\p_1+x_2\p_2-x_3\p_3$ & \\
& 7 %10
&$\p_1$, $\p_2$, $x_2\p_1$, $x_3\p_1+x_2\p_2$ &
\dotline
$b\not=\pm1$
& 6 %11
&$\p_1$, $x_2\p_1$, $-\p_2$, $(1+b)x_1\p_1+bx_2\p_2+\p_3$ & \\
& 7 %12
&$\p_1$, $x_2\p_1$, $-\p_2$, $(1+b)x_1\p_1+bx_2\p_2$ &
\dotline
$b=0$
& 8 %3
&$\p_1$, $\p_2$, $x_2\p_1+\p_3$, $x_1\p_1+x_2\p_2+x_4\p_3$ & \\
& 9 %5
&$\p_1$, $\p_2$, $x_2\p_1+\p_3$, $x_1\p_1+x_2\p_2+C\p_3$ & $(*)$
\myhline
$A_{4.9}^a$,\ $a\geq 0$
& 1 &$\p_1$, $\p_2$, $x_2\p_1+\p_3$, $\frac12(4ax_1+x_3^2-x_2^2)\p_1+(ax_2+x_3)\p_2+(-x_2+ax_3)\p_3+\p_4$ & \\
$[e_2,e_3]=e_1$
&&&\\
$[e_1,e_4]=2ae_1$
& 2 &$\p_1$, $\p_2$, $x_2\p_1+\p_3$, $\frac12(4ax_1+x_3^2-x_2^2)\p_1+(ax_2+x_3)\p_2+(-x_2+ax_3)\p_3$ & \\
$[e_2,e_4]=ae_2-e_3$
&&& \\
$[e_3,e_4]=e_2+ae_3$
& 3 %4
&$\p_1$, $\p_2$, $x_2\p_1+x_3\p_2$, $(2ax_1-\frac12 x_2^2)\p_1+(a-x_3)x_2\p_2-(1+x_3^2)\p_3$ &
\dotline
$a=0$
& 4 %3
&$\p_1$, $\p_2$, $x_2\p_1+\p_3$, $\frac 12(x_3^2-x_2^2+2x_4)\p_1+x_3\p_2-x_2\p_3$&
\myhline
$A_{4.10}$
& 1 &$\p_1$, $\p_2$, $x_1\p_1+x_2\p_2+\p_3$, $x_2\p_1-x_1\p_2+\p_4$ &\\
$[e_1,e_3]=e_1$
& 2 &$\p_1$, $\p_2$, $x_1\p_1+x_2\p_2+\p_3$, $x_2\p_1-x_1\p_2+x_4\p_3$ & \\
$[e_2,e_3]=e_2$
& 3 &$\p_1$, $\p_2$, $x_1\p_1+x_2\p_2+\p_3$, $x_2\p_1-x_1\p_2+C\p_3$& $(*)$\\
$[e_1,e_4]=-e_2$
& 4 &$\p_1$, $x_2\p_1$, $x_1\p_1+\p_3$, $-x_1x_2\p_1-(1+x_2^2)\p_2$ & \\
$[e_2,e_4]=e_1$
& 5 &$\p_1$, $\p_2$,  $x_1\p_1+x_2\p_2$, $x_2\p_1-x_1\p_2+\p_3$& \\
& 6 &$\p_1$, $\p_2$, $x_1\p_1+x_2\p_2$, $x_2\p_1-x_1\p_2$ & \\
& 7 &$\p_1$, $x_2\p_1$, $x_1\p_1$, $-x_1x_2\p_1-(1+x_2^2)\p_2$& \vspaceafter\\
\hline
\end{tabular}
\end{center}

\noindent
{\bf Remarks for Table 5.}
\begin{description}\renewcommand{\itemsep}{0ex}\vspace{-1ex}

\item[$R(A_{4.5}^{1,1,1},N,\varphi),$ $N=6,7.$]  $\varphi=\varphi(x_3).$
The realizations $R(A_{4.5}^{1,1,1},N,\varphi)$ and $R(A_{4.5}^{1,1,1},N,\tilde\varphi)$
are equivalent iff condition~(\ref{equiv.cond.r3a1.3}) is satisfied
($\tilde \varphi=\tilde \varphi(\tilde x_3),$ $\alpha_{41}=\alpha_{42}=\alpha_{43}=0$).

\item[$R(A_{4.5}^{1,1,1},N,\varphi),$ $N=9,10.$]  $\varphi=\varphi(x_2),$  $\varphi''\ne 0.$
The realizations $R(A_{4.5}^{1,1,1},N,\varphi)$ and $R(A_{4.5}^{1,1,1},N,\tilde\varphi)$
are equivalent iff condition~(\ref{equiv.cond.r3a1.5}) is satisfied
($\tilde \varphi=\tilde \varphi(\tilde x_2),$ $\alpha_{41}=\alpha_{42}=\alpha_{43}=0$).

\item[$R(A_{4.5}^{a,b,c},5,(\varepsilon_1,\varepsilon_2)),$]  where $-1\le a<b<c=1,$  $b>0$ if $a=-1.$
$\varepsilon_i\in\{0;1\}$, $(\varepsilon_1,\varepsilon_2)\not=(0,0)$
(three different variants are possible). All the variants are inequivalent.

\item[$R(A_{4.8}^0,9,C).$] $C\not=0$ (since $R(A_{4.8}^0,9,0)=R(A_{4.8}^0,2)$).

\item[$R(A_{4.10},3,C).$] $C$ is an arbitrary constant.

\end{description}

\newpage

\centerline{\small{\bf Table 6.} Realizations of real unsolvable three- and
four-dimensional Lie algebras}

\vspace{-2mm}

\begin{center}\small
\begin{tabular}{|p{3cm}|r|p{11cm}|p{0.5cm}|}
\hline\vspacebefore
\hfil Algebra & $N$ &\hfil  Realization&\hfil  $(*)$\\
\hline\vspacebefore
$sl(2,{\mathbb R})$
& 1 & $\p_1$, $x_1\p_1 +x_2\p_2$, $x_1^2\p_1+2x_1x_2\p_2+x_2\p_3$ & \\
$[e_1,e_2]=e_1$
& 2 & $\p_1$, $x_1\p_1 +x_2\p_2$, $(x_1^2-x_2^2)\p_1+2x_1x_2\p_2$ & \\
$[e_2,e_3]=e_3$
& 3 & $\p_1$, $x_1\p_1 +x_2\p_2$, $(x_1^2+x_2^2)\p_1+2x_1x_2\p_2$ & \\
$[e_1,e_3]=2e_2$
& 4 & $\p_1$, $x_1\p_1 +x_2\p_2$, $x_1^2\p_1+2x_1x_2\p_2$ & \\
& 5 & $\p_1$, $x_1\p_1$, $x_1^2\p_1$ & \myhline
$sl(2,{\mathbb R})\oplus A_1$
& 1 & $\p_1$, $x_1\p_1+x_2\p_2$, $x_1^2\p_1+2x_1x_2\p_2+x_2\p_3$, $\p_4$ & \\
$[e_1,e_2]=e_1$
& 2 & $ \p_1$, $x_1\p_1 +x_2\p_2$, $x_1^2\p_1+2x_1x_2\p_2+x_2\p_3$, $x_2\p_1+2x_2x_3\p_2+(x_3^2+x_4)\p_3$ & \\
$[e_2,e_3]=e_3$
& 3 & $\p_1$, $x_1\p_1 +x_2\p_2$, $x_1^2\p_1+2x_1x_2\p_2+x_2\p_3$, $x_2\p_1+2x_2x_3\p_2+(x_3^2+c)\p_3$, \quad $c\in \{-1;0;1\}$ & \\
$[e_1,e_3]=2e_2$
& 4 & $\p_1$, $x_1\p_1 +x_2\p_2$, $(x_1^2+x_2^2)\p_1+2x_1x_2\p_2$, $\p_3$ & \\
& 5 & $\p_1$, $x_1\p_1 +x_2\p_2$, $(x_1^2-x_2^2)\p_1+2x_1x_2\p_2$, $\p_3$ & \\
& 6 & $\p_1$, $x_1\p_1 +x_2\p_2$, $x_1^2\p_1+2x_1x_2\p_2$, $\p_3$ & \\
& 7 & $\p_1$, $x_1\p_1 +x_2\p_2$, $x_1^2\p_1+2x_1x_2\p_2$, $x_2x_3\p_2$ & \\
& 8 & $\p_1$, $x_1\p_1 +x_2\p_2$, $x_1^2\p_1+2x_1x_2\p_2$, $x_2\p_2$ & \\
& 9 & $\p_1$, $x_1\p_1$, $x_1^2\p_1$, $\p_2$ &
\myhline
$so(3)$
& 1 & $-\sin x_1 \tan x_2\p_1-\cos x_1\p_2$, $-\cos x_1\tan x_2 \p_1 +\sin x_1 \p_2$, $\p_1$ & \\
$[e_2,e_3]=e_1$
& 2 & $-\sin x_1 \tan x_2 \p_1-\cos x_1\p_2+\sin x_1 \sec x_2\p_3$,& \\
$[e_3,e_1]=e_2$
&    & $-\cos x_1\tan x_2 \p_1 +\sin x_1 \p_2+\cos x_1\sec x_2\p_3$, $\p_1$ & \\
$[e_1,e_2]=e_3$&&&
\myhline
$so(3)\oplus A_1$
& 1 & $-\sin x_1 \tan x_2\p_1-\cos x_1\p_2$, $-\cos x_1\tan x_2 \p_1 +\sin x_1 \p_2$, $\p_1$, $\p_3$ & \\
$[e_2,e_3]=e_1$
& 2 & $-\sin x_1 \tan x_2 \p_1-\cos x_1\p_2+\sin x_1\sec x_2\p_3$, & \\
$[e_3,e_1]=e_2$
&    & $-\cos x_1\tan x_2 \p_1 +\sin x_1 \p_2+\cos x_1\sec x_2\p_3$, $\p_1$, $\p_3$ & \\
$[e_1,e_2]=e_3$
& 3 & $-\sin x_1 \tan x_2 \p_1-\cos x_1\p_2+\sin x_1\sec x_2\p_3$, & \\
&    & $-\cos x_1\tan x_2 \p_1 +\sin x_1 \p_2+\cos x_1\sec x_2\p_3$, $\p_1$, $x_4\p_3$ & \\
& 4 & $-\sin x_1 \tan x_2 \p_1-\cos x_1\p_2+\sin x_1\sec x_2\p_3$, & \\
&    & $-\cos x_1\tan x_2 \p_1 +\sin x_1 \p_2+\cos x_1\sec x_2\p_3$, $\p_1$, $\p_4$& \vspaceafter\\
\hline
\end{tabular}
\end{center}

\medskip

\noindent
{\bf Remark.}
The realizations $R(so(3),1)$ and $R(so(3),2)$ are well-known.
At the best of our knowledge, completeness of the list of these realizations
was first proved
in~\cite{zhdanov&lahno&fushchych2000}.
We do not assert that the adduced forms of realizations are optimal for all applications
and that the classification from Table 6 is canonical.

Consider the realization $R(so(3),1)$ of rank 2 in more details.
The corresponding transformation group acts transitively on the manifold~$S^2$.
With the stereographic projection $\tan{x_1}=t/x$, ${\rm cotan}\ x_2=\sqrt{x^2+t^2}$
it can be reduced to the well known realization on the plane~\cite{Gonzales&Kamran&Olver1992a}:
\[
(1+t^2)\partial_t+xt\partial_x,\quad\ x\partial_t-t\partial_x, \quad
-xt\partial_t-(1+x^2)\partial_x.
\]
If dimension of the $x$-space is not smaller than $3$, the variables $x_1$, $x_2$ and
the implicit variable $x_3$ in $R(so(3),1)$ can be interpreted as the angles and the radius
of the spherical coordinates that is equivalent to imbedding $S^2$ in ${\mathbb R}^3$.
Then, in the corresponding Cartesian coordinates the realization $R(so(3),1)$ has the well-known form:
\[
x_2\partial_3-x_3\partial_2,\qquad x_3\partial_1-x_1\partial_3,\qquad x_1\partial_2-x_2\partial_1,
\]
which is generated by the standard representation of $SO(3)$ as the rotation group in ${\mathbb R}^3$.

\newpage

\section{Comparison of our results and those of \cite{soh&mahomed2001jphysa}}
\label{secWafo}

The results of this paper include, as a particular case,
realizations in three variables $x=(x_1,x_2,x_3)$,
which were considered in~\cite{soh&mahomed2001jphysa}.
That is why it is interesting for us to compare the lists of realizations.

In general, a result of classification
may contain errors of two types:
\begin{itemize}
\itemsep=0pt
\item missing some inequivalent cases and
\item including mutually equivalent cases.
\end{itemize}
Summarizing the comparison given below, we can state
that errors of both types are in~\cite{soh&mahomed2001jphysa}.
Namely, for three-dimensional algebras 3 cases are missing,
1 case is equivalent to other case, and 1 case can be reduced
to 3 essentially simpler cases. For four-dimensional algebras
34 cases are missing, 13 cases are equivalent to other cases.
Such errors are usually caused by incorrect performing of changes of variables
and also shortcomings in the algorithms employed.
See other errors in the comparison list.

Below we keep notations of~\cite{soh&mahomed2001jphysa}
(${\cal L}_{\dots},$ ${\cal L}_{\dots}^{\dots},$ $X_{\dots}$)
and our notations ($A_{\dots},$ $R(A_{\dots},\dots),$ $e_{\dots}$) for algebras, realizations and their basis elements.
We list the pairs of equivalent realizations ${\cal L}_{r.m_1}^{k_1}$ and
$R(A_{r.m_2},k_2)$ using the shorthand notation  $k_1\sim k_2$
as well as all the differences of classifications.
In the cases when equivalence of realizations is not obvious
we give the necessary transformations of variables and basis changes.

\begin{description}

\item[Three-dimensional algebras]

\item[${\cal L}_{3.1}\sim 3A_1.$]
$1\sim 1;$
$2\sim 3$ (one of the parameter-functions of ${\cal L}_{3.1}$ can be made equal to $t$);
$3\sim 4;$
the realization $R(3A_1,5)$ is missing in~\cite{soh&mahomed2001jphysa}.

\item[${\cal L}_{3.2}\sim A_{2.1}\oplus A_1$] ($X_1=-e_2,$ $X_2=e_1,$ $X_3=e_3$).
$1\sim 3;$
$2\sim 1;$
the series of realizations ${\cal L}_{3.2}^3$ with two  parameter-functions $f$ and $g$ can be reduced to
three realizations: \\
$R(A_{2.1}\oplus A_1,2)$ if $f'\not=0$ ($x_1=y-xg(t)/f(t),$ $x_2=\ln|x|/f(t),$ $x_3=1/f(t)$),\\
$R(A_{2.1}\oplus A_1,3)$ if $f'=0$ and $f\not=0$  ($x_1=y-xg(t)/f,$ $x_2=\ln|x|/f,$ $x_3=t,$
$X_1=-e_2-(1/f)e_3,$ $X_2=e_1,$ $X_3=e_3$), which coincides with ${\cal L}_{3.2}^1,$\\
$R(A_{2.1}\oplus A_1,4)$ if $f=0$ and, therefore, $g\not=0$ ($x_1=g(t)x,$ $x_2=y,$ $x_3=t$).
\item[${\cal L}_{3.3}\sim A_{3.1}.$]
$1\sim 3;$
$2\sim 2;$
$3\sim 1;$
$4\sim {\cal L}_{3.3}^1.$
\item[${\cal L}_{3.4}\sim A_{3.2}.$]
$1\sim 2;$
$2\sim 1;$
$3\sim 3.$
\item[${\cal L}_{3.5}\sim A_{3.3}.$]
$1\sim 2;$
$2\sim 1;$
$3\sim 4;$
$4\sim 3.$
\item[${\cal L}_{3.6}^a\sim A_{3.4}^a.$]
$1\sim 2;$
$2\sim 1;$
$3\sim 3.$
\item[${\cal L}_{3.7}^a\sim A_{3.5}^a.$]
$1\sim 2;$
$2\sim 1;$
$3\sim 3.$
\item[${\cal L}_{3.8}\sim sl(2,\R).$]
$1\sim 5;$
$2\sim 1;$
$3\sim 3$ ($x_1=(x+t)/2$, $x_2=(x-t)/2$);
$4\sim 4$ ($x_1=-x/t$, $x_2=1/t^2$, $x_3=y$);
the realization $R(sl(2,\R),2)$ is missing in~\cite{soh&mahomed2001jphysa}.
\item[${\cal L}_{3.9}\sim so(3).$]
$1\sim 1$ ($x_1=\arctan t/x,$ $x_2=\mathop{\rm arccot}\nolimits\sqrt{x^2+t^2},$ $e_1=X_3,$ $e_2=-X_1,$ $e_3=X_2$);
the realization $R(so(3),2)$ is missing in~\cite{soh&mahomed2001jphysa}.

\item[Four-dimensional algebras]

\item[${\cal L}_{4.1}\sim 4A_1.$]
$1\sim 8$ (one of the parameter-functions of ${\cal L}_{3.1}$ can be made equal to $t$);
$2\sim 10;$
the realization $R(4A_1,11)$ is missing in~\cite{soh&mahomed2001jphysa}.

\item[${\cal L}_{4.2}\sim A_{2.1}\oplus 2A_1$] ($X_1=-e_2,$ $X_2=e_1,$ $X_3=e_3,$ $X_4=e_4$).
$1\sim 10;$
$2\sim 4;$
$3\sim {\cal L}_{4.2}^2$ ($\tilde x=\ln|t|,$ $\tilde y=y,$ $\tilde t=x/t$);
$4\sim {\cal L}_{4.2}^5$ ($\tilde x=x/t,$ $\tilde y=y,$ $\tilde t=1/t$);
$5\sim 6;$
$6\sim 9;$
$7\sim {\cal L}_{4.2}^1$ if $f=0$ ($\tilde x=ye^{-x}/g(t),$ $\tilde y=e^{-x}/g(t),$ $\tilde t=te^{-x}/g(t)$)
or $7\sim {\cal L}_{4.2}^6$ if $f\not=0$ ($\tilde x=-e^{-x}/f(t),$ $\tilde y=y-xg(t)/f(t),$ $\tilde t=t$).

\item[${\cal L}_{4.3}\sim 2A_{2.1}$] ($X_1=-e_2,$ $X_2=e_1,$ $X_3=-e_4,$ $X_4=e_3$).
$1\sim {\cal L}_{4.3}^3$
($\tilde x=t,$ $\tilde y=x,$ $\tilde t=y;$ $\tilde X_1=X_3,$ $\tilde X_2=X_4,$ $\tilde X_3=X_1,$ $\tilde X_4=X_2$);
$2\sim 7;$
$3\sim 3^{C=0};$
$4\sim 6$ ($x_1=y,$ $x_2=t,$ $x_3=\ln|x/t|$);
$5\sim {\cal L}_{4.3}^3$ ($\tilde x=1/x,$ $\tilde y=y/x,$ $\tilde t=t$);
$6\sim 3^{C=1}$ ($x_1=y,$ $x_2=x/t,$ $x_3=\ln|t|$);
$7\sim 4$ ($x_1=y,$ $x_2=x/t,$ $x_3=1/t$);
$8\sim 5;$
the realization $R(2A_{2.1},3,C)$ ($C\not=0,1$) is missing in~\cite{soh&mahomed2001jphysa}.

\item[${\cal L}_{4.4}\sim A_{3.1}\oplus A_1.$]
The realization ${\cal L}_{4.4}^1$ is a particular case of ${\cal
L}_{4.4}^4;$ $2\sim 5;$ the basis operators of ${\cal L}_{4.4}^3$
do not satisfy the commutative relations of ${\cal L}_{4.4};$
$4\sim 8$.

\item[${\cal L}_{4.5}\sim A_{3.2}\oplus A_1.$]
$1\sim 8$ ($x_1=x,$ $x_2=t,$ $x_3=ye^t$);
$2\sim 6$ ($x_1=x,$ $x_2=y,$ $x_3=\ln|t|$);
$3\sim 5$ ($x_1=x-tye^{-t},$ $x_2=ye^{-t},$ $x_3=-t$);
$4\sim 3$ ($x_1=t,$ $x_2=x,$ $x_3=y$);
the realizations $R(A_{3.2}\oplus A_1,7)$ and $R(A_{3.2}\oplus A_1,9)$ are missing in~\cite{soh&mahomed2001jphysa}.

\item[${\cal L}_{4.6}^1\sim A_{3.3}\oplus A_1.$]
$1\sim 9$ ($x_1=x,$ $x_2=t,$ $x_3=\ln|y|$);
$2\sim 5$ ($x_1=x,$ $x_2=y,$ $x_3=\ln|t|$);
$3\sim {\cal L}_{4.6}^{1,2}$
($\tilde x=y,$ $\tilde y=x,$ $\tilde t=t;$ $\tilde X_1=X_2,$ $\tilde X_2=X_1,$ $\tilde X_3=-X_3,$ $\tilde X_4=X_4$);
$4\sim 3;$
$5\sim {\cal L}_{4.6}^{1,2}$
($\tilde x=x,$ $\tilde y=ty,$ $\tilde t=t;$ $\tilde X_1=X_1+X_2,$ $\tilde X_2=X_2,$ $\tilde X_3=X_3,$ $\tilde X_4=X_4$);
the series of realizations $R(A_{3.3}\oplus A_1,8)$ are missing in~\cite{soh&mahomed2001jphysa}.

\item[${\cal L}_{4.6}^a\sim A_{3.4}^a\oplus A_1$] ($-1\le a<1$, $a\not=0$).
$1\sim 8$ ($x_1=x,$ $x_2=t,$ $x_3=y|t|^{-\frac1{1-a}}$);
$2\sim 6$ ($x_1=x,$ $x_2=y,$ $x_3=\ln|t|$);
$3\sim 10$ if $a\not=-1$ ($x_1=x,$ $x_2=y,$ $x_3=\frac1a\ln|t|$),
$3\sim {\cal L}_{4.6}^{-1,2}$ for $a=-1$
($\tilde x=y,$ $\tilde y=x,$ $\tilde t=t;$ $\tilde X_1=X_2,$ $\tilde X_2=X_1,$ $\tilde X_3=X_3,$ $\tilde X_4=X_4$);
$4\sim 3;$
$5\sim 5$ ($x_1=x,$ $x_2=yt^a+xt^{a-1},$ $x_3=\ln|t|$);
the realizations $R(A_{3.4}^a\oplus A_1,7)$ and $R(A_{3.4}^a\oplus A_1,9)$
are missing in~\cite{soh&mahomed2001jphysa}.

\item[${\cal L}_{4.7}\sim A_{3.5}^0\oplus A_1.$]
The basis operators of ${\cal L}_{4.7}^1$ do not satisfy the
commutative relations of ${\cal L}_{4.7};$ $2\sim 3;$ $3\sim 5;$
the realizations $R(A_{3.5}^0\oplus A_1,6),$ $R(A_{3.5}^0\oplus
A_1,7),$ and  $R(A_{3.5}^0\oplus A_1,8)$ are missing
in~\cite{soh&mahomed2001jphysa}; the zero value of the parameter
of algebra series $A_{3.5}^a\oplus A_1$ is not special with
respect to constructing of inequivalent realizations.

\item[${\cal L}_{4.8}^a\sim A_{3.5}^a\oplus A_1$] ($a>0$).
$1\sim 3;$
$2\sim 5$ (the notation of $X_4$ contains some misprints);
the realizations $R(A_{3.5}^a\oplus A_1,6),$ $R(A_{3.5}^a\oplus A_1,7),$ and  $R(A_{3.5}^a\oplus A_1,8)$
are missing in~\cite{soh&mahomed2001jphysa}.

\item[${\cal L}_{4.9}\sim sl(2,\R)\oplus A_1$] ($e_1=X_1,$ $e_2=X_2,$ $e_3=-X_3,$ $e_4=X_4$).
$1\sim 3^{c=0}$ ($x_1=t+x/(1+y),$ $x_2=t/(1+y),$  $x_3=-y(1+y)$);
$2\sim 4$ ($x_1=(t+x)/2,$ $x_2=(t-x)/2,$ $x_3=y$);
$3\sim 6$ ($x_1=-x/t,$ $x_2=-1/t^2,$ $x_3=y$);
$4\sim 9;$
the realizations $R(sl(2,\R)\oplus A_1,3,c)$ ($c=\pm1$), $R(sl(2,\R)\oplus A_1,5),$ $R(sl(2,\R)\oplus A_1,7),$
and $R(sl(2,\R)\oplus A_1,8)$ are missing in~\cite{soh&mahomed2001jphysa}.

\item[${\cal L}_{4.10}\sim so(3)\oplus A_1.$]
$1\sim 1$ ($x_1=\arctan t/x,$ $x_2=\mathop{\rm arccot}\nolimits\sqrt{x^2+t^2},$ $x_3=y,$
$e_1=X_3,$ $e_2=-X_1,$ $e_3=X_2,$ $e_4=X_4$);
the realization $R(so(3)\oplus A_1,2)$ is missing in~\cite{soh&mahomed2001jphysa}.

\item[${\cal L}_{4.11}\sim A_{4.1}.$]
$1\sim 7;$
$2\sim 5;$
$3\sim 3;$
the realizations $R(A_{4.1},6)$ and $R(A_{4.1},8)$ are missing in~\cite{soh&mahomed2001jphysa}.

\item[${\cal L}_{4.12}^a\sim A_{4.2}^a$] ($a\not=0$).
$1\sim 6;$
$2\sim 4;$
$3\sim 2;$
$4\sim 7$ for $a\not=1$ and $4\sim {\cal L}_{4.12}^{a,2}$ for $a=1;$
the realizations $R(A_{4.2}^a,5),$ $R(A_{4.2}^a,7)$ ($a=1$), and $R(A_{4.2}^a,8)$ ($a\not=1$)
are missing in~\cite{soh&mahomed2001jphysa}.

\item[${\cal L}_{4.13}\sim A_{4.3}.$]
$1\sim 7;$
$2\sim 3;$
$3\sim 5;$
the realizations $R(A_{4.3},6)$ and $R(A_{4.3},8)$ are missing in~\cite{soh&mahomed2001jphysa}.

\item[${\cal L}_{4.14}\sim A_{4.4}.$]
$1\sim 6;$
$2\sim 2;$
$3\sim 4;$
the realizations $R(A_{4.4},5)$ and $R(A_{4.4},7)$ are missing in~\cite{soh&mahomed2001jphysa}.

\item[ ${\cal L}_{4.15}^{a,b}\sim A_{4.5}^{a,b,1}$]
($-1\le a<b<1,$ $ab\not=0$, $e_1=-X_2,$ $e_2=X_3,$ $e_3=X_1,$ $e_4=X_4$).
$1\sim 4$ ($x_1=-x/t,$ $x_2=-y/t,$ $x_3=-1/t$);
$2\sim 2;$
$3\sim 5^{\varepsilon_1=0}$ ($x_1=-y,$ $x_2=x/t,$ $x_3=(1-b)^{-1}\ln|t|$);
$4\sim 6$ ($x_1=-y,$ $x_2=t,$ $x_3=x$);
$5\sim 5^{\varepsilon_1=\varepsilon_2=1}$ ($x_1=-y+e^{(a-1)t}x,$ $x_2=e^{(b-1)t}x,$ $x_3=t$);
the realizations $R(A_{4.5}^{a,b,1},5^{\varepsilon_2=0})$ and $R(A_{4.5}^{a,b,1},7)$
are missing in~\cite{soh&mahomed2001jphysa}.

\item[${\cal L}_{4.15}^{a,a}\sim A_{4.5}^{1,1,a^{-1}}$]
($-1< a<1,$ $a\not=0$, $e_1=X_3,$ $e_2=X_2,$ $e_3=X_1,$ $e_4=X_4,$
${\cal L}_{4.15}^{-1,-1}\sim{\cal L}_{4.15}^{-1,1}$).
$1\sim 4$ ($x_1=x/y,$ $x_2=t/y,$ $x_3=1/y$);
$2\sim 2;$
$3\sim 7$ ($x_1=x/t,$ $x_2=y,$ $x_3=a(a-1)^{-1}\ln|t|$);
$4\sim 6$ ($x_1=y/t,$ $x_2=1/t,$ $x_3=x$).

\item[${\cal L}_{4.15}^{a,1}\sim A_{4.5}^{1,1,a}$]
($-1\le a<1,$ $a\not=0$, $e_1=X_1,$ $e_2=X_3,$ $e_3=X_2,$ $e_4=X_4$).
$1\sim 4;$
$2\sim 2;$
$3\sim 6;$
$4\sim 7$ ($x_1=x,$ $x_2=y/t,$ $x_3=(a-1)^{-1}\ln|t|$).

\item[${\cal L}_{4.16}\sim A_{4.5}^{1,1,1}.$]
$1\sim 2;$
$2\sim 4;$
$3\sim 7$ (the function $f(t)$ can be made equal to $t$);
the realizations $R(A_{4.5}^{1,1,1},9)$ and
$R(A_{4.5}^{1,1,1},10)$ are missing in~\cite{soh&mahomed2001jphysa}.

\item[${\cal L}_{4.17}^{a,b}\sim A_{4.6}^{a,b}.$]
$1\sim 5;$
$2\sim 2;$
$3\sim 4;$
the realizations $R(A_{4.6}^{a,b},6)$ is missing in~\cite{soh&mahomed2001jphysa}.

\item[${\cal L}_{4.18}\sim A_{4.7}.$]
$1\sim 5$ ($x_1=x/2,$ $x_2=t,$ $x_3=y$);
$2\sim 4$ ($x_1=x/2,$ $x_2=t,$ $x_3=y$);
$3\sim 2$ ($x_1=x/2,$ $x_2=t,$ $x_3=y$);
$4\sim 3$ ($x_1=y,$ $x_2=x,$ $x_3=-t$).

\item[${\cal L}_{4.19}\sim A_{4.8}^{-1}.$]
$1\sim 7;$
$2\sim {\cal L}_{4.19}^8$ and
$3\sim {\cal L}_{4.19}^1$
($\tilde x=t,$ $\tilde y=x,$ $\tilde t=-y,$ $\tilde X_1=X_1,$ $\tilde X_2=-X_3,$ $\tilde X_3=X_2,$ $\tilde X_4=-X_4$);
$4\sim {\cal L}_{4.19}^5$
($\tilde x=t,$ $\tilde y=x,$ $\tilde t=e^{-2y},$ $\tilde X_1=X_1,$ $\tilde X_2=-X_3,$ $\tilde X_3=X_2,$ $\tilde X_4=-X_4$);
$5\sim 4$ ($x_1=y,$ $x_2=x,$ $x_3=\frac12\ln|t|$);
$6\sim 3;$
$7\sim 2;$
$8\sim 5.$

\item[${\cal L}_{4.20}^b\sim A_{4.8}^b$] ($-1<b\le 1$).
$1\sim 5;$
$2\sim 7$ and
$3\sim 6$
(these realizations can be inscribed in the list of inequivalent realizations iff $b\not=\pm1$);
$4\sim 4$ for $b\not=1$ ($x_1=y,$ $x_2=x,$ $x_3=(1-b)^{-1}\ln|t|$) and $4\sim
{\cal L}_{4.20}^{b.1}$ if $b=1;$
$5\sim 3;$
$6\sim 2;$
the realizations $R(A_{4.8}^{0},9,C)$ is missing in~\cite{soh&mahomed2001jphysa}.

\item[${\cal L}_{4.21}^a\sim A_{4.9}^a$] ($a\ge0$).
$1\sim 2;$
$2\sim 3;$
the realizations $R(A_{4.9}^0,4)$ is missing in~\cite{soh&mahomed2001jphysa}.

\item[${\cal L}_{4.22}\sim A_{4.10}.$]
$1\sim 7;$
$2\sim 6;$
$3\sim 4;$
$4\sim 5;$
$5\sim 3.$

\end{description}

\section{Conclusion}\label{secConl}
We plan to extend this study by including the results of classifying realizations with respect to
the strong equivalence and more detailed description of
algebraic properties of low-dimen\-sio\-nal Lie algebras and the classification technique.
We have also begun the investigations on
complete description of differential invariants and operators of invariant differentiation
for all the constructed realizations, as well as ones on applications of the obtained results.
(Let us note that the complete system of differential invariants for all the Lie groups,
from Lie's classification, of point and contact transformations acting on a two-dimensional complex space
was determined in~\cite{olver1994}.
The differential invariants of one-parameter groups of local transformations were exhaustively
described in~\cite{popovych&boyko2002} in the case of arbitrary number of independent and dependent  variables.)
Using the above classification of inequivalent realizations
of real Lie algebras of dimension no greater than four, one can solve, in a quite clear way,
the group classification problems for  the following classes of differential equations with real variables:
\begin{itemize}\itemsep=0ex
\item
ODEs of order up to four;
\item
systems of two second-order ODEs;
\item
systems of two, three and four first-order ODEs;
\item
general systems of two hydrodynamic-type equations with two independent variables;
\item
first-order PDEs with two independent variables;
\item second-order evolutionary PDEs.
\end{itemize}
All the above classes of differential equations occur frequently
in applications (classical, fluid and quantum mechanics, general
relativity, mathematical biology, etc). Third- and fourth-order ODEs and the
second class were investigated, in some way, in
\cite{Cerquetelli&Ciccoli&Nucci2002,mahomed&leach1989,Schmucker&Czichowski1998,soh&mahomed2001jphysa}.
Now we perform group classification for the third and fourth
classes and fourth-order ODEs. Solving the group classification
problem for the last class is necessary in order to construct first-order
differential constraints being compatible with well-known
nonlinear second-order PDEs.

Our results can be also applied to solving the interesting and important problem of
studying finite-dimensional Lie algebras of first-order differential operators.
(There are considerable number of papers devoted to this problem,
see e.g.~\cite{baranovskii&shirokov2003,Gonzales&Kamran&Olver1992b,Milsom1995}.)

It is obvious that our classification can be transformed to classification of
realizations of complex Lie algebras of dimension no greater than four in vector
fields on a space of an arbitrary (finite) number of complex variables.
We also hope to solve the analogous problem for five-dimensional
algebras in the near future.

\medskip

{\bf Acknowledgments.} The authors are grateful to Profs.~A.~Nikitin and R.~Zhdanov
and Drs.~A.~Sergyeyev,  I.~Yehorchenko and A.~Zhalij,  for useful discussions.
The authors thank Profs. G.~Post,
I.~Shirokov, A.~Turbiner and P.~Wintenitz for interesting comments to the previous versions of this preprint.
The authors specially appreciate Prof. P.~Wintenitz for the kindly provided manuscript~\cite{winternitz1990}
and Prof. J.~Patera for very stimulated lectures for us and permanent support of our research on the
subject.

The research of VMB was partially supported by STCU, Project N U039k.
VMB and MON were partially supported by National Academy of Sciences of Ukraine
in the form of the grant for young scientists.

\appendix
\newcommand{\Int}{\mathop{\rm Int}\nolimits}
\newcommand{\Aut}{\mathop{\rm Aut}\nolimits}
\newcommand{\Der}{\mathop{\rm Der}\nolimits}
\newcommand{\Megaideals}{\mathop{{\rm M}_0}\nolimits}
\newcommand{\CharIdeals}{\mathop{{\rm Ch}_0}\nolimits}
\newcommand{\Ideals}{\mathop{{\rm I}_0}\nolimits}
\newcommand{\Subalgebras}{{\rm S}}

\section{Low-dimensional real Lie algebras}\label{SectionOfResultsOnLowDimRealLieAlgebras}

In this appendix different results on low-dimensional real Lie algebras are collected.
Below we follow the Mubarakzyanov's classification and numeration of these algebras.
For each algebra~$A$ we adduce
\begin{itemize}\itemsep=0ex
\item
only non-zero commutators between basis elements,
\item
the group $\Int$ of inner automorphisms (namely, the general form of matrices that form $\Int$
in the chosen basis of $A$),
\item
the complete automorphism group $\Aut$
(more precisely, the general form of matrices that form $\Aut$
in the chosen basis of $A$, these matrices are always supposed non-singular),
\item
a basis of the differentiation algebra $\Der$,
\item
the set~$\Megaideals$ of proper megaideals
(excluding improper megaideals $\{0\}$ and the whole algebra~$A$ which exist for any algebra),
\item
the set~$\CharIdeals$ of proper characteristic ideals which are not
megaideals,
\item
the set~$\Ideals$ of proper ideals which are not characteristic,
\item
the complete set~$\Subalgebras$ of $\Int$-inequivalent
proper subalgebras arranged according to dimensions and structures.
\end{itemize}

We use following parameters:
\begin{enumerate}
\item[] $\varepsilon=\pm1$, \ \ $\varkappa,\gamma,\nu,\zeta,p,q,
\alpha_{ij}, \theta_i \in {\mathbb R}$, \ \ $p^2+q^2=1$, % $p'^2+q'^2=1$
\end{enumerate}
and the notations:
\begin{enumerate}
\item[] $m_{j_1 \dots j_k }^{i_1 \dots i_k }$
(for the fixed values $i_1, \dots, i_k$ and $j_1, \dots, j_k$) denotes
the determinant of the matrix \linebreak $(a_{ij})_{j\in\{j_1 \dots j_k \}}^{i\in\{i_1 \dots i_k \}}$;

\item[]
$s_{j_1 j_2 }^{i_1 \ldots i_k }$ denotes the scalar product of subcolumns of the $j_1$-th
and $j_2$-th columns, namely\linebreak
$s_{j_1 j_2 }^{i_1\ldots i_k }=a_{i_1j_1}a_{i_1j_2}
+a_{i_2j_1}a_{i_2j_2}+ \dots +a_{i_kj_1}a_{i_kj_2}$;
\item[]
$E_{ij}$ (for the fixed values $i$ and $j$) denotes
the matrix $(\delta_{ii'}\delta_{jj'})$ with $i'$ and $j'$
running the numbers of rows and column correspondingly,
i.e. the matrix with the unit on the cross of the $i$-th
row and the $j$-th column and the zero otherwise;
\item[]
indices $i$, $j$, $k$, $i'$, $j'$ and $k'$ run from 1 to $\dim A$;
$\delta_{ik}$ is the Kronecker delta.
\end{enumerate}

The lists of inequivalent subalgebras of three- and
four-dimensional Lie algebras were taken from the Patera's and
Winternitz's classification~\cite{patera&winternitz1977}, where 
the basis elements of subalgebras are adduced in
such manner that the basis elements of the corresponding derived algebra are written to the right
of a semicolon. For Abelian subalgebras the semicolon which should be
on the extreme right is omitted.

\subsection{Three-dimensional real Lie algebras}

\noindent
{\mathversion{bold}$3A_{1}$}\quad (Abelian, Bianchi~I)\\
Any subspace of $3A_1$ as a usual vector space
is a subalgebra  and, moreover, an ideal of $3A_1$, and $\Aut(3A_1)=GL(3,\R)$.
\\[1ex]
$\Int\colon
\left(\begin{array}{ccc}
1 & 0 & 0  \\
0 & 1 & 0  \\
0 & 0 & 1
\end{array}\right)$
\qquad
$\Aut\colon
\left(\begin{array}{ccc}
\alpha_{11} & \alpha_{12} & \alpha_{13}  \\
\alpha_{21} & \alpha_{22} & \alpha_{23}  \\
\alpha_{31} & \alpha_{32} & \alpha_{33}
\end{array}\right)$
\\[1ex]
\noindent
$\Der\colon\, E_{11},\ E_{12},\ E_{13},\ E_{21},\ E_{22},\ E_{23},\ E_{31},\ E_{32},\ E_{33}$
\\[1ex]
$\Megaideals\colon$\ ---
\\
$\CharIdeals\colon$\ ---
\\
$\Ideals=S$
%$\Ideals\colon$\ $\langle e_1+\varkappa e_2+ \gamma e_3\rangle$,\ $\langle e_2+\varkappa e_3 \rangle$,\
%$\langle e_3 \rangle$,\ $\langle e_1+\varkappa e_3, e_2+\gamma e_3 \rangle$,\
%$\langle e_1+\varkappa e_2, e_3 \rangle$,\ $\langle e_2, e_3 \rangle$.
\\
$\Subalgebras\colon$\
1-dim\ $\sim A_1$:\
$\langle e_1+\varkappa e_2+ \gamma e_3\rangle$, \
$\langle e_2+\varkappa e_3 \rangle$, \
$\langle e_3 \rangle$
\\
$\phantom{\Subalgebras\colon}$\
2-dim\ $\sim 2A_1$:\
$\langle e_1+\varkappa e_3, e_2+\gamma e_3 \rangle$, \
$\langle e_1+\varkappa e_2, e_3 \rangle$, \
$\langle e_2, e_3 \rangle$
\\[1ex]
Note, that an another normalization of parameters is possible for the
1-dimensional subalgebras:
$\langle \alpha_1 e_1+\alpha_2 e_2+ \alpha_3 e_3\rangle$,\
$0<\alpha_1^2+\alpha_2^2+\alpha_3^2\le1$.

\bigskip

\noindent
{\mathversion{bold}$A_{2.1}\oplus A_{1}$}:
$[e_1,e_2]=e_1$ \quad (decomposable solvable, Bianchi~III)
\\[1ex]
$\Int\colon
\left(\begin{array}{ccc}
e^{\theta_2} & \theta_1 & 0  \\
0 & 1 & 0  \\
0 & 0 & 1
\end{array}\right)$
\qquad
$\Aut\colon
\left(\begin{array}{ccc}
\alpha_{11} & \alpha_{12} & 0  \\
0 & 1 & 0  \\
0 & \alpha_{32} & \alpha_{33}
\end{array}\right)$
\qquad
$\Der\colon\, E_{11},\ E_{12},\ E_{32},\ E_{33}$
\\[1ex]
$\Megaideals\colon$\ $\langle e_1 \rangle$, $\langle e_3 \rangle$, $\langle e_1, e_3 \rangle$
\\
$\CharIdeals\colon$\ ---
\\
$\Ideals\colon$\ $\langle
e_2+\varkappa e_3; e_1 \rangle$
\\
\noindent
$\Subalgebras\colon$\
1-dim\ $\sim A_1$:\
$\langle pe_3+qe_2 \rangle$, \
$\langle e_1+\varepsilon e_3 \rangle$, \
$\langle e_1 \rangle$
\\
$\phantom{\Subalgebras\colon}$\ 2-dim\ $\sim 2A_1$:\
$\langle e_2, e_3 \rangle$, \
$\langle e_1, e_3 \rangle$
\\
\hphantom{$\phantom{\Subalgebras\colon}$\ 2-dim} $\sim A_{2.1}$:\
$\langle e_2+\varkappa e_3; e_1 \rangle$

\bigskip

\noindent
{\mathversion{bold}$A_{3.1}$}: $[e_2,e_3]=e_1$ \quad (Weyl algebra, nilpotent, Bianchi~II)
\\[1ex]
$\Int\colon
\left(\begin{array}{ccc}
1 & \theta_3 & \theta_2  \\
0 & 1 & 0  \\
0 & 0 & 1
\end{array}\right)$
\qquad
$\Aut\colon
\left(\begin{array}{ccc}
m_{23}^{23} & \alpha_{12} & \alpha_{13}  \\
0 & \alpha_{22} & \alpha_{23}  \\
0 & \alpha_{32} & \alpha_{33}
\end{array}\right)$
\\[1ex]
$\Der\colon\, E_{11}+E_{22},\ E_{12},\ E_{13},\ E_{23},\ E_{32},\ -E_{22}+E_{33}$
\\[1ex]
$\Megaideals\colon$\ $\langle e_1 \rangle$
\\
$\CharIdeals\colon$\ ---
\\
$\Ideals\colon$\ $\langle e_1, pe_2+qe_3 \rangle$
\\
\noindent
$\Subalgebras\colon$\
1-dim\ $\sim A_1$:\
$\langle e_1 \rangle$, \
$\langle pe_2+qe_3 \rangle$
\\
$\phantom{\Subalgebras\colon}$\ 2-dim\ $\sim 2A_1$:\
$\langle e_1, pe_2+qe_3 \rangle$

\bigskip

\noindent
{\mathversion{bold}$A_{3.2}$}: $[e_1,e_3]=e_1,\  [e_2,e_3]=e_1+e_2$  \quad (solvable, Bianchi~IV)
\\[1ex]
$\Int\colon
\left(\begin{array}{ccc}
e^{\theta_3} & \theta_3 e^{\theta_3} & \theta_1+\theta_2  \\
0 & e^{\theta_3} & \theta_2  \\
0 & 0 & 1
\end{array}\right)$
\qquad
$\Aut\colon
\left(\begin{array}{ccc}
\alpha_{11} & \alpha_{12} & \alpha_{13}  \\
0 & \alpha_{11} & \alpha_{23}  \\
0 & 0 & 1
\end{array}\right)$
\qquad
$\Der\colon\, E_{11}+E_{22},\ E_{12},\ E_{13},\ E_{23}$
\\[1ex]
$\Megaideals\colon$\ $\langle e_1 \rangle$, \ $\langle e_1, e_2 \rangle$
\\
$\CharIdeals\colon$\ ---
\\
$\Ideals\colon$\ ---
\\
$\Subalgebras\colon$\
1-dim\ $\sim A_1$:\
$\langle e_1 \rangle$, \
$\langle e_2 \rangle$, \
$\langle e_3 \rangle$
\\
$\phantom{\Subalgebras\colon}$\ 2-dim\ $\sim 2A_1$:\ $\langle e_1,
e_2\rangle$
\\
\hphantom{$\phantom{\Subalgebras\colon}$\ 2-dim} $\sim A_{2.1}$:\
$\langle e_3; e_1 \rangle$

\bigskip

\noindent
{\mathversion{bold}$A_{3.3}$}: $[e_1,e_3]=e_1,\  [e_2,e_3]=e_2$  \quad (solvable, Bianchi~V)
\\[1ex]
$\Int\colon
\left(\begin{array}{ccc}
e^{\theta_3} & 0 & \theta_1  \\
0 & e^{\theta_3} & \theta_2  \\
0 & 0 & 1
\end{array}\right)$
\qquad
$\Aut\colon
\left(\begin{array}{ccc}
\alpha_{11} & \alpha_{12} & \alpha_{13}  \\
\alpha_{21} & \alpha_{11} & \alpha_{23}  \\
0 & 0 & 1
\end{array}\right)$
\qquad
$\Der\colon\, E_{11},\ E_{12},\ E_{13},\ E_{21},\ E_{22},\ E_{23}$
\\[1ex]
$\Megaideals\colon$\ $\langle e_1, e_2 \rangle$
\\
$\CharIdeals\colon$\ ---
\\
$\Ideals\colon$\ $\langle pe_1+qe_2 \rangle$
\\
\noindent
$\Subalgebras\colon$\
1-dim\ $\sim A_1$:\
$\langle e_3 \rangle$, \
$\langle pe_1+qe_2 \rangle$
\\
$\phantom{\Subalgebras\colon}$\ 2-dim\ $\sim 2A_1$:\
$\langle e_1, e_2 \rangle$
\\
\hphantom{$\phantom{\Subalgebras\colon}$\ 2-dim} $\sim A_{2.1}$:\
$\langle e_3; pe_1+qe_2 \rangle$

\bigskip

\noindent
{\mathversion{bold}$A_{3.4}^{a}$}: $[e_1,e_3]=e_1,\
[e_2,e_3]=a e_2,\ -1\leq a < 1,\ a\ne 0$
 \quad (solvable, Bianchi~VI)
\\[1ex]
$\Int\colon
\left(\begin{array}{ccc}
e^{\theta_3} & 0 & \theta_1  \\
0 & e^{a\theta_3} & \theta_2  \\
0 & 0 & 1
\end{array}\right)$\\[1ex]
$-1< a< 1,\ a\ne 0\colon$ %%% A3.4
\\[1ex]
$\Aut\colon
\left(\begin{array}{ccc}
\alpha_{11} & 0 & \alpha_{13}  \\
0 & \alpha_{22} & \alpha_{23}  \\
0 & 0 & 1
\end{array}\right)$
\qquad
$\Der\colon\, E_{11},\ E_{13},\ E_{22},\ E_{23}$
\\[1ex]
$\Megaideals\colon$\ $\langle e_1 \rangle$, \
$\langle e_2 \rangle$, \ $\langle e_1, e_2 \rangle$
\\
$\CharIdeals\colon$\ ---
\\
$\Ideals\colon$\ ---
\\
\noindent
$\Subalgebras\colon$\ 1-dim\ $\sim A_1$:\
$\langle e_1 \rangle$, \
$\langle e_2 \rangle$, \
$\langle e_3 \rangle$, \
$\langle e_1+\varepsilon e_2 \rangle$
\\
$\phantom{\Subalgebras\colon}$\ 2-dim\ $\sim 2A_1$:\
$\langle e_1, e_2 \rangle$
\\
\hphantom{$\phantom{\Subalgebras\colon}$\ 2-dim} $\sim A_{2.1}$:\
$\langle e_3; e_1 \rangle$, \ $\langle e_3; e_2 \rangle$\\[1ex]
$a=-1$                                                          %%% A3.4
\quad (${}\sim p(1,1)$, i.e. the Poincar\'e algebra in~$\mathbb{R}^{1,1}$):\\[1ex]
$\Aut\colon
\left(\begin{array}{ccc}
\alpha_{11} & 0 & \alpha_{13}  \\
0 & \alpha_{22} & \alpha_{23}  \\
0 & 0 & 1
\end{array}\right)$,\
$\left(\begin{array}{ccc}
0 & \alpha_{12} & \alpha_{13}  \\
\alpha_{21} & 0 & \alpha_{23}  \\
0 & 0 & -1
\end{array}\right)$
\qquad
$\Der\colon\, E_{11},\ E_{13},\ E_{22},\ E_{23}$
\\[1ex]
$\Megaideals\colon$\ $\langle e_1, e_2 \rangle$
\\
$\CharIdeals\colon$\ $\langle e_1 \rangle$, \ $\langle e_2 \rangle$
\\
$\Ideals\colon$\ ---
\\
\noindent $\Subalgebras\colon$\ 1-dim\ $\sim A_1$:\
$\langle e_1 \rangle$, \ $\langle e_2 \rangle$, \ $\langle e_3 \rangle$, \
$\langle e_1+\varepsilon e_2 \rangle$
\\
$\phantom{\Subalgebras\colon}$\ 2-dim\ $\sim 2A_1$:\
$\langle e_1, e_2 \rangle$
\\
\hphantom{$\phantom{\Subalgebras\colon}$\ 2-dim} $\sim A_{2.1}$:\
$\langle e_3; e_1 \rangle$, \ $\langle e_3; e_2 \rangle$

\bigskip

\noindent
{\mathversion{bold}$A_{3.5}^{b}$}:
$[e_1,e_3]=be_1-e_2,\ [e_2,e_3]=e_1+be_2,\ b\ge 0$  \quad (solvable, Bianchi~VII)
\\
The canonical commutation relations of the Lie algebra of Bianchi type~VII:
\,
$[e'_1,e'_2]=0$, $[e'_1,e'_3]=e'_2$, $[e'_2,e'_3]=-e'_1+he'_2$, $h^2<4$
can be reduced to the canonical commutation relations
of the algebra $A_{3.5}^{b}$ by the basis transformation:
$e_1=-\frac{2b}{h}e'_1+be'_2$, $e_2=e'_2$, $e_3=\frac{2b}{h}e'_3$, $b=\frac{h}{\sqrt{4-h^2}} $.
\\[1ex]
$\Int\colon
\left(\begin{array}{ccc}
\cos\theta_3 e^{b\theta_3} & \sin\theta_3 e^{b\theta_3} & b\theta_1+\theta_2  \\
-\sin\theta_3 e^{b\theta_3} & \cos\theta_3 e^{b\theta_3} & -\theta_1+b\theta_2  \\
 0 & 0 & 1
\end{array}\right)$
\\[1ex]
$b> 0\colon$ %%% A3.5
\\[1ex]
$\Aut\colon
\left(\begin{array}{ccc}
\alpha_{11} & \alpha_{12} & \alpha_{13}  \\
-\alpha_{12} & \alpha_{11} & \alpha_{23}  \\
0 & 0 & 1
\end{array}\right)$
\qquad
$\Der\colon\, E_{11}+E_{22},\ E_{12}-E_{21},\ E_{13},\ E_{23}$
\\[1ex]
$\Megaideals\colon$\ $\langle e_1, e_2 \rangle$
\\
$\CharIdeals\colon$\ ---
\\
$\Ideals\colon$\ ---
\\
\noindent
$\Subalgebras\colon$\
1-dim\ $\sim A_1$:\
$\langle e_1 \rangle$, \
$\langle e_3 \rangle$
\\
$\phantom{\Subalgebras\colon}$\ 2-dim\ $\sim 2A_1$:\
$\langle e_1, e_2 \rangle$

\bigskip
\noindent
$b=0$ %%% A3.5
\quad (${}\sim e(2)$, i.e. the Euclidian algebra in~$\mathbb{R}^2$):\\[1ex]
$\Aut\colon
\left(\begin{array}{ccc}
\alpha_{11} & \alpha_{12} & \alpha_{13}  \\
-\alpha_{12} & \alpha_{11} & \alpha_{23}  \\
0 & 0 & 1
\end{array}\right)$,\
$\left(\begin{array}{ccc}
\alpha_{11} & \alpha_{12} & \alpha_{13}  \\
\alpha_{12} & -\alpha_{11} & \alpha_{23}  \\
0 & 0 & -1
\end{array}\right)$
\qquad
$\Der\colon\, E_{11}+E_{22},\ E_{12}-E_{21},\ E_{13},\ E_{23}$
\\
$\Megaideals\colon$\ $\langle e_1, e_2 \rangle$
\\
$\CharIdeals\colon$\ ---
\\
$\Ideals\colon$\ ---
\\
\noindent
$\Subalgebras\colon$\
1-dim\ $\sim A_1$:\
$\langle e_1 \rangle$, \
$\langle e_3 \rangle$
\\
$\phantom{\Subalgebras\colon}$\ 2-dim\ $\sim 2A_1$:\
$\langle e_1, e_2 \rangle$

\bigskip

\noindent
{\mathversion{bold}$sl(2,\R)$}: $[e_1,e_2]=e_1,\ [e_2,e_3]=e_3,\ [e_1,e_3]=2e_2$\\ %%%%  sl(2,R)
\phantom{{\mathversion{bold}$sl(2,\R)$}:}
({}$\sim su(1,1)\sim so(1,2)\sim sp(1,\mathbb{R})$, simple, Bianchi~VIII)
\\
The canonical commutation relations of the Lie algebra $sl(2,\R)$
can be reduced to the canonical commutation relations $[e'_1,e'_2]=-e'_3$, $[e'_2,e'_3]=e'_1$, $[e'_3,e'_1]=e'_2$
of the algebra $so(1,2)$ by the basis transformation $e'_i=e_j\beta_{ji}$, where
\\[1ex]
$(\beta_{ij})=\left(\begin{array}{crr}
%-1 & -\sqrt{2} & -1\\ -1 & 0 & 1\\ \sqrt{2} & 1 & \sqrt{2}
0 & \frac12 & \frac12\\[0.4ex] 1 & 0 & 0\\[0.4ex] 0 & \!\!-\frac12 & \,\,\frac12
\end{array}\right)$.
\\[1ex]
In the basis~$\{e'_i\}$ $\Int$ coincides with the matrix group~${\rm Lor}(1,2)$ of
$3\times 3$ Lorentz matrices, i.e.\\[1ex]
$\Int(so(1,2))\colon\ J^{1}(\varphi_1)J^{2}(\varphi_2)J^{3}(\varphi_3)$,  where\\[1ex]
$J^{1}(\varphi_1)=
\left(\begin{array}{ccc}
1 & 0 & 0\\
0 & \cosh\varphi_1 & \sinh\varphi_1\\
0 & \sinh\varphi_1 & \cosh\varphi_1
\end{array}\right)$,\qquad
$J^{2}(\varphi_2)=
\left(\begin{array}{ccc}
\cosh\varphi_2 & 0 & \sinh\varphi_2\\
0 & 1 & 0 \\
\sinh\varphi_2 & 0 & \cosh\varphi_2
\end{array}\right)$,\\[1ex]
$J^{3}(\varphi_3)=
\left(\begin{array}{ccc}
\cos\varphi_3 & \sin\varphi_3 & 0  \\
-\sin\varphi_3 & \cos\varphi_3 & 0  \\
0 & 0 & 1
\end{array}\right)$.
\\
Therefore,
\quad
$\Int(sl(2,\R))\colon\,
(\beta_{ij})J^{1}(\varphi_1)J^{2}(\varphi_2)J^{3}(\varphi_3)(\beta_{ij})^{-1}$
\\[1ex]
$\Aut(so(1,2))$ coincides with the matrix group~$SO(1,2)$ of
$3\times 3$ special pseudo-orthogonal matrices, which is generated by
the elements of $\Int(so(1,2))={\rm Lor}(1,2)$
and of the additional discrete transformation:\\[1ex]
$\mathcal{E}=\left(\begin{array}{ccc}
1 & 0 & 0\\
0 & -1 & 0\\
0 & 0 & -1
\end{array}\right)$.
\\[1ex]
Hence, the automorphism group of the algebra $sl(2,\R)$ can be written in the form:\\[1ex]
$\Aut(sl(2,\R))\colon\,
(\beta_{ij})J^{1}(\varphi_1)J^{2}(\varphi_2)J^{3}(\varphi_3)(\beta_{ij})^{-1},\
(\beta_{ij})\mathcal{E}J^{1}(\varphi_1)J^{2}(\varphi_2)J^{3}(\varphi_3)(\beta_{ij})^{-1}$
\\
$\Der\colon\, E_{12}+2E_{23},\ 2E_{21}+E_{32},\ E_{11}-E_{33}$
\\
$\Megaideals\colon$\ ---
\\
$\CharIdeals\colon$\ ---
\\
$\Ideals\colon$\ ---
\\
\noindent
$\Subalgebras\colon$\
1-dim\ $\sim A_1$:\
$\langle e_1 \rangle$, \
$\langle e_2 \rangle$, \
$\langle e_1+e_3 \rangle$
\\
$\phantom{\Subalgebras\colon}$\ 2-dim\ $\sim A_{2.1}$:\ $\langle e_2;e_1 \rangle$

\bigskip

\noindent
{\mathversion{bold}$so(3)$}: $[e_2,e_3]=e_1,\ [e_3,e_1]=e_2,\ [e_1,e_2]=e_3$ %%% so(3)
\quad (${}\sim su(2)\sim sp(1)$, simple, Bianchi~IX)
\\
For the algebra~$so(3)$,
$\Aut$ coincides with $\Int$ and is the matrix group~$SO(3)$ of $3\times 3$ special orthogonal matrices, i.e.

%\\[1ex]
\noindent
$\Aut=\Int\colon \left(\begin{array}{ccc}
\alpha_{11} & \alpha_{12} & \alpha_{13}  \\
\alpha_{21} & \alpha_{22} & \alpha_{23}  \\
\alpha_{31} & \alpha_{32} & \alpha_{33}
\end{array}\right)$, \quad where $\alpha_{ij}\alpha_{kj}=\delta_{ik}$ and $\det\,(\alpha_{ij})=1$.\\[1ex]
In an explicit form any $3\times 3$ special orthogonal matrix~$(\alpha_{ij})$ can be presented via
the Euler angles~$\varphi_i$:
$(\alpha_{ij})=J^1(\varphi_1)J^2(\varphi_2)J^3(\varphi_3)$, where $J^i(\varphi_i)$ is the matrix of rotation
on the angle~$\varphi_i$ with respect to the $i$-th axis:
\\[1ex]
$J^1(\varphi_1)=\left(\begin{array}{ccc}
1 & 0 & 0\\ 0 & \cos\varphi_1 & \sin\varphi_1 \\ 0 & -\sin\varphi_1& \cos\varphi_1
\end{array}\right)$,\qquad
$J^2(\varphi_2)=
\left(\begin{array}{ccc}
\cos\varphi_2 & 0 & -\sin\varphi_2\\ 0 & 1 & 0  \\ \sin\varphi_2 & 0 & \cos\varphi_2
\end{array}\right)$,\\[1ex]
$J^3(\varphi_3)=
\left(\begin{array}{ccc}
\cos\varphi_3 & \sin\varphi_3 & 0  \\ -\sin\varphi_3 & \cos\varphi_3 & 0  \\ 0 & 0 & 1
\end{array}\right)$.
\\[1ex]
$\Der\colon\, E_{23}+E_{32},\ E_{31}+E_{13},\ E_{12}-E_{21}$
\\[1ex]
$\Megaideals\colon$\ ---
\\
$\CharIdeals\colon$\ ---
\\
$\Ideals\colon$\ ---
\\
\noindent
$\Subalgebras\colon$\
1-dim\ $\sim A_1$:\
$\langle e_1 \rangle$

\subsection{Four-dimensional real Lie algebras}

{\mathversion{bold}$4A_{1}$}: (Abelian)    %%% 4A1\\
\\
Any subspace of $4A_1$ as a usual vector space
is a subalgebra  and, moreover, an ideal of $4A_1$, and $\Aut(4A_1)=GL(4,\R)$.
\\[1ex]
$\Int\colon
\left(\begin {array}{cccc}
1&0&0&0\\
0&1&0&0\\
0&0&1&0\\
0&0&0&1\end {array}\right)$
\qquad
$\Aut\colon
\left(\begin{array}{cccc}
\alpha_{11} & \alpha_{12} & \alpha_{13} & \alpha_{14}\\
\alpha_{21} & \alpha_{22} & \alpha_{23} & \alpha_{24}\\
\alpha_{31} & \alpha_{32} & \alpha_{33} & \alpha_{34}\\
\alpha_{41} & \alpha_{42} & \alpha_{43} & \alpha_{44}\\
\end{array}\right)$
\\[1ex]
\noindent
$\Der\colon\, E_{11},\ E_{12},\ E_{13},\ E_{14},\
              E_{21},\ E_{22},\ E_{23},\ E_{24},\
              E_{31},\ E_{32},\ E_{33},\ E_{34},\
              E_{41},\ E_{42},\ E_{43},\ E_{44}$
\\[1ex]
$\Megaideals\colon$\ ---
\\
$\CharIdeals\colon$\ ---
\\
$\Ideals=S$
%$\Ideals\colon$\
%$\langle e_1+\varkappa e_2+\gamma e_3+\zeta e_4 \rangle$,\
%$\langle e_2+\varkappa e_3+\gamma e_4 \rangle$,\
%$\langle e_3+\varkappa e_4 \rangle$,\
%$\langle e_4 \rangle$,\
%$\langle e_1+\varkappa e_3+\gamma e_4, e_2+\zeta e_3+ \nu e_4 \rangle$,\
%$\langle e_3, e_4 \rangle$,\
%$\langle e_1+\varkappa e_2+\gamma e_4, e_3+\zeta e_4 \rangle$,\
%$\langle e_2+\varkappa e_4, e_3+\gamma e_4 \rangle$,\
%$\langle e_1+\varkappa e_2+\gamma e_3, e_4 \rangle$,\
%$\langle e_2+\varkappa e_3, e_4 \rangle$,\
%$\langle e_1+\varkappa e_4, e_2+\gamma e_4, e_3+\zeta e_4 \rangle$,\
%$\langle e_1+\varkappa e_2, e_3, e_4 \rangle$,\
%$\langle e_2, e_3, e_4 \rangle$,\
%$\langle e_1+\varkappa e_3, e_2+\gamma e_3, e_4 \rangle$.
\\
$\Subalgebras\colon$\
1-dim\ $\sim A_1$:\
$\langle e_1+\varkappa e_2+\gamma e_3+\zeta e_4 \rangle$,
$\langle e_2+\varkappa e_3+\gamma e_4 \rangle$,
$\langle e_3+\varkappa e_4 \rangle$,
$\langle e_4 \rangle$
\\
$\phantom{\Subalgebras\colon}$\ 2-dim\ $\sim 2A_1$:\
$\langle e_1+\varkappa e_3+\gamma e_4, e_2+\zeta e_3+ \nu e_4 \rangle$,
$\langle e_1+\varkappa e_2+\gamma e_4, e_3+\zeta e_4 \rangle$,
$\langle e_2+\varkappa e_4, e_3+\gamma e_4 \rangle$, \linebreak
$\phantom{\Subalgebras\colon {\mbox 2-dim}\sim 3A_{1}}$
$\langle e_1+\varkappa e_2+\gamma e_3, e_4 \rangle$,
$\langle e_2+\varkappa e_3, e_4 \rangle$,
$\langle e_3, e_4 \rangle$.
\\
$\phantom{\Subalgebras\colon}$\ 3-dim\ $\sim 3A_{1}$:\
$\langle e_1+\varkappa e_4, e_2+\gamma e_4, e_3+\zeta e_4 \rangle$,
$\langle e_1+\varkappa e_2, e_3, e_4 \rangle$,
$\langle e_2, e_3, e_4 \rangle$,
$\langle e_1+\varkappa e_3, e_2+\gamma e_3, e_4 \rangle$

\bigskip

\noindent
{\mathversion{bold}$A_{2.1}\oplus 2A_1$}: $[e_1,e_2]=e_1$
(decomposable solvable)%% A2.1+2A1
\\[1ex]
$\Int\colon
\left(\begin {array}{cccc}
e^{\theta_2} &\theta_1 & 0 & 0 \\
0 & 1 & 0 & 0 \\
0 & 0 & 1 & 0 \\
0 & 0 & 0 & 1\end {array}\right)$
\qquad
$\Aut\colon
\left(\begin{array}{cccc}
\alpha_{11}& \alpha_{12}& 0& 0 \\
0 & 1 & 0& 0\\
0 & \alpha_{32} & \alpha_{33}& \alpha_{34}\\
0 & \alpha_{42} & \alpha_{43} & \alpha_{44}\\
\end{array}\right)$
\\[1ex]
\noindent
$\Der\colon\, E_{11},\ E_{12},\ E_{32},\ E_{33},\ E_{34},\ E_{42},\ E_{43},\ E_{44}$
\\[1ex]
$\Megaideals\colon$\ $\langle e_1 \rangle$, \
$\langle e_3, e_4 \rangle$, \ $\langle e_1, e_3, e_4 \rangle$
\\
$\CharIdeals\colon$\ ---
\\
$\Ideals\colon$\ $\langle pe_3+qe_4 \rangle$, \ $\langle
e_2+\varkappa e_3+\gamma e_4; e_1 \rangle$, \
$\langle e_1, pe_3+qe_4 \rangle$, \
$\langle e_2+\varkappa (q e_3-p  e_4), pe_3+q e_4; e_1 \rangle$
\\
$\Subalgebras\colon$\
1-dim\ $\sim A_1$:\
$\langle e_1 \rangle$,
$\langle pe_3+qe_4 \rangle$, \
$\langle e_2+\varkappa e_3+\gamma e_4 \rangle$, \
$\langle e_1+ pe_3+qe_4 \rangle$
\\
$\phantom{\Subalgebras\colon}$\ 2-dim\ $\sim 2A_1$:\
$\langle e_2+\varkappa (q e_3-p  e_4), pe_3+q e_4 \rangle$, \
$\langle e_3, e_4 \rangle$, \
$\langle e_1+qe_3-pe_4, pe_3+q e_4 \rangle$, \
$\langle e_1, pe_3+qe_4 \rangle$
\\
\hphantom{$\phantom{\Subalgebras\colon}$\ 2-dim}
$\sim A_{2.1}$:\
$\langle e_2+\varkappa e_3+\gamma e_4; e_1 \rangle$
\\
$\phantom{\Subalgebras\colon}$\ 3-dim\ $\sim 3A_{1}$:\
$\langle e_2, e_3, e_4 \rangle$, \
$\langle e_1, e_3, e_4 \rangle$
\\
\hphantom{$\phantom{\Subalgebras\colon}$\ 3-dim}
$\sim A_{2.1}\oplus A_1$:\
$\langle e_2+\varkappa (q e_3-p  e_4), pe_3+q e_4; e_1 \rangle$

\bigskip

\noindent
{\mathversion{bold}$2A_{2.1}$}: $[e_1,e_2]=e_1,\ [e_3,e_4]=e_3$ (decomposable solvable)%% 2A2.1
\\[1ex]
$\Int\colon
\left(\begin {array}{cccc}
e^{\theta_2} &\theta_1 & 0 & 0 \\
0 & 1 & 0 & 0 \\
0 & 0 & e^{\theta_4} & \theta_3 \\
0 & 0 & 0 & 1\end {array}\right)$
\qquad
$\Aut\colon
\left(\begin{array}{cccc}
\alpha_{11}& \alpha_{12}& 0& 0 \\
0 & 1 & 0& 0\\
0 & 0 & \alpha_{33}& \alpha_{34}\\
0 & 0 & 0 & 1\\
\end{array}\right)$,\
$\left(\begin{array}{cccc}
0& 0& \alpha_{13}& \alpha_{14} \\
0 & 0 & 0& 1\\
\alpha_{31} & \alpha_{32} & 0 & 0\\
0 & 1 & 0 & 0\\
\end{array}\right)$
\\[1ex]
\noindent
$\Der\colon\, E_{11},\ E_{12},\ E_{33},\ E_{34}$
\\[1ex]
$\Megaideals\colon$\ $\langle e_1, e_3 \rangle$, \
$\langle e_2+\varepsilon e_4; e_1,e_3\rangle$
\\
$\CharIdeals\colon$\ $\langle e_1 \rangle$,\ $\langle e_3 \rangle$,\
$\langle e_2; e_1 \rangle$,\
$\langle e_4; e_3 \rangle$,\
$\langle e_2, e_3; e_1 \rangle$,\ $\langle e_1, e_4; e_3 \rangle$,
\ $\langle e_2+\varkappa e_4; e_1,e_3 \rangle$, \ $\varkappa \not=0,\pm1$
\\
$\Ideals\colon$\ ---
\\
$\Subalgebras\colon$\
1-dim\ $\sim A_1$:\
$\langle e_1 \rangle$, \
$\langle e_3 \rangle$, \
$\langle e_4 \rangle$, \
$\langle e_2+\varkappa e_4 \rangle$, \
$\langle e_2+\varepsilon e_3 \rangle$, \
$\langle e_1+\varepsilon e_3 \rangle$, \
$\langle e_1+\varepsilon e_4 \rangle$
\\
$\phantom{\Subalgebras\colon}$\ 2-dim\ $\sim 2A_1$:\
$\langle e_2, e_4 \rangle$, \
$\langle e_2, e_3 \rangle$, \
$\langle e_1, e_4 \rangle$, \
$\langle e_1, e_3 \rangle$
\\
\hphantom{$\phantom{\Subalgebras\colon}$\ 2-dim}
$\sim A_{2.1}$:\
$\langle e_2+\varkappa e_4; e_1 \rangle$, \
$\langle e_4+\varkappa e_2; e_3 \rangle$, \
$\langle e_2+\varepsilon e_3; e_1 \rangle$, \
$\langle e_4+\varepsilon e_1; e_3 \rangle$, \
$\langle e_2+e_4; e_1+\varepsilon e_3 \rangle$
\\
$\phantom{\Subalgebras\colon}$\ 3-dim\ $\sim A_{2.1}\oplus A_{1}$:\
$\langle e_2, e_4; e_1 \rangle$, \
$\langle e_2, e_3; e_1 \rangle$, \
$\langle e_2, e_4; e_3 \rangle$, \
$\langle e_1, e_4; e_3 \rangle$
\\
\hphantom{$\phantom{\Subalgebras\colon}$\ 3-dim}
$\sim A_{3.3}$:\
$\langle e_2+e_4; e_1, e_3 \rangle$
\\
\hphantom{$\phantom{\Subalgebras\colon}$\ 3-dim}
$\sim A^a_{3.4}$:\
$\langle e_2+\varkappa e_4; e_1, e_3 \rangle$
\\[1ex]
The parameter $a$ in the algebra $A_{3.4}^{a}$ should satisfy condition:\\[1ex]
$a=\left\{\begin{array}{ll} \varkappa, & -1 \leq \varkappa<1, \   \varkappa\not= 0,\\
                                    \frac{1}{\varkappa},& |\varkappa|>1 \end{array}\right.$
\bigskip

\noindent
{\mathversion{bold}$A_{3.1}\oplus A_1$}: $[e_2,e_3]=e_1$ (decomposable nilpotent)%% A3.1+A1
\\[1ex]
$\Int\colon
\left(\begin {array}{cccc}
1 & \theta_3 & \theta_2 & 0 \\
0 & 1 & 0 & 0 \\
0 & 0 & 1 & 0 \\
0 & 0 & 0 & 1\end {array}\right)$
\qquad
$\Aut\colon
\left(\begin{array}{cccc}
m_{23}^{23}& \alpha_{12} & \alpha_{13} & \alpha_{14} \\
0 & \alpha_{22} & \alpha_{23} & 0\\
0 & \alpha_{32} & \alpha_{33} & 0\\
0 & \alpha_{42} & \alpha_{43} & \alpha_{44}\\
\end{array}\right)$
\\[1ex]
$\Der\colon\, E_{12},\ E_{13},\ E_{14},\ E_{11}+E_{22},\
E_{23},\ E_{32},\ E_{11}+E_{33},\ E_{42},\ E_{43},\ E_{44}$
\\[1ex]
$\Megaideals\colon$\ $\langle e_1 \rangle$, \ $\langle e_1, e_4 \rangle$
\\
$\CharIdeals\colon$\ ---
\\
$\Ideals\colon$\ $\langle e_1+\varkappa e_4 \rangle$, \  $\langle e_4
\rangle$, \ $\langle e_1, e_4+\varkappa(pe_2+qe_3) \rangle$, \ $\langle
e_1, pe_2+qe_3 \rangle$, \ $\langle e_1, pe_2+qe_3, e_4
\rangle$,\\
$\phantom{\Ideals\colon}$\ $\langle e_2+\nu e_4, e_3+\gamma e_4; e_1 \rangle$, \
$\varkappa \ne 0$
\\
$\Subalgebras\colon$\
1-dim\ $\sim A_1$:\
$\langle e_1+\varkappa e_4 \rangle$, \
$\langle e_4 \rangle$, \
$\langle pe_2+qe_3+\varkappa e_4 \rangle$
\\
$\phantom{\Subalgebras\colon}$\ 2-dim\ $\sim 2A_1$:\
$\langle e_1, e_4+\varkappa(pe_2+qe_3) \rangle$, \
$\langle e_4, pe_2+qe_3 \rangle$, \
$\langle e_1+\varkappa e_4, pe_2+qe_3 \rangle$
\\
$\phantom{\Subalgebras\colon}$\ 3-dim\ $\sim 3A_{1}$:\
$\langle e_1, pe_2+qe_3, e_4 \rangle$
\\
\hphantom{$\phantom{\Subalgebras\colon}$\ 3-dim} $\sim A_{3.1}$:\
$\langle e_2+\varkappa e_4, e_3+\gamma e_4; e_1 \rangle$

\bigskip

\noindent
{\mathversion{bold}$A_{3.2}\oplus A_1$}: $[e_1,e_3]=e_1,\ [e_2,e_3]=e_1+e_2$
(decomposable solvable)                                                %% A3.2+A1
\\[1ex]
$\Int\colon
\left(\begin {array}{cccc}
e^{\theta_3} &\theta_3e^{\theta_3} & \theta_1+\theta_2 & 0 \\
0 & e^{\theta_3} & \theta_2 & 0 \\
0 & 0 & 1 & 0\\
0 & 0 & 0 & 1 \end {array}\right)$
\qquad
$\Aut\colon
\left(\begin{array}{cccc}
\alpha_{11}& \alpha_{12}& \alpha_{13} & 0 \\
0 & \alpha_{11} & \alpha_{23}& 0\\
0 & 0 & 1 & 0\\
0 & 0 & \alpha_{43} & \alpha_{44}\\
\end{array}\right)$
\\[1ex]
$\Der\colon\, E_{12},\ E_{13},\ E_{11}+E_{22},\ E_{23},\ E_{43},\ E_{44}$
\\[1ex]
$\Megaideals\colon$\ $\langle e_1 \rangle$, \ $\langle e_4 \rangle$, \
$\langle e_1, e_2 \rangle$,\
$\langle e_1, e_4 \rangle$, \ $\langle e_1, e_2, e_4 \rangle$
\\
$\CharIdeals\colon$\ ---
\\
$\Ideals\colon$\  $\langle
e_3+\varkappa e_4; e_1, e_2 \rangle$
\\
$\Subalgebras\colon$\
1-dim\ $\sim A_1$:\
$\langle e_1 \rangle$, \
$\langle e_1+\varepsilon e_4 \rangle$, \
$\langle e_2+\varkappa e_4 \rangle$, \
$\langle e_3+\varkappa e_4 \rangle$, \
$\langle e_4 \rangle$
\\
$\phantom{\Subalgebras\colon}$\ 2-dim\ $\sim 2A_1$:\
$\langle e_1+\varkappa e_4, e_2 \rangle$, \
$\langle e_1, e_2+\varepsilon e_4 \rangle$, \
$\langle e_1, e_4 \rangle$, \
$\langle e_2, e_4 \rangle$, \
$\langle e_3, e_4 \rangle$
\\
\hphantom{$\phantom{\Subalgebras\colon}$\ 2-dim}
$\sim A_{2.1}$:\
$\langle e_3+\varkappa e_4; e_1 \rangle$
\\
$\phantom{\Subalgebras\colon}$\ 3-dim\ $\sim 3A_{1}$:\
$\langle e_1, e_2, e_4  \rangle$
\\
\hphantom{$\phantom{\Subalgebras\colon}$\ 3-dim}
$\sim A_{2.1}\oplus A_{1}$:\
$\langle e_3, e_4; e_1 \rangle$
\\
\hphantom{$\phantom{\Subalgebras\colon}$\ 3-dim}
$\sim A_{3.2}$:\
$\langle e_3+\varkappa e_4; e_1, e_2 \rangle$

\bigskip

\noindent
{\mathversion{bold}$A_{3.3}\oplus A_1$}: $[e_1,e_3]=e_1,\ [e_2,e_3]=e_2$
(decomposable solvable)                                                %% A3.3+A1
\\[1ex]
$\Int\colon
\left(\begin {array}{cccc}
e^{\theta_3} & 0 & \theta_1 & 0 \\
0 & e^{\theta_3} & \theta_2 & 0 \\
0 & 0 & 1 & 0\\
0 & 0 & 0 & 1 \end {array}\right)$
\qquad
$\Aut\colon
\left(\begin{array}{cccc}
\alpha_{11}& \alpha_{12}& \alpha_{13} & 0 \\
\alpha_{21} & \alpha_{22} & \alpha_{23}& 0\\
0 & 0 & 1 & 0\\
0 & 0 & \alpha_{43} & \alpha_{44}\\
\end{array}\right)$
\\[1ex]
\noindent
$\Der\colon\, E_{11},\   E_{12},\ E_{13},\ E_{21},\ E_{22},\ E_{23},\ E_{43},\ E_{44}$
\\[1ex]
$\Megaideals\colon$\ $\langle e_4 \rangle$, \ $\langle e_1, e_2 \rangle$, \
$\langle e_1, e_2, e_4 \rangle$
\\
$\CharIdeals\colon$\ ---
\\
$\Ideals\colon$\ $\langle pe_1+qe_2 \rangle$, \ $\langle p e_1+qe_2, e_4
\rangle$, \ $\langle e_3+\varkappa e_4; e_1, e_2 \rangle$
\\
$\Subalgebras\colon$\
1-dim\ $\sim A_1$:\
$\langle e_4 \rangle$, \
$\langle pe_1+qe_2 \rangle$, \
$\langle e_3+\varkappa e_4 \rangle$, \
$\langle pe_1+qe_2+\varepsilon e_4 \rangle$
\\
$\phantom{\Subalgebras\colon}$\ 2-dim\ $\sim 2A_1$:\
$\langle e_1, e_2 \rangle$, \
$\langle e_3, e_4 \rangle$, \
$\langle pe_1+qe_2, e_4 \rangle$, \
$\langle e_1, e_2+\varepsilon e_4 \rangle$, \
$\langle e_1+\varepsilon e_4, e_2+\varkappa e_1 \rangle$
\\
\hphantom{$\phantom{\Subalgebras\colon}$\ 2-dim}
$\sim A_{2.1}$:\
$\langle e_3+\varkappa e_4; pe_1+qe_2 \rangle$
\\
$\phantom{\Subalgebras\colon}$\ 3-dim\ $\sim 3A_{1}$:\
$\langle e_1, e_2, e_4  \rangle$
\\
\hphantom{$\phantom{\Subalgebras\colon}$\ 3-dim}
$\sim A_{2.1}\oplus A_{1}$:\
$\langle e_3, e_4; pe_1+qe_2 \rangle$
\\
\hphantom{$\phantom{\Subalgebras\colon}$\ 3-dim}
$\sim A_{3.2}$:\
$\langle e_3+\varkappa e_4; e_1, e_2 \rangle$

\bigskip

\noindent
{\mathversion{bold}$A_{3.4}^a\oplus A_1$}: $[e_1,e_3]=e_1,\ [e_2,e_3]=ae_2,\ -1\le a<1,\ a\ne0$  %% A3.4+A1
(decomposable solvable)
\\[1ex]
$\Int\colon
\left(\begin {array}{cccc}
e^{\theta_3} & 0 & \theta_1 & 0 \\
0 & e^{a\theta_3} & \theta_2 & 0 \\
0 & 0 & 1 & 0\\
0 & 0 & 0 & 1 \end {array}\right)$

\bigskip

\noindent
$-1<a<1,\ a\ne0\colon$                       %% A3.4+A1
\\[1ex]
$\Aut\colon
\left(\begin{array}{cccc}
\alpha_{11}& 0& \alpha_{13} & 0 \\
0 & \alpha_{22} & \alpha_{23}& 0\\
0 & 0 & 1 & 0\\
0 & 0 & \alpha_{43} & \alpha_{44}\\
\end{array}\right)$
\qquad
$\Der\colon\, E_{11},\  E_{13},\ E_{22},\ E_{23},\ E_{43},\ E_{44}$
\\[1ex]
$\Megaideals\colon$\ $\langle e_1 \rangle$, \ $\langle e_2 \rangle$, \ $\langle e_4 \rangle$, \
$\langle e_1, e_2 \rangle$, \ $\langle e_1, e_4 \rangle$, \
$\langle e_2, e_4 \rangle$, \ $\langle e_1, e_2, e_4 \rangle$
\\
$\CharIdeals\colon$\ ---
\\
$\Ideals\colon$\ $\langle e_3+\varkappa e_4; e_1, e_2 \rangle$
\\
$\Subalgebras\colon$\ 1-dim\ $\sim A_1$:\ $\langle e_1 \rangle$, \
$\langle e_2 \rangle$, \ $\langle e_4 \rangle$, \ $\langle
e_1+\varepsilon e_4 \rangle$, \ $\langle e_2+\varepsilon e_4 \rangle$, \
$\langle e_3+\varkappa e_4 \rangle$, \ $\langle e_1+\varepsilon
e_2+\varkappa e_4 \rangle$
\\
$\phantom{\Subalgebras\colon}$\ 2-dim\ $\sim 2A_1$:\ $\langle e_1,
e_2 \rangle$, \ $\langle e_1, e_4 \rangle$, \ $\langle e_2, e_4
\rangle$, \ $\langle e_3, e_4 \rangle$, \ $\langle e_1+\varepsilon e_2,
e_4 \rangle$, \ $\langle e_1, e_2+\varepsilon e_4 \rangle$, \ $\langle
e_1+\varepsilon e_4,  e_2+\varkappa e_4 \rangle$
\\
\hphantom{$\phantom{\Subalgebras\colon}$\ 2-dim} $\sim A_{2.1}$:\
$\langle e_3+\varkappa e_4; e_1\rangle$, \ $\langle e_3+\varkappa e_4;
e_2\rangle$
\\
$\phantom{\Subalgebras\colon}$\ 3-dim\ $\sim 3A_{1}$:\ $\langle e_1,
e_2, e_4  \rangle$
\\
\hphantom{$\phantom{\Subalgebras\colon}$\ 3-dim} $\sim A_{2.1}\oplus
A_{1}$:\ $\langle e_3, e_4; e_1  \rangle$, \  $\langle e_3, e_4; e_2
\rangle$
\\
\hphantom{$\phantom{\Subalgebras\colon}$\ 3-dim} $\sim A_{3.4}^a$:\
$\langle e_3+\varkappa e_4; e_1, e_2 \rangle$
\\[1ex]
$a=-1\colon$                                           %% A3.4+A1
\\[1ex]
\bigskip
$\Aut\colon
\left(\begin{array}{cccc}
\alpha_{11}& 0& \alpha_{13} & 0 \\
0 & \alpha_{22} & \alpha_{23}& 0\\
0 & 0 & 1 & 0\\
0 & 0 & \alpha_{43} & \alpha_{44}\\
\end{array}\right)$,\
$\left(\begin{array}{cccc}
0 & \alpha_{12} & \alpha_{13} & 0 \\
\alpha_{21} & 0 & \alpha_{23}& 0\\
0 & 0 & -1 & 0\\
0 & 0 & \alpha_{43} & \alpha_{44}\\
\end{array}\right)$
\qquad
$\Der\colon\, E_{11},\  E_{13},\ E_{22},\ E_{23},\ E_{43},\ E_{44}$
\\
$\Megaideals\colon$\ $\langle e_4 \rangle$, \ $\langle e_1, e_2 \rangle$, \
$\langle e_1, e_2, e_4 \rangle$
\\
$\CharIdeals\colon$\ $\langle e_1 \rangle$, \ $\langle e_2 \rangle$, \
$\langle e_1, e_4 \rangle$, \ $\langle e_2, e_4 \rangle$
\\
$\Ideals\colon$\
 $\langle
e_3+\varkappa e_4; e_1, e_2 \rangle$
\\
$\Subalgebras\colon$\
1-dim\ $\sim A_1$:\
$\langle e_1 \rangle$, \ $\langle e_2 \rangle$, \ $\langle e_4 \rangle$, \
$\langle e_1+\varepsilon e_4 \rangle$, \
$\langle e_2+\varepsilon e_4 \rangle$, \
$\langle e_3+\varkappa e_4 \rangle$, \
$\langle e_1+\varepsilon e_2+\varkappa e_4 \rangle$
\\
$\phantom{\Subalgebras\colon}$\ 2-dim\ $\sim 2A_1$:\
$\langle e_1, e_2 \rangle$, \
$\langle e_1, e_4 \rangle$, \
$\langle e_2, e_4 \rangle$, \
$\langle e_3, e_4 \rangle$, \
$\langle e_1+\varepsilon e_2, e_4 \rangle$, \
$\langle e_1, e_2+\varepsilon e_4 \rangle$, \
$\langle e_1+\varepsilon e_4,  e_2+\varkappa e_4 \rangle$
\\
\hphantom{$\phantom{\Subalgebras\colon}$\ 2-dim}
$\sim A_{2.1}$:\
$\langle e_3+\varkappa e_4; e_1\rangle$, \
$\langle e_3+\varkappa e_4; e_2\rangle$
\\
$\phantom{\Subalgebras\colon}$\ 3-dim\ $\sim 3A_{1}$:\
$\langle e_1, e_2, e_4  \rangle$
\\
\hphantom{$\phantom{\Subalgebras\colon}$\ 3-dim}
$\sim A_{2.1}\oplus A_{1}$:\
$\langle e_3, e_4; e_1  \rangle$, \
$\langle e_3, e_4; e_2  \rangle$
\\
\hphantom{$\phantom{\Subalgebras\colon}$\ 3-dim}
$\sim A_{3.4}^a$:\
$\langle e_3+\varkappa e_4; e_1, e_2 \rangle$

\bigskip

\noindent
{\mathversion{bold}$A_{3.5}^b\oplus A_1$}: $[e_1,e_3]=be_1-e_2,\ [e_2,e_3]=e_1+be_2,\ b\geq 0$  %% A3.5+A1
(decomposable solvable)
\\[1ex]
$\Int\colon
\left(\begin {array}{cccc}
\cos\theta_3 e^{b\theta_3} & \sin\theta_3 e^{b\theta_3} & b\theta_1+\theta_2 & 0 \\
-\sin\theta_3 e^{b\theta_3} & \cos\theta_3 e^{b\theta_3} & -\theta_1+b\theta_2 & 0 \\
0 & 0 & 1 & 0\\
0 & 0 & 0 & 1 \end {array}\right)$\\[1ex]
$b>0\colon$                       %% A3.5+A1
\\[1ex]
\noindent
$\Aut\colon
\left(\begin{array}{cccc}
\alpha_{11}& \alpha_{12} & \alpha_{13} & 0 \\
-\alpha_{12} & \alpha_{11} & \alpha_{23}& 0\\
0 & 0 & 1 & 0\\
0 & 0 & \alpha_{43} & \alpha_{44}\\
\end{array}\right)$
\qquad
$\Der\colon\, E_{12}-E_{21},\  E_{13},\ E_{11}+E_{22},\ E_{23},\ E_{43},\ E_{44}$
\\[1ex]
$\Megaideals\colon$\ $\langle e_4 \rangle$, \ $\langle e_1, e_2 \rangle$, \
$\langle e_1, e_2, e_4 \rangle$
\\
$\CharIdeals\colon$\ ---
\\
$\Ideals\colon$\ $\langle e_3+\varkappa e_4; e_1, e_2 \rangle$
\\
$\Subalgebras\colon$\ 1-dim\ $\sim A_1$:\ $\langle e_4 \rangle$, \
$\langle e_1+\varkappa e_4 \rangle$, \ $\langle e_3+\gamma e_4
\rangle$, \ $\varkappa \ge 0$
\\
$\phantom{\Subalgebras\colon}$\ 2-dim\ $\sim 2A_1$:\ $\langle
e_1+\varkappa e_4, e_2 \rangle$, \ $\langle e_1, e_4 \rangle$, \
$\langle e_3, e_4 \rangle$, \ $\varkappa \ge 0$
\\
$\phantom{\Subalgebras\colon}$\ 3-dim\ $\sim 3A_{1}$:\ $\langle e_1,
e_2, e_4  \rangle$
\\
\hphantom{$\phantom{\Subalgebras\colon}$\ 3-dim} $\sim A_{3.5}^b$:\
$\langle e_3+\varkappa e_4; e_1, e_2 \rangle$
\\[1ex]
$b=0\colon$                                           %% A3.5+A1
\\
$\Aut\colon
\left(\begin{array}{cccc}
\alpha_{11}& \alpha_{12} & \alpha_{13} & 0 \\
-\alpha_{12} & \alpha_{11} & \alpha_{23}& 0\\
0 & 0 & 1 & 0\\
0 & 0 & \alpha_{43} & \alpha_{44}
\end{array}\right)$,\
$\left(\begin{array}{cccc}
\alpha_{11} & \alpha_{12} & \alpha_{13} & 0 \\
\alpha_{12} & -\alpha_{11} & \alpha_{23}& 0\\
0 & 0 & -1 & 0\\
0 & 0 & \alpha_{43} & \alpha_{44}
\end{array}\right)$
\\[1ex]
$\Der\colon\, E_{12}-E_{21},\  E_{13},\ E_{11}+E_{22},\ E_{23},\ E_{43},\ E_{44}$
\\[1ex]
$\Megaideals\colon$\ $\langle e_4 \rangle$, \ $\langle e_1, e_2 \rangle$, \
$\langle e_1, e_2, e_4 \rangle$
\\
$\CharIdeals\colon$\ ---
\\
$\Ideals\colon$\ $\langle e_3+\varkappa e_4; e_1, e_2 \rangle$
\\
$\Subalgebras\colon$\
1-dim\ $\sim A_1$:\
$\langle e_4 \rangle$, \
$\langle e_1+\varkappa e_4 \rangle$, \
$\langle e_3+\gamma e_4 \rangle$, \ $\varkappa \ge 0$
\\
$\phantom{\Subalgebras\colon}$\ 2-dim\ $\sim 2A_1$:\
$\langle e_1+\varkappa e_4, e_2 \rangle$, \
$\langle e_1, e_4 \rangle$, \
$\langle e_3, e_4 \rangle$, \
$\varkappa \ge 0$
\\
$\phantom{\Subalgebras\colon}$\ 3-dim\ $\sim 3A_{1}$:\
$\langle e_1, e_2, e_4  \rangle$
\\
\hphantom{$\phantom{\Subalgebras\colon}$\ 3-dim}
$\sim A_{3.5}^0$:\
$\langle e_3+\varkappa e_4; e_1, e_2 \rangle$

\bigskip

\noindent
{\mathversion{bold}$sl(2,\R)\oplus A_1$}: $[e_1,e_2]=e_1,\ [e_2,e_3]=e_3,\ [e_1,e_3]=2e_2$ (unsolvable)%% sl(2,R)+A1
\\[1ex]
$\Int\colon
\left(\begin{array}{cccc}
\alpha_{11}& \alpha_{12} & \alpha_{13} & 0 \\
\alpha_{21} & \alpha_{22} & \alpha_{23}& 0\\
\alpha_{31} & \alpha_{32} & \alpha_{33}& 0\\
0 & 0 & 0 & 1
\end{array}\right)$,
\
where the submatrix $(\alpha_{ij})_{i,j=\overline{1,3}}$
belongs to $\Int(sl(2,\R))$
\\[1ex]
$\Aut\colon
\left(\begin{array}{cccc}
\alpha_{11}& \alpha_{12} & \alpha_{13} & 0 \\
\alpha_{21} & \alpha_{22} & \alpha_{23}& 0\\
\alpha_{31} & \alpha_{32} & \alpha_{33}& 0\\
0 & 0 & 0 & \!\!\alpha_{44}\!\!
\end{array}\right)$,
\
where the submatrix $(\alpha_{ij})_{i,j=\overline{1,3}}$
belongs to $\Aut(sl(2,\R))$
\\[1ex]
$\Der\colon\, E_{11}-E_{33},\ E_{12}+2E_{23},\ 2E_{21}+E_{32},\ E_{44}$
\\[1ex]
$\Megaideals\colon$\ $\langle e_4 \rangle$, \ $\langle ; e_1, e_2, e_3 \rangle$
\\
$\CharIdeals\colon$\ ---
\\
$\Ideals\colon$\ ---
\\
$\Subalgebras\colon$\
1-dim\ $\sim A_1$:\
$\langle e_1 \rangle$, \ $\langle e_4 \rangle$, \
$\langle e_2+\varkappa e_4 \rangle$, \
$\langle e_1+\varepsilon e_4 \rangle$, \
$\langle e_1+e_3+\gamma e_4 \rangle$, \
$\varkappa \ge 0$
\\
$\phantom{\Subalgebras\colon}$\ 2-dim\ $\sim 2A_1$:\
$\langle e_1, e_4 \rangle$, \
$\langle e_2, e_4 \rangle$, \
$\langle e_1+e_3, e_4 \rangle$
\\
\hphantom{$\phantom{\Subalgebras\colon}$\ 2-dim}
$\sim A_{2.1}$:\
$\langle e_2+\varkappa e_4; e_1\rangle$
\\
$\phantom{\Subalgebras\colon}$\ 3-dim\ $\sim A_{2.1}\oplus A_{1}$:\
$\langle e_2, e_4; e_1  \rangle$
\\
\hphantom{$\phantom{\Subalgebras\colon}$\ 3-dim}
$\sim sl(2,\R)$:\
$\langle ;e_1, e_2, e_3  \rangle$

\bigskip

\noindent
{\mathversion{bold}$so(3)\oplus A_1$}: $[e_1,e_2]=e_3,\ [e_2,e_3]=e_1,\ [e_3,e_1]=e_2$ (unsolvable)%% so(3)+A1
\\[1ex]
$\Int\colon
\left(\begin{array}{cccc}
\alpha_{11}& \alpha_{12} & \alpha_{13} & 0 \\
\alpha_{21} & \alpha_{22} & \alpha_{23}& 0\\
\alpha_{31} & \alpha_{32} & \alpha_{33}& 0\\
0 & 0 & 0 & 1
\end{array}\right)$,
\
where the submatrix $(\alpha_{ij})_{i,j=\overline{1,3}}$
belongs to $SO(3)$
\\[1ex]
$\Aut\colon
\left(\begin{array}{cccc}
\alpha_{11}& \alpha_{12} & \alpha_{13} & 0 \\
\alpha_{21} & \alpha_{22} & \alpha_{23}& 0\\
\alpha_{31} & \alpha_{32} & \alpha_{33}& 0\\
0 & 0 & 0 & \!\!\alpha_{44}\!\!
\end{array}\right)$,
\
where the submatrix $(\alpha_{ij})_{i,j=\overline{1,3}}$
belongs to $SO(3)$
 \\[1ex]
$\Der\colon\, E_{12}-E_{21},\ E_{13}-E_{31},\ E_{32}-E_{23},\ E_{44}$
\\[1ex]
$\Megaideals\colon$\ $\langle e_4 \rangle$, \ $\langle ; e_1, e_2, e_3 \rangle$
\\
$\CharIdeals\colon$\ ---
\\
$\Ideals\colon$\ ---
\\
$\Subalgebras\colon$\
1-dim\ $\sim A_1$:\
$\langle e_4 \rangle$, \
$\langle e_1+\varkappa e_4 \rangle$, \
$\varkappa \ge 0$
\\
$\phantom{\Subalgebras\colon}$\ 2-dim\ $\sim 2A_1$:\
$\langle e_1, e_4 \rangle$
\\
$\phantom{\Subalgebras\colon}$\ 3-dim\ $\sim so(3)$:\
$\langle ;e_1, e_2, e_3  \rangle$

\bigskip

\noindent
{\mathversion{bold}$A_{4.1}$}: $[e_2,e_4]=e_1,\  [e_3,e_4]=e_2$ (indecomposable nilpotent) %%% A4.1
\\[1ex]
$\Int\colon
\left(\begin {array}{cccc}
1 &\theta_4 &\frac12\theta_4^2 & \theta_2 \\
0&1&\theta_4&\theta_3 \\
0&0&1&0\\
0&0&0&1\end {array}\right)$
\qquad
$\Aut\colon
\left(\begin{array}{cccc}
\alpha_{33}\alpha_{44}^2& \alpha_{23}\alpha_{44}& \alpha_{13}& \alpha_{14} \\
0 & \alpha_{33}\alpha_{44} & \alpha_{23}& \alpha_{24}\\
0&0&\alpha_{33}&\alpha_{34}\\
0 & 0 & 0 & \alpha_{44}\\
\end{array}\right)$
\\[1ex]
\noindent
$\Der\colon\, E_{13},\ E_{14},\ E_{12}+E_{23},\ E_{24},\ E_{11}+E_{22}+E_{33},\ E_{34},\ 2E_{11}+E_{22}+E_{44}$
\\[1ex]
$\Megaideals\colon$\ $\langle e_1 \rangle$, \
$\langle e_1, e_2 \rangle$, \ $\langle e_1, e_2, e_3 \rangle$
\\
$\CharIdeals\colon$\ ---
\\
$\Ideals\colon$\ $\langle e_4+\varkappa e_3, e_2; e_1 \rangle$
\\
$\Subalgebras\colon$\
1-dim\ $\sim A_1$:\
$\langle e_1 \rangle$, \
$\langle e_2 \rangle$, \
$\langle e_3+\varkappa e_1 \rangle$, \
$\langle e_4+\varkappa e_3 \rangle$
\\
$\phantom{\Subalgebras\colon}$\ 2-dim\ $\sim 2A_1$:\
$\langle e_1, e_2 \rangle$, \
$\langle e_1, e_3 \rangle$, \
$\langle e_2, e_3+\varkappa e_1 \rangle$, \
$\langle e_1, e_4+\varkappa e_3 \rangle$
\\
$\phantom{\Subalgebras\colon}$\ 3-dim\ $\sim 3A_1$:\
$\langle e_1, e_2, e_3  \rangle$
\\
\hphantom{$\phantom{\Subalgebras\colon}$\ 3-dim}
$\sim A_{3.1}$:\
$\langle e_4+\varkappa e_3, e_2; e_1 \rangle$

\bigskip

\noindent                                                                                      %%% A4.2^b
{\mathversion{bold}$A_{4.2}^b$}: $[e_1,e_4]=be_1,\  [e_2,e_4]=e_2,\  [e_3,e_4]=e_2+e_3,\ b\ne 0$
(indecomposable solvable)
\\[1ex]
$\Int\colon
\left(\begin {array}{cccc}
e^{b\theta_4} & 0 & 0 & \theta_1 \\
0& e^{\theta_4} &\theta_4e^{\theta_4}& \theta_2 +\theta_3\\
0&0&e^{\theta_4}& \theta_3\\
0&0&0&1\end {array}\right)$\\[1ex]
$b\ne 0,\ b\ne1\colon$                                                                             %%% A4.2
\\
$\Aut\colon
\left(\begin{array}{cccc}
\alpha_{11}& 0 & 0 & \alpha_{14} \\
0 & \alpha_{22} & \alpha_{23}& \alpha_{24}\\
0 & 0 & \alpha_{22} & \alpha_{34}\\
0 & 0 & 0 & 1\\
\end{array}\right)$
\\[1ex]
$\Der\colon\, E_{11},\ E_{14},\ E_{23},\ E_{24},\ E_{22}+E_{33},\ E_{34}$
\\[1ex]
$\Megaideals\colon$\ $\langle e_1 \rangle$, \ $\langle e_2 \rangle$, \
$\langle e_1, e_2 \rangle$, \ $\langle e_2, e_3 \rangle$, \ $\langle
e_1, e_2, e_3 \rangle$
\\
$\CharIdeals\colon$\ ---
\\
$\Ideals\colon$\ ---
\\
$\Subalgebras\colon$\
1-dim\ $\sim A_1$:\
$\langle e_1 \rangle$, \
$\langle e_2 \rangle$, \
$\langle e_4 \rangle$, \
$\langle e_3+\varkappa e_1 \rangle$, \
$\langle e_1+\varepsilon e_2 \rangle$
\\
$\phantom{\Subalgebras\colon}$\ 2-dim\ $\sim 2A_1$:\
$\langle e_1, e_2 \rangle$, \
$\langle e_1+\varkappa e_2, e_3 \rangle$, \
$\langle e_2, e_3 \rangle$, \
$\langle e_1+\varepsilon e_3, e_2 \rangle$
\\
\hphantom{$\phantom{\Subalgebras\colon}$\ 2-dim}
$\sim A_{2.1}$:\
$\langle e_4; e_1 \rangle$, \
$\langle e_4; e_2 \rangle$
\\
$\phantom{\Subalgebras\colon}$\ 3-dim\ $\sim 3A_1$:\
$\langle e_1, e_2, e_3  \rangle$
\\
\hphantom{$\phantom{\Subalgebras\colon}$\ 3-dim}
$\sim A_{3.2}$:\
$\langle e_4; e_2, e_3 \rangle$
\\
\hphantom{$\phantom{\Subalgebras\colon}$\ 3-dim}
$\sim A^a_{3.4}$:\
$\langle e_4; e_1, e_2 \rangle$
\\
The parameter $a$ in the algebra $A_{3.4}^{a}$ should satisfy condition:
$a=\left\{\begin{array}{ll} b, & -1\leq b <1, \ b\not=0,\\ \frac{1}{b}, &
|b|> 1 \end{array}\right.$\\[1ex]
$b=1\colon$                                                                              %%% A4.2
\\
$\Aut\colon
\left(\begin{array}{cccc}
\alpha_{11}& 0 & \alpha_{13}& \alpha_{14} \\
\alpha_{21} & \alpha_{22} & \alpha_{23}& \alpha_{24}\\
0 & 0 & \alpha_{22} & \alpha_{34}\\
0 & 0 & 0 & 1\\
\end{array}\right)$
\\[1ex]
\noindent
$\Der\colon\, E_{11},\ E_{13},\ E_{14},\ E_{21},\ E_{23},\ E_{24},\ E_{22}+E_{33},\ E_{34}$
\\[1ex]
$\Megaideals\colon$\ $\langle e_2 \rangle$, \ $\langle e_1, e_2
\rangle$, \ $\langle e_1, e_2, e_3 \rangle$
\\
$\CharIdeals\colon$\ ---
\\
$\Ideals\colon$\ $\langle e_1+\varkappa e_2\rangle$, \ $\langle e_2, e_3+\varkappa e_1 \rangle$
\\
$\Subalgebras\colon$\
1-dim\ $\sim A_1$:\
$\langle pe_1+qe_2 \rangle$, \
$\langle e_3+\varkappa e_1 \rangle$, \
$\langle e_4 \rangle$
\\
$\phantom{\Subalgebras\colon}$\ 2-dim\ $\sim 2A_1$:\
$\langle e_1, e_2 \rangle$, \
$\langle e_1+\varkappa e_2, e_3 \rangle$, \
$\langle e_2, e_3+\varkappa e_1 \rangle$
\\
\hphantom{$\phantom{\Subalgebras\colon}$\ 2-dim}
$\sim A_{2.1}$:\
$\langle e_4; pe_1+qe_2 \rangle$
\\
$\phantom{\Subalgebras\colon}$\ 3-dim\ $\sim 3A_1$:\
$\langle e_1, e_2, e_3 \rangle$
\\
\hphantom{$\phantom{\Subalgebras\colon}$\ 3-dim}
$\sim A_{3.2}$:\
$\langle e_4; e_2, e_3+\varkappa e_1 \rangle$
\\
\hphantom{$\phantom{\Subalgebras\colon}$\ 3-dim}
$\sim A_{3.3}$:\
$\langle e_4; e_1, e_2 \rangle$

\bigskip

\noindent
{\mathversion{bold}$A_{4.3}$}: $[e_1,e_4]=e_1,\  [e_3,e_4]=e_2$
(indecomposable solvable)                                                                     %%% A4.3
\\[1ex]
$\Int\colon
\left(\begin {array}{cccc}
e^{\theta_4} & 0 & 0 & \theta_1 \\
0 & 1 &\theta_4 & \theta_3 \\
0 & 0 & 1 & 0\\
0 & 0 & 0 & 1\end {array}\right)$
\qquad
$\Aut\colon
\left(\begin{array}{cccc}
\alpha_{11}& 0 & 0 & \alpha_{14} \\
0 & \alpha_{22} & \alpha_{23} & \alpha_{24}\\
0 & 0 & \alpha_{22} & \alpha_{34}\\
0 & 0 & 0 & 1\\
\end{array}\right)$
\\[1ex]
\noindent
$\Der\colon\, E_{11},\ E_{14},\ E_{23},\ E_{24},\ E_{22}+E_{33},\ E_{34}$
\\[1ex]
$\Megaideals\colon$\ $\langle e_1 \rangle$, \
$\langle e_2 \rangle$, \ $\langle e_1, e_2 \rangle$, \
$\langle e_2, e_3 \rangle$, \ $\langle e_1, e_2, e_3 \rangle$
\\
$\CharIdeals\colon$\ ---
\\
$\Ideals\colon$\ $\langle e_4+\varkappa e_3, e_2; e_1 \rangle$
\\
$\Subalgebras\colon$\
1-dim\ $\sim A_1$:\
$\langle e_1\rangle$, \
$\langle e_2\rangle$, \
$\langle e_1+\varepsilon e_2 \rangle$, \
$\langle e_3+\varkappa e_1 \rangle$, \
$\langle e_4+\varkappa e_3 \rangle$
\\
$\phantom{\Subalgebras\colon}$\ 2-dim\ $\sim 2A_1$:\
$\langle e_1, e_2 \rangle$, \
$\langle e_1+\varkappa e_2, e_3 \rangle$, \
$\langle e_2, e_3 \rangle$, \
$\langle e_2, e_3+\varepsilon e_1 \rangle$, \
$\langle e_2, e_4+\varkappa e_3 \rangle$
\\
\hphantom{$\phantom{\Subalgebras\colon}$\ 2-dim}
$\sim A_{2.1}$:\
$\langle e_4+\varkappa e_3; e_1 \rangle$
\\
$\phantom{\Subalgebras\colon}$\ 3-dim\ $\sim 3A_1$:\
$\langle e_1, e_2, e_3 \rangle$
\\
\hphantom{$\phantom{\Subalgebras\colon}$\ 3-dim}
$\sim A_{2.1}\oplus A_1$:\
$\langle e_4+\varkappa e_3, e_2; e_1 \rangle$
\\
\hphantom{$\phantom{\Subalgebras\colon}$\ 3-dim}
$\sim A_{3.1}$:\
$\langle e_3, e_4; e_2 \rangle$

\bigskip

\noindent
{\mathversion{bold}$A_{4.4}$}: $[e_1,e_4]=e_1,\  [e_2,e_4]=e_1+e_2,\  [e_3,e_4]=e_2+e_3$
(indecomposable solvable)                                                                         %%% A4.4
\\[1ex]
$\Int\colon
\left(\begin {array}{cccc}
e^{\theta_4} & \theta_4e^{\theta_4} & \frac12\theta_4^2e^{\theta_4} & \theta_1 +\theta_2\\
0 & e^{\theta_4} & \theta_4e^{\theta_4} & \theta_2+\theta_3 \\
0 & 0 & e^{\theta_4} & \theta_3 \\
0 & 0 & 0 & 1\end {array}\right)$
\qquad
$\Aut\colon
\left(\begin{array}{cccc}
\alpha_{11}& \alpha_{12} & \alpha_{13} & \alpha_{14} \\
0 & \alpha_{11} & \alpha_{12} & \alpha_{24}\\
0 & 0 & \alpha_{11} & \alpha_{34}\\
0 & 0 & 0 & 1\\
\end{array}\right)$
\\[1ex]
\noindent
$\Der\colon\, E_{11}+E_{22}+E_{33},\ E_{13},\ E_{14},\ E_{24},\ E_{12}+E_{23},\ E_{34}$
\\[1ex]
$\Megaideals\colon$\ $\langle e_1 \rangle$, \
$\langle e_1, e_2 \rangle$, \ $\langle e_1, e_2, e_3 \rangle$
\\
$\CharIdeals\colon$\ ---
\\
$\Ideals\colon$\ ---
\\
$\Subalgebras\colon$\
1-dim\ $\sim A_1$:\
$\langle e_1+\varkappa e_3 \rangle$, \
$\langle e_2 \rangle$, \
$\langle e_3 \rangle$, \
$\langle e_4 \rangle$
\\
$\phantom{\Subalgebras\colon}$\ 2-dim\ $\sim 2A_1$:\
$\langle e_1+\varkappa e_3, e_2 \rangle$, \
$\langle e_1, e_3 \rangle$, \
$\langle e_2, e_3 \rangle$
\\
\hphantom{$\phantom{\Subalgebras\colon}$\ 2-dim}
$\sim A_{2.1}$:\
$\langle e_4; e_1 \rangle$
\\
$\phantom{\Subalgebras\colon}$\ 3-dim\ $\sim 3A_1$:\
$\langle e_1, e_2, e_3 \rangle$
\\
\hphantom{$\phantom{\Subalgebras\colon}$\ 3-dim}
$\sim A_{3.2}$:\
$\langle e_4; e_1, e_2 \rangle$

\bigskip

\noindent
{\mathversion{bold}$A_{4.5}^{a,b,c}$}: $[e_1,e_4]=ae_1,\  [e_2,e_4]=be_2,\
 [e_3,e_4]=ce_3,\ abc\ne 0$
(indecomposable solvable)                                                                         %%% A4.5^a,b,c
\\[1ex]
$\Int\colon
\left(\begin {array}{cccc}
e^{a\theta_4} & 0 & 0 & \theta_1 \\
0 & e^{b\theta_4} & 0 & \theta_2 \\
0 & 0 & e^{c\theta_4} & \theta_3 \\
0 & 0 & 0 & 1\end {array}\right)$\\[1ex]
$-1\leq a<b<c=1$, \ $b>0$ if $a=-1$:                                                              %%% A4.5^1,b,c
\\[1ex]
$\Aut\colon
\left(\begin{array}{cccc}
\alpha_{11}& 0 & 0 & \alpha_{14} \\
0 & \alpha_{22} & 0 & \alpha_{24}\\
0 & 0 & \alpha_{33} & \alpha_{34}\\
0 & 0 & 0 & 1\\
\end{array}\right)$
\\[1ex]
\noindent
$\Der\colon\, E_{11},\ E_{14},\ E_{22},\ E_{24},\ E_{33},\ E_{34}$
\\[1ex]
$\Megaideals\colon$\ $\langle e_1 \rangle$, \ $\langle e_2 \rangle$, \
$\langle e_3 \rangle$, \ $\langle e_1, e_2 \rangle$, \ $\langle e_2, e_3
\rangle$, \ $\langle e_1, e_3 \rangle$, \ $\langle e_1, e_2, e_3 \rangle$
\\
$\CharIdeals\colon$\ ---
\\
$\Ideals\colon$\ ---
\\
$\Subalgebras\colon$\
1-dim\ $\sim A_1$:\
$\langle e_1 \rangle$, \ $\langle e_2 \rangle$, \
$\langle e_3 \rangle$, \ $\langle e_4 \rangle$, \
$\langle e_1+\varepsilon e_3 \rangle$, $\langle e_2+\varepsilon e_3 \rangle$, \
$\langle e_1+\varepsilon e_2+\varkappa e_3 \rangle$, \
 $\varkappa\ne 0$
\\
$\phantom{\Subalgebras\colon}$\ 2-dim\ $\sim 2A_1$:\
$\langle e_1, e_2 \rangle$, \ $\langle e_1, e_3 \rangle$, \
$\langle e_2, e_3 \rangle$, \
$\langle e_1, e_2+\varepsilon e_3 \rangle$, \
$\langle e_2, e_1+\varepsilon e_3 \rangle$, \
$\langle e_3, e_1+\varepsilon e_2 \rangle$,
\\
\hphantom{$\phantom{\Subalgebras\colon}$\ 2-dim\ $\sim 2A_1$:}
$\langle e_1+\varepsilon e_3, e_2+\varkappa e_3 \rangle$, \
$\varkappa\ne 0$
\\
\hphantom{$\phantom{\Subalgebras\colon}$\ 2-dim}
$\sim A_{2.1}$:\
$\langle e_4; e_1 \rangle$, \ $\langle e_4; e_2 \rangle$, \ $\langle e_4; e_3 \rangle$
\\
$\phantom{\Subalgebras\colon}$\ 3-dim\ $\sim 3A_{1}$:\
$\langle e_1, e_2, e_3 \rangle$
\\
\hphantom{$\phantom{\Subalgebras\colon}$\ 3-dim} $\sim
A_{3.4}^{v}$:\ $\langle e_4; e_1, e_3 \rangle^{*}$, \ $\langle e_4;
e_2, e_3 \rangle^{**}$, \ $\langle e_4; e_1, e_2 \rangle^{***}$
\\[1ex]
The parameter $v$ in the algebra $A_{3.4}^{v}$ should satisfy the
condition $v=a$ in the case *,  $v=b$ in the case~**
and $v=\left\{\begin{array}{ll} a{/}b, &  |a{/}b|< 1,\\
b{/}a, & |a{/}b|>1
\end{array}\right.$
in the case ***.\\[1ex]
$a=b=1$, $c\ne 1$:                                                              %%% A4.5^1,1,c
\\[1ex]
$\Aut\colon
\left(\begin{array}{cccc}
\alpha_{11} & \alpha_{12} & \alpha_{13} & \alpha_{14} \\
\alpha_{21} & \alpha_{22} & 0 & \alpha_{24}\\
0 & 0 & \alpha_{33} & \alpha_{34}\\
0 & 0 & 0 & 1\\
\end{array}\right)$
\\[1ex]
\noindent
$\Der\colon\, E_{11},\ E_{12},\ E_{14},\ E_{21},\ E_{22}, \ E_{24},\
 E_{33},\ E_{34}$
\\[1ex]
$\Megaideals\colon$\ $\langle e_3 \rangle$, \
$\langle e_1, e_2 \rangle$, \ $\langle e_1, e_2, e_3 \rangle$
\\
$\CharIdeals\colon$\ ---
\\
$\Ideals\colon$\ $\langle pe_1+qe_2 \rangle$, \ $\langle e_3,
pe_1+qe_2\rangle$
\\
$\Subalgebras\colon$\
1-dim\ $\sim A_1$:\
$\langle e_3 \rangle$, \
$\langle pe_1+qe_2 \rangle$, \
$\langle e_4 \rangle$, \
$\langle e_2+\varepsilon e_3 \rangle$, \
$\langle e_1+\varkappa e_2+\varepsilon e_3 \rangle$
\\
$\phantom{\Subalgebras\colon}$\ 2-dim\ $\sim 2A_1$:\
$\langle e_3, pe_1+qe_2 \rangle$, \
$\langle e_1, e_2 \rangle$, \
$\langle e_2, e_1+\varepsilon e_3 \rangle$, \
$\langle e_1+\varkappa e_2, e_2+\varepsilon e_3 \rangle$
\\
\hphantom{$\phantom{\Subalgebras\colon}$\ 2-dim}
$\sim A_{2.1}$:\
$\langle e_4; e_3 \rangle$, \
$\langle e_4; pe_1+qe_2 \rangle$
\\
$\phantom{\Subalgebras\colon}$\ 3-dim\ $\sim 3A_{1}$:\
$\langle e_1, e_2, e_3 \rangle$
\\
\hphantom{$\phantom{\Subalgebras\colon}$\ 3-dim}
$\sim A_{3.3}$:\
$\langle e_4; e_1, e_2 \rangle$
\\
\hphantom{$\phantom{\Subalgebras\colon}$\ 3-dim}
$\sim A_{3.4}^{c}$:\
$\langle e_4; e_3, pe_1+qe_2 \rangle$\\[1ex]
$a=b=c=1$:                                                                       %%% A4.5^1,1,1
\\[1ex]
$\Aut\colon
\left(\begin{array}{cccc}
\alpha_{11}& \alpha_{12} & \alpha_{13} & \alpha_{14} \\
\alpha_{21} & \alpha_{22} & \alpha_{23} & \alpha_{24}\\
\alpha_{31} & \alpha_{32} & \alpha_{33} & \alpha_{34}\\
0 & 0 & 0 & 1\\
\end{array}\right)$
\\[1ex]
\noindent
$\Der\colon\, E_{11},\ E_{12},\ E_{13},\ E_{14},\ E_{21},\ E_{22},\ E_{23},\ E_{24},\
E_{31},\ E_{32},\ E_{33},\ E_{34}$
\\[1ex]
$\Megaideals\colon$\ $\langle e_1, e_2, e_3 \rangle$
\\
$\CharIdeals\colon$\ ---
\\
$\Ideals\colon$\ $\langle e_1+\varkappa e_2+\gamma e_3 \rangle$, \
$\langle e_2+\varkappa e_3 \rangle$, \ $\langle e_3 \rangle$, \
$\langle e_1+\varkappa e_3, e_2+\gamma e_3 \rangle$,
\ $\langle e_1+\varkappa e_2, e_3 \rangle$, \
$\langle e_2, e_3 \rangle$
\\
$\Subalgebras\colon$\
1-dim\ $\sim A_1$:\
$\langle e_1+\varkappa e_2+\gamma e_3 \rangle$, \
$\langle e_2+\varkappa e_3 \rangle$, \
$\langle e_3 \rangle$, \ $\langle e_4 \rangle$
\\
$\phantom{\Subalgebras\colon}$\ 2-dim\ $\sim 2A_1$:\
$\langle e_1+\varkappa e_3, e_2+\gamma e_3 \rangle$, \
$\langle e_1+\varkappa e_2, e_3 \rangle$, \ $\langle e_2, e_3 \rangle$
\\
\hphantom{$\phantom{\Subalgebras\colon}$\ 2-dim}
$\sim A_{2.1}$:\
$\langle e_4; e_1+\varkappa e_2+\gamma e_3 \rangle$, \
$\langle e_4; e_2+\varkappa e_3 \rangle$, \
$\langle e_4; e_3 \rangle$
\\
$\phantom{\Subalgebras\colon}$\ 3-dim\ $\sim 3A_{1}$:\
$\langle e_1, e_2, e_3 \rangle$
\\
\hphantom{$\phantom{\Subalgebras\colon}$\ 3-dim}
$\sim A_{3.3}$:\
$\langle e_4; e_1+\varkappa e_3, e_2+\gamma e_3\rangle$, \
$\langle e_4; e_1+\varkappa e_2, e_3\rangle$, \
$\langle e_4; e_2, e_3\rangle$

\bigskip

\noindent
{\mathversion{bold}$A_{4.6}^{a,b}$}: $[e_1,e_4]=ae_1,\  [e_2,e_4]=be_2-e_3,\  [e_3,e_4]=e_2+be_3,\ a>0$
(indecomposable solvable)                                                                 %%% A4.6^a,b
\\[1ex]
$\Int\colon
\left(\begin {array}{cccc}
e^{a\theta_4} & 0 & 0 & a\theta_1 \\
0 & e^{b\theta_4}\cos{\theta_4} & e^{b\theta_4}\sin{\theta_4} & b\theta_2+\theta_3 \\
0 & -e^{b\theta_4}\sin{\theta_4} & e^{b\theta_4}\cos{\theta_4} & -\theta_2+b\theta_3 \\
0 & 0 & 0 & 1\end {array}\right)$ \qquad $\Aut\colon
\left(\begin{array}{cccc}
\alpha_{11}& 0 & 0 & \alpha_{14} \\
0 & \alpha_{22} & \alpha_{23} & \alpha_{24}\\
0 & -\alpha_{23} & \alpha_{22} & \alpha_{34}\\
0 & 0 & 0 & 1\\
\end{array}\right)$
\\[1ex]
\noindent
$\Der\colon\, E_{11},\ E_{14},\ E_{24},\ E_{34},\ E_{22}+E_{33},\ E_{23}- E_{32}$
\\[1ex]
$\Megaideals\colon$\ $\langle e_1 \rangle$, \
$\langle e_2, e_3 \rangle$, \ $\langle e_1, e_2, e_3 \rangle$
\\
$\CharIdeals\colon$\ ---
\\
$\Ideals\colon$\ ---
\\
$\Subalgebras\colon$\
1-dim\ $\sim A_1$:\
$\langle e_1+\varkappa e_3 \rangle$, \
$\langle e_3 \rangle$, \
$\langle e_4 \rangle$, \ $\varkappa\ge 0$
\\
$\phantom{\Subalgebras\colon}$\ 2-dim\ $\sim 2A_1$:\
$\langle e_1+\varkappa e_3, e_2 \rangle$, \
$\langle e_2, e_3 \rangle$, \ $\varkappa\ge 0$
\\
\hphantom{$\phantom{\Subalgebras\colon}$\ 2-dim}
$\sim A_{2.1}$:\
$\langle e_4; e_1 \rangle$
\\
$\phantom{\Subalgebras\colon}$\ 3-dim\ $\sim 3A_{1}$:\
$\langle e_1, e_2, e_3 \rangle$
\\
\hphantom{$\phantom{\Subalgebras\colon}$\ 3-dim}
$\sim A_{3.5}^{b}$:\
$\langle e_4; e_2, e_3 \rangle$

\bigskip

\noindent
{\mathversion{bold}$A_{4.7}$}: $[e_2,e_3]=e_1,\  [e_1,e_4]=2e_1,\
[e_2,e_4]=e_2,\  [e_3,e_4]=e_2+e_3$
(indecomposable solvable)                                                                %%% A4.7
\\[1ex]
$\Int\colon
\left(\begin {array}{cccc}
e^{2\theta_4} & -\theta_3 e^{\theta_4}
& -\theta_3\theta_4 e^{\theta_4}+\theta_2 e^{\theta_4} &
\theta_1+\theta_2\theta_3-\frac 12\theta_3^2 \\
0 & e^{\theta_4} & \theta_4e^{\theta_4} & \theta_2+\theta_3 \\
0 & 0 & e^{\theta_4} & \theta_3 \\
0 & 0 & 0 & 1\end {array}\right)$
\\[1ex]
$\Aut\colon
\left(\begin{array}{cccc}
\alpha_{33}^2& -\alpha_{33}\alpha_{34} & -m^{23}_{34} -\alpha_{33}\alpha_{34}
%\alpha_{33}(\alpha_{24}-\alpha_{34})-\alpha_{23}\alpha_{34}
& \alpha_{14} \\
0 & \alpha_{33} & \alpha_{23} & \alpha_{24}\\
0 & 0 & \alpha_{33} & \alpha_{34}\\
0 & 0 & 0 & 1\\
\end{array}\right)$
\\[1ex]
\noindent
$\Der\colon\, 2E_{11}+E_{22}+E_{33},\ E_{14},\ E_{12}+E_{13}-E_{34},\ E_{23},\ E_{13}+E_{24}$
\\[1ex]
$\Megaideals\colon$\ $\langle e_1 \rangle$, \
$\langle e_1, e_2 \rangle$, \ $\langle e_2, e_3; e_1 \rangle$
\\
$\CharIdeals\colon$\ ---
\\
$\Ideals\colon$\ ---
\\
$\Subalgebras\colon$\
1-dim\ $\sim A_1$:\
$\langle e_1 \rangle$, \
$\langle e_2 \rangle$, \
$\langle e_3 \rangle$, \
$\langle e_4 \rangle$
\\
$\phantom{\Subalgebras\colon}$\ 2-dim\ $\sim 2A_1$:\
$\langle e_1, e_2 \rangle$, \
$\langle e_1, e_3 \rangle$
\\
\hphantom{$\phantom{\Subalgebras\colon}$\ 2-dim}
$\sim A_{2.1}$:\
$\langle e_4; e_1 \rangle$, \
$\langle e_4; e_2 \rangle$
\\
$\phantom{\Subalgebras\colon}$\ 3-dim\ $\sim A_{3.1}$:\
$\langle e_2, e_3; e_1 \rangle$
\\
\hphantom{$\phantom{\Subalgebras\colon}$\ 3-dim}
$\sim A_{3.4}^{\frac12}$:\
$\langle e_4; e_1, e_2 \rangle$

\bigskip

\noindent
{\mathversion{bold}$A_{4.8}^{b}$}: $[e_2,e_3]=e_1,\  [e_1,e_4]=(1+b)e_1,\    %%% A4.8
[e_2,e_4]=e_2,\ [e_3,e_4]=be_3,\ |b|\leq 1$\ (indecomposable solvable)
\\[1ex]
$\Int\colon
\left(\begin {array}{cccc}
e^{(1+b)\theta_4} &-\theta_3 e^{\theta_4} &\theta_2 e^{b \theta_4} & (1+b){\theta_1}+b{\theta_2}{\theta_3} \\
0&e^{\theta_4}&0&\theta_2 \\
0&0&e^{b \theta_4}& b \theta_3 \\
0&0&0&1\end {array}\right)$

\bigskip

\noindent $|b|<1,\ b\ne 0\colon$                                               %%% A4.8 |b|<1 b<>0
\\[1ex]
$\Aut\colon \left(\begin{array}{cccc}
\alpha_{22}\alpha_{33}& \alpha_{12}& \alpha_{33}\alpha_{24}& \alpha_{14} \\
0 & \alpha_{22} & 0& \alpha_{24}\\
0&0&\alpha_{33}& -b\alpha_{12}\alpha_{22}^{-1}\\
0 & 0 & 0 & 1\\
\end{array}\right)$
\\[1ex]
\noindent $\Der\colon\, E_{12}-bE_{34},\ E_{14},\ E_{11}+E_{22},\
E_{13}+E_{24},\ E_{11}+E_{33}$
\\[1ex]
$\Megaideals\colon$\ $\langle e_1 \rangle$, \ $\langle e_1, e_2
\rangle$, \ $\langle e_1, e_3 \rangle$, \
$\langle e_2, e_3; e_1 \rangle$
\\
$\CharIdeals\colon$\ ---
\\
$\Ideals\colon$\ ---
\\
$\Subalgebras\colon$\ 1-dim\ $\sim A_1$:\ $\langle e_1 \rangle$, \
$\langle e_2 \rangle$, \ $\langle e_3 \rangle$, $\langle e_4\rangle$, \
$\langle e_2+\varepsilon e_3 \rangle$
\\
$\phantom{\Subalgebras\colon}$\ 2-dim\ $\sim 2A_1$:\ $\langle e_1,
e_2 \rangle$, \ $\langle e_1, e_3 \rangle$, \ $\langle e_1,
e_2+\varepsilon e_3 \rangle$
\\
\hphantom{$\phantom{\Subalgebras\colon}$\ 2-dim} $\sim A_{2.1}$:\
$\langle e_4; e_1 \rangle$, \ $\langle e_4; e_2 \rangle$, \ $\langle
e_4; e_3 \rangle$
\\
$\phantom{\Subalgebras\colon}$\ 3-dim\ $\sim A_{3.1}$:\ $\langle
e_2, e_3; e_1 \rangle$
\\
\hphantom{$\phantom{\Subalgebras\colon}$\ 3-dim} $\sim A_{3.4}^a$:\
$\langle e_4; e_1, e_2 \rangle^{*}$, \ $\langle e_4; e_1, e_3
\rangle^{**}$
\\[1ex]
The parameter $a$ in the algebra $A_{3.4}^{a}$ should satisfy the
conditions:\\[1ex]
$a=\left\{\begin{array}{ll} 1+b, & |1+b|< 1,\\
1{/}(1+b), & |1+b|>1 \end{array}\right.$ \quad in the case * \quad and\\[1ex]
$a=\left\{\begin{array}{ll} b{/}(1+b), & |b{/}(1+b)|<1\  \rm{or}\  b=-\frac12,\\
(1+b){/}b, & |b{/}(1+b)|>1
\end{array}\right.$\quad
in the case **.

\bigskip

\noindent
$b=0\colon$                                                      %%% A4.8 b=0
\\[1ex]
$\Aut\colon
\left(\begin{array}{cccc}
\alpha_{22}\alpha_{33}& \alpha_{12}& \alpha_{33}\alpha_{24}& \alpha_{14} \\
0 & \alpha_{22} & 0& \alpha_{24}\\
0&0&\alpha_{33}&0\\
0 & 0 & 0 & 1\\
\end{array}\right)$
\\[1ex]
\noindent $\Der\colon\, E_{12},\ E_{14},\ E_{11}+E_{22},\
E_{13}+E_{24},\ E_{11}+E_{33}$
\\[1ex]
This automorphism group can be obtained from the general case $|b|<1$, $b\ne 0$ by substitution $b=0$,
but these two cases can not be combined into one because of existence of additional (marked with *)
megaideal  in the case $b=0$.
\\[1ex]
$\Megaideals\colon$\ $\langle e_1 \rangle$, \
$\langle e_1, e_2 \rangle$, \ $\langle e_1, e_3 \rangle$, \
$\langle e_2, e_3; e_1 \rangle$, \ $\langle e_4; e_1, e_2 \rangle$*
\\
$\CharIdeals\colon$\ ---
\\
$\Ideals\colon$\ $\langle e_4+\varkappa e_3; e_1, e_2 \rangle$, \ $\varkappa\ne 0$
\\
$\Subalgebras\colon$\ 1-dim\ $\sim A_1$:\ $\langle e_1 \rangle$, \
$\langle e_2 \rangle$, \ $\langle e_3 \rangle$, \ $\langle
e_2+\varepsilon e_3 \rangle$, \ $\langle e_4+\varkappa e_3 \rangle$
\\
$\phantom{\Subalgebras\colon}$\ 2-dim\ $\sim 2A_1$:\ $\langle e_1,
e_2 \rangle$, \ $\langle e_1, e_3 \rangle$, \ $\langle e_1,
e_2+\varepsilon e_3 \rangle$, \ $\langle e_3, e_4 \rangle$
\\
\hphantom{$\phantom{\Subalgebras\colon}$\ 2-dim} $\sim A_{2.1}$:\
$\langle e_4+\varkappa e_3; e_1 \rangle$, \ $\langle e_4; e_2 \rangle$
\\
$\phantom{\Subalgebras\colon}$\ 3-dim\ $\sim A_{2.1}\oplus A_1$:\
$\langle e_3, e_4; e_1 \rangle$
\\
\hphantom{$\phantom{\Subalgebras\colon}$\ 3-dim}
$\sim A_{3.1}$:\
$\langle e_2, e_3; e_1 \rangle$
\\
\hphantom{$\phantom{\Subalgebras\colon}$\ 3-dim}
$\sim A_{3.2}$:\
$\langle e_4+\varkappa e_3; e_1, e_2 \rangle$, \ $\varkappa\ne 0$
\\
\hphantom{$\phantom{\Subalgebras\colon}$\ 3-dim}
$\sim A_{3.3}$:\
$\langle e_4; e_1, e_2 \rangle$

\bigskip

\noindent
$b=1\colon$                                                          %%% A4.8 b=1
\\[1ex]
$\Aut\colon$
$\left(\begin{array}{cccc}
m_{23}^{23} & -m_{24}^{23} & -m_{34}^{23} & \alpha_{14} \\
0& \alpha_{22}& \alpha_{23}& \alpha_{24}\\
0& \alpha_{32}& \alpha_{33}& \alpha_{34}\\
0& 0& 0& 1\\
\end{array}\right)$
\\[1ex]
\noindent $\Der\colon\, E_{12}-E_{34},\ E_{14},\ E_{11}+E_{22},\
E_{23},\ E_{13}+E_{24},\ E_{32},\ E_{11}+E_{33}$
\\[1ex]
$\Megaideals\colon$\ $\langle e_1 \rangle$, \ $\langle e_2, e_3; e_1 \rangle$
\\
$\CharIdeals\colon$\ ---
\\
$\Ideals\colon$\ $\langle e_1, pe_2+qe_3 \rangle$
\\
$\Subalgebras\colon$\
1-dim\ $\sim A_1$:\
$\langle e_1 \rangle$, \
$\langle pe_2+qe_3 \rangle$, \
$\langle e_4 \rangle$
\\
$\phantom{\Subalgebras\colon}$\ 2-dim\ $\sim 2A_1$:\
$\langle e_1, pe_2+qe_3 \rangle$
\\
\hphantom{$\phantom{\Subalgebras\colon}$\ 2-dim} $\sim A_{2.1}$:\
$\langle e_4; e_1 \rangle$, \ $\langle e_4; pe_2+qe_3 \rangle$
\\
$\phantom{\Subalgebras\colon}$\ 3-dim\ $\sim A_{3.1}$:\
$\langle e_2, e_3; e_1 \rangle$
\\
\hphantom{$\phantom{\Subalgebras\colon}$\ 3-dim}
$\sim A_{3.4}^{1/2}$:\
$\langle e_4; e_1, pe_2+qe_3 \rangle$

\bigskip

\noindent
$b=-1\colon$                           %%% A4.8 b=-1
\\[1ex]
$\Aut\colon$\ $\left(\begin{array}{cccc}
\alpha_{22}\alpha_{33}& \alpha_{22}\alpha_{34}& \alpha_{33}\alpha_{24}& \alpha_{14} \\
0 & \alpha_{22} & 0& \alpha_{24}\\
0&0&\alpha_{33}&\alpha_{34}\\
0 & 0 & 0 & 1\\
\end{array}\right)$,\quad
$\left(\begin{array}{cccc}
-\alpha_{32}\alpha_{23}& -\alpha_{32}\alpha_{24}& -\alpha_{23}\alpha_{34}& \alpha_{14} \\
0 & 0 & \alpha_{23} & \alpha_{24}\\
0& \alpha_{32}& 0& \alpha_{34}\\
0 & 0 & 0 & -1\\
\end{array}\right)$
\\[1ex]
\noindent $\Der\colon\, E_{12}+E_{34},\ E_{14},\ E_{11}+E_{22},\
E_{13}+E_{24},\ E_{11}+E_{33}$
\\[1ex]
$\Megaideals\colon$\ $\langle e_1 \rangle$, \ $\langle e_2, e_3; e_1 \rangle$
\\
$\CharIdeals\colon$\ $\langle e_1, e_3 \rangle$, \ $\langle e_1, e_2 \rangle$
\\
$\Ideals\colon$\ ---
\\
$\Subalgebras\colon$\
1-dim\ $\sim A_1$:\
$\langle e_1 \rangle$, \
$\langle e_2 \rangle$, \
$\langle e_2+\varepsilon e_3 \rangle$, \
$\langle e_3 \rangle$, \
$\langle e_4+\varkappa e_1 \rangle$
\\
$\phantom{\Subalgebras\colon}$\ 2-dim\ $\sim 2A_1$:\
$\langle e_1, e_2 \rangle$, \
$\langle e_1, e_3 \rangle$, \
$\langle e_1, e_2+\varepsilon e_3 \rangle$, \
$\langle e_4, e_1 \rangle$
\\
\hphantom{$\phantom{\Subalgebras\colon}$\ 2-dim} $\sim A_{2.1}$:\
$\langle e_4+\varkappa e_1; e_2 \rangle$, \ $\langle e_4+\varkappa
e_1; e_3 \rangle$
\\
$\phantom{\Subalgebras\colon}$\ 3-dim\ $\sim A_{2.1}\oplus A_{1}$:\
$\langle e_4, e_1; e_2 \rangle$, \
$\langle e_4, e_1; e_3 \rangle$
\\
\hphantom{$\phantom{\Subalgebras\colon}$\ 3-dim}
$\sim A_{3.1}$:\
$\langle e_2, e_3; e_1 \rangle$

\bigskip

\noindent
{\mathversion{bold}$A_{4.9}^{a}$}: $[e_2,e_3]=e_1,\  [e_1,e_4]=2ae_1,\  %%%% A4.9
[e_2,e_4]=ae_2-e_3,\ [e_3,e_4]=e_2+ae_3,\ a\ge 0$\\[1ex] (indecomposable solvable)
\\[1ex]
$\Int\colon
\left(\begin {array}{c@{\,\,\,\,}c@{\,\,\,\,}c@{\,\,\,\,}c}
e^{2a\theta_4} & -(\theta_2\sin\theta_4+\theta_3\cos\theta_4)e^{a\theta_4}
&(\theta_2\cos\theta_4-\theta_3\sin\theta_4)e^{a\theta_4} &
2a \theta_1+a\theta_2\theta_3-\frac 12\theta_2^2-\frac 12\theta_3^2
\\
0&\cos\theta_4 e^{a\theta_4}&\sin\theta_4 e^{a\theta_4}&a\theta_2+\theta_3\\
0&-\sin\theta_4 e^{a\theta_4}&\cos\theta_4 e^{a\theta_4}&-\theta_2+a\theta_3  \\
0&0&0&1\end {array}\right)$

\bigskip

\noindent
$a> 0\colon$          %%%% A4.9
\\[1ex]
$\Aut\colon
\left(\begin{array}{cccc}
s^{23}_{33} & (a^2+1)^{-1}(m_{34}^{23}-as_{34}^{23}) & -(a^2+1)^{-1}(am_{34}^{23}+s_{34}^{23}) & \alpha_{14} \\
0 & \alpha_{33} & \alpha_{23} & \alpha_{24}\\
0 & -\alpha_{23} & \alpha_{33} & \alpha_{34}\\
0 & 0 & 0 & 1\\
\end{array}\right)$
\\[1ex]
\noindent
$\Der\colon\, aE_{12}+E_{13}-(1+a^2)E_{34},\ E_{14},\ -E_{12}+E_{24}+aE_{34},\ -E_{23}+E_{32},\ 2E_{11}+E_{22}+E_{33}$
\\[1ex]
$\Megaideals\colon$\ $\langle e_1 \rangle$, \ $\langle e_2, e_3; e_1 \rangle$
\\
$\CharIdeals\colon$\ ---
\\
$\Ideals\colon$\ ---
\\
$\Subalgebras\colon$\
1-dim\ $\sim A_1$:\
$\langle e_1 \rangle$, \
$\langle e_2 \rangle$, \
$\langle e_4 \rangle$
\\
$\phantom{\Subalgebras\colon}$\ 2-dim\ $\sim 2A_1$:\
$\langle e_1, e_2 \rangle$
\\
\hphantom{$\phantom{\Subalgebras\colon}$\ 2-dim}
$\sim A_{2.1}$:\
$\langle e_4; e_1 \rangle$
\\
$\phantom{\Subalgebras\colon}$\ 3-dim\ $\sim A_{3.1}$:\
$\langle e_2, e_3; e_1 \rangle$

\bigskip

\noindent
$a=0\colon$                                                                        %%%% A4.9 a=0
\\[1ex]
$\Aut\colon
\left(\begin{array}{cccc}
\pm s_{33}^{23} & m_{34}^{23} & \mp s_{34}^{23} & \alpha_{14} \\
0 & \pm \alpha_{33} & \alpha_{23} & \alpha_{24}\\
0 & \mp \alpha_{23} & \alpha_{33} & \alpha_{34}\\
0 & 0 & 0 & \pm 1
\end{array}\right)$
%,\
%$\phantom{\Aut\colon}$
%$\left(\begin{array}{cccc}
%-s_{33}^{23} & m_{34}^{23} & s_{34}^{23} & \alpha_{14} \\
%0 & -\alpha_{33} & \alpha_{23} & \alpha_{24}\\
%0 & \alpha_{23} & \alpha_{33} & \alpha_{34}\\
%0 & 0 & 0 & -1\\
%\end{array}\right)$
\\[1ex]
\noindent
$\Der\colon\, E_{13}-E_{34},\ E_{14},\ -E_{12}+E_{24},\ -E_{23}+E_{32},\ 2E_{11}+E_{22}+E_{33}$
\\[1ex]
$\Megaideals\colon$\ $\langle e_1 \rangle$, \ $\langle e_2, e_3; e_1 \rangle$
\\
$\CharIdeals\colon$\ ---
\\
$\Ideals\colon$\ ---
\\
$\Subalgebras\colon$\
1-dim\ $\sim A_1$:\
$\langle e_1 \rangle$, \
$\langle e_2 \rangle$, \
$\langle e_4+\varkappa e_1 \rangle$
\\
$\phantom{\Subalgebras\colon}$\ 2-dim\ $\sim 2A_1$:\
$\langle e_1, e_2 \rangle$, \
$\langle e_1, e_4 \rangle$
\\
$\phantom{\Subalgebras\colon}$\ 3-dim\ $\sim A_{3.1}$:\
$\langle e_2, e_3; e_1 \rangle$

%\bigskip
\pagebreak

\noindent {\mathversion{bold}$A_{4.10}$}: $[e_1,e_3]=e_1,\
[e_2,e_3]=e_2,\  [e_1,e_4]=-e_2,\  [e_2,e_4]=e_1$
(indecomposable solvable)                                                        %%% A4.10
\\[1ex]
$\Int\colon
\left(\begin {array}{cccc}
\cos\theta_4 e^{\theta_3} & \sin\theta_4 e^{\theta_3} & \theta_1 & \theta_2 \\
-\sin\theta_4 e^{\theta_3} & \cos\theta_4 e^{\theta_3} &\theta_2 & -\theta_1 \\
0 & 0 & 1 & 0\\
0 & 0 & 0 & 1\end {array}\right)$
\qquad
$\Aut\colon
\left(\begin{array}{cccc}
\pm \alpha_{22}& \alpha_{12} & \alpha_{13} & \pm\alpha_{23} \\
\mp \alpha_{12} & \alpha_{22} & \alpha_{23} & \mp \alpha_{13}\\
0 & 0 & 1 & 0\\
0 & 0 & 0 & \pm 1
\end{array}\right)$
\\[1ex]
\noindent
$\Der\colon\, E_{11}+E_{22},\ E_{14}+E_{23},\ E_{13}-E_{24},\ E_{12}-E_{21}$
\\[1ex]
$\Megaideals\colon$\ $\langle e_1, e_2 \rangle$, \
$\langle e_2, e_3;  e_1\rangle$,
$\langle e_4; e_1, e_2 \rangle$
\\
$\CharIdeals\colon$\ $\langle e_4+\varkappa e_3; e_1, e_2 \rangle$, \ $\varkappa \ne 0$
\\
$\Ideals\colon$\ ---
\\
$\Subalgebras\colon$\
1-dim\ $\sim A_1$:\
$\langle e_1\rangle$, \
$\langle e_3\rangle$, \
$\langle e_4+\varkappa e_3 \rangle$
\\
$\phantom{\Subalgebras\colon}$\ 2-dim\ $\sim 2A_1$:\
$\langle e_1, e_2 \rangle$, \
$\langle e_3, e_4\rangle$
\\
\hphantom{$\phantom{\Subalgebras\colon}$\ 2-dim}
$\sim A_{2.1}$:\
$\langle e_3; e_1 \rangle$
\\
$\phantom{\Subalgebras\colon}$\ 3-dim\ $\sim A_{3.3}$:\
$\langle e_3; e_1, e_2 \rangle$
\\
\hphantom{$\phantom{\Subalgebras\colon}$\ 3-dim}
$\sim A_{3.5}^{|\varkappa|}$:\
$\langle e_4+\varkappa e_3; e_1, e_2 \rangle$

\end{document}